\documentclass[twoside,reqno]{article}
\usepackage{amsmath}
\usepackage{amsthm}
\usepackage{enumerate}
\usepackage{amssymb}
\usepackage{amsmath}
\usepackage{amsfonts}
\usepackage{bm} 

\usepackage{verbatim}
\usepackage{url}
\usepackage[latin1]{inputenc}

\usepackage{caption}
\usepackage{graphicx,color}
\usepackage{pdflscape}
\usepackage{float}
\usepackage[final]{pdfpages}

\usepackage[dvipdfm]{geometry}
\usepackage{multirow}

\usepackage{xr}
\usepackage{algorithmic}
\usepackage{algorithm}
\usepackage{listings}

\newcommand{\ud}{\mathrm{d}}
\newcommand{\bit}{\begin{itemize}}
\newcommand{\eit}{\end{itemize}}
\newcommand{\E}[1]{\operatorname{E}\left[ #1 \right]}
\newcommand{\Var}[1]{\operatorname{Var}\left[ #1 \right]}
\newcommand{\cov}[2]{\operatorname{Cov}\left[ #1,#2 \right]}

\newcommand{\Cov}[1]{\operatorname{Var}\left[ #1 \right]}

\newcommand{\ePotegObl}[1]{\left[\begin{array}{c|c} e^{#1 \mathbf{A}} & (e^{#1 \mathbf{A}}-\mathbf{I})\mathbf{
A}^{-1}\mathbf{B} \\ \hline \mathbf{0} & \mathbf{I} \end{array}\right]}
\newcommand{\ePotegOblT}[1]{\left[\begin{array}{c|c} e^{#1 \mathbf{A}^{T}} & \mathbf{0}  \\ \hline \mathbf{B}^
{T}\mathbf{A}^{-T}(e^{#1 \mathbf{A}^{T}}-\mathbf{I}) & \mathbf{I} \end{array}\right]}

\newcommand\independent{\protect\mathpalette{\protect\independenT}{\perp}}
\def\independenT#1#2{\mathrel{\rlap{$#1#2$}\mkern2mu{#1#2}}}

\def\be{\begin{equation}}
\def\ee{\end{equation}}
\def\bea{\begin{equation*}}
\def\eea{\end{equation*}}

\newtheorem{theo}{Theorem}[section]
\newtheorem{Lemma}[theo]{Lemma}

\newtheorem{prop}[theo]{Proposition}

\theoremstyle{definition}
\newtheorem{definition}[theo]{Definition}

\theoremstyle{remark}
\newtheorem{preremark}[theo]{Remark}
\newtheorem{preex}[theo]{Example}

\numberwithin{equation}{section}

\date{\today}
\author{Krzysztof Bartoszek}
\def\authorname{Krzysztof Bartoszek}

\def\longtitle{Multivariate Aspects of Phylogenetic Comparative Methods}

\def\keywords{\noindent{\bf Keywords:}
General Linear Model, Ornstein--Uhlenbeck process, Multivariate phylogenetic comparative method, Evolutionary model, Adaptation, Optimality,
Measurement error, Regression, Adaptation, Major-axis regression, Reduced major-axis regression, Structural equation, Allometry, Phylogenetic inertia.
}
\def\degree{Licentiate of Philosophy}
\def\division{Division of Mathematical Statistics}
\def\department{Department of Mathematical Sciences}
\def\cth{Chalmers University of Technology\\and University of Gothenburg}
\def\gbg{G\"oteborg, Sweden 2011}
\def\post{SE-412 96 G\"OTEBORG, Sweden}
\def\phone{Phone: +46 (0)31-772 10 00}
\def\authorsmail{krzbar@chalmers.se}
\def\printservice{Typeset with \LaTeX.\\\department\\Printed in G\"oteborg, Sweden 2011}

\begin{document}

%\frontmatter
\pagenumbering{roman}
{\setlength{\parindent}{0pt}
\thispagestyle{empty}

\begin{center}
{\sc Thesis for the Degree of \degree}

\vspace{4cm}

\parbox{\textwidth}{\center\LARGE\bf\longtitle}

\vspace{1.5cm}

{\Large\sc\authorname}

\end{center}

\vfill
\begin{center}
\begin{figure}[htbp]
\includegraphics[width=0.9\textwidth]{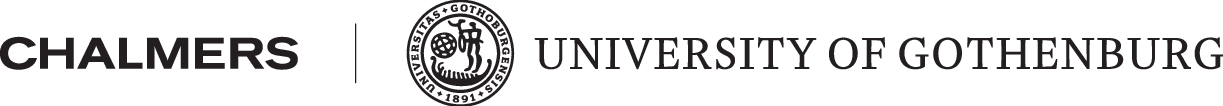}
\end{figure}
{\it\division}\\
{\it\department}\\
{\sc\cth}\\
\gbg
\end{center}

\pagebreak

\thispagestyle{empty}

{\bf{\small\longtitle}}\\
{\it \authorname}\\

\vspace{3mm}

Copyright \copyright\ \authorname, 2011.

\vspace{5mm}

ISSN 1652-9715\\
Report 2011:21\\

\vspace{3mm}

\vfill

\department \\
\division\\
\cth\\
\post\\
\phone

\vspace{\baselineskip}

Author e-mail: {\tt \authorsmail}
%{\bf Cover:}\\ 
%\coverpicture

\vspace{\baselineskip}

\printservice
}

\noindent
{\bf{\small\longtitle}}\\
{\it \authorname}\\

\begin{center}
{\bf Abstract}
\end{center}
This thesis concerns multivariate phylogenetic comparative methods. 
We investigate two aspects of them. The first is the bias caused by measurement error
in regression studies of comparative data. We calculate the formula for the bias and show
how to correct for it. We also study whether it is always
advantageous to correct for the bias as correction can increase 
the mean square error of the estimate. We propose a criterion,
which depends on the observed data, that indicates whether it is 
beneficial to correct or not. Accompanying the results 
is an \texttt{R} program that offers the bias correction tool.

The second topic is a multivariate model for trait
evolution which is based on an Ornstein--Uhlenbeck type stochastic process,
often used for studying trait adaptation, co--evolution, allometry or
trade--offs. 
Alongside the description of the 
model and presentation of its most important features
we present an \texttt{R} program estimating the model's parameters.
To the best of our knowledge our program is the first program
that allows for nearly all combinations of key model parameters 
providing the biologist with a flexible tool for studying
multiple interacting traits in the Ornstein--Uhlenbeck framework.
There are numerous packages 
available that include the Ornstein--Uhlenbeck process
but their multivariate capabilities seem limited.

\vfill
\keywords

\cleardoublepage

\begin{center}
{\bf Acknowledgment}
\end{center}
I would like to thank Petter Mostad, Olle Nerman and Serik Sagitov for their support
in writing this thesis and Patrik Albin, and Thomas Ericsson for many helpful comments.
~\\~\\
\noindent
I would also like to thank Staffan Andersson, Mohammad Asadzadeh, Jakob Augustin, Wojciech Bartoszek, Joachim Domsta, 
Peter Gennemark, Thomas F. Hansen, Micha\l\ Krze\-mi\'n\-ski, Pietro Li\'o, Jason Pienaar, Maria Prager, 
Ma\l gorzata Pu\l ka, Jaros\l aw Sko\-kow\-ski, Anil Sorathiya, Franciska Steinhoff, Anna Stokowska and 
Kjetil Voje for fruitful collaborations.
~\\~\\
\noindent
I finally thank my colleagues at the Department of Mathematical Sciences for a wonderful
working environment and my parents for their constant support.

\begin{flushright}
\authorname\\
G\"oteborg, \today
\end{flushright}

\cleardoublepage

\begin{center}
{\bf List of Papers}
\end{center}
The licenciate thesis is based on the following papers.
\begin{enumerate}[I.]
\item Hansen T. and \textbf{Bartoszek K.} (2011).
Interpreting the evolutionary regression: the interplay between observational
and biological errors in phylogenetic comparative studies. 
Systematic Biology \emph{in press}
\item \textbf{Bartoszek K.}, Pienaar J., Mostad P., Andersson S., Hansen T. (2011).
A comparative method for studying multivariate adaptation. \emph{Working paper}
\end{enumerate}

\cleardoublepage

\tableofcontents
%\listoffigures

%\mainmatter
\pagenumbering{arabic}
\setcounter{page}{2}

\section{Introduction}\label{secIntro}
In evolutionary biology one of the fundamental concerns  is how different inherited traits depend  on each
other and react to the changing environment. The natural approach to answer questions related to these problems
is to record trait measurements and environmental conditions, and look how they 
relate to each other, \emph{e.g.} via a regression analysis. 

There is however a problem 
that is not taken into account with just this approach. Namely, the sample is not independent as the different
species are related by a common evolutionary history. 
We assume throughout this thesis that 
the underlying phylogeny is fully known. This means that
we know the topology and branch lengths of the phylogenetic tree
describing the common evolutionary history of the species.

The immediate consequence of this common evolutionary history is 
that closely related species will have similar trait values. As it is very probable
that closely related species will be living in similar environments 
we could
observe also a false dependency of traits and environment. One needs to correct for the this common
phylogenetic history to be able to distinguish between effects stemming from similarity
and evolutionary effects. 
This is of particular importance when studying the adaptation of traits towards
environmental niches. Are species living in a similar environment similar because
they have adapted to it or is it because they are closely related?

In statistical terms forgetting about the inter--species dependency
means that we would be analyzing data  under a wrong model.
Such an analysis from the perspective of the correct model 
could lead to biased estimates and misleading parameter interpretations.
It would certainly result in
wrong confidence intervals and p--values. 

The thesis presented here does not focus on the biological motivations for comparative analysis
but on the mathematics and computations involved in estimating parameters of a multivariate
model for continuous comparative data. Biological background for the subject can be found in \emph{e.g.}
\cite{PHarMPag} which also discusses methodologies for discrete data and simple continuous settings.
Evolution of discrete characters is also discussed in \cite{MPag1994} or \cite{MPag19991}. 
Another
approach to studying comparative data is 
using general estimating equations, see \emph{e.g.} \cite{AGra1989}, 
\cite{TGarADicCJanJJon93}, \cite{MPag1993}
or \cite{EParJCla2002}. 

One of the underlying assumptions is that the different time scales match. This means that the
speed of losing the ancestral effects is not orders of magnitude faster than
species diversification. If it would be so then we would essentially have an independent 
sample with no need to take the phylogeny into account.
This problem is considered in \emph{e.g.} 
\cite{RFrePHarMPag2002} or \cite{MPag19992}. 
The book \cite{EPar2006} is about various \texttt{R} \cite{R} packages for analyzing data connected to phylogenies
including comparative data. 

This thesis is based on two papers. With each of the papers is 
an \texttt{R} program related to it.

Paper I which follows this thesis concerns the problem of measurement error 
(or observational variance, as most comparative studies take average values from a number
of individuals) in phylogenetic comparative analysis studies. The problem of measurement error
is widely addressed in the literature (see \emph{e.g.} \cite{JBuo} or \cite{WFul1987}) and in the multivariate
case well studied (\emph{e.g.} \cite{LGle1992}). However in the general case of dependent
errors in the predictor variables or dependently evolving predictor variables we didn't come
across any results in the literature. In Paper I we consider general covariance
matrices $\mathbf{V}_{d}$ and $\mathbf{V}_{u}$ relating the predictors and their measurement
errors respectively between observations. Also we mention that if one assumes some
structure on these matrices then substantial simplifications can be made in the bias
formula. In chapter \ref{secPapIA} of this thesis we consider different covariance structures 
resulting in the bias formulas
(\ref{eqMultIndepBias}),  (\ref{eqMultIndepPredBias}), (\ref{eqBiasIndepPredII}), (\ref{eqBiasIndepPredIII}). 

We also study in Paper I whether it is always
advantageous to correct for the bias, as correcting can increase
the mean square error of the estimate. We propose a criterion,
which depends on the observed data, that indicates whether it is
beneficial to correct or not. 
In chapter \ref{secMSEbias} of the thesis we 
look in more detail at conditions in the single predictor case for when it is better to 
correct, see equation (\ref{eq1predMSEdiff}). 
We also derive similar conditions for the multipredictor case.
We study how the situation changes when one introduces
fixed effects into the linear model, see equations (\ref{eqFnFbias}) and (\ref{eqFnFnMnFE}).
Lastly we see what can be said about the unconditional (of the observed with error
design matrix) bias in regression estimates.

Paper II presents and develops a multivariate model of trait
evolution which is based on an Ornstein--Uhlenbeck type stochastic process,
often used for studying trait adaptation, co--evolution, allometry or
trade--offs.
Alongside the description of the
model and discussion of its most important features is a software package
to estimate its parameters. 
We are not limited to looking at interactions between
traits and environment. Within the presented multivariate framework it
is possible to pose hypotheses about adaptation or trade--offs
which can be rigorously tested with the correction for the phylogeny.

The thesis itself is made up of eight chapters. 
The first is this one, the introduction. 
Chapter \ref{secReg} reminds the reader of the general linear regression model which
is fundamental to the findings of the two papers. Chapters \ref{secIntroPapI}
and \ref{secIntroPapII} are introductions to Paper I and Paper II, respectively. 
Possible directions of future development are outlined in chapter \ref{secFuture}.
Chapter \ref{secPapIA} contains detailed derivations concerning bias due to measurement error
for specific covariance structures in predictor variables and their measurement errors. 
Chapters \ref{secPapIIB} and \ref{secPapIIC} concern 
properties of the stochastic processes considered in Paper II and 
matrix parametrizations needed for the estimation
procedures in the software package accompanying Paper II.

In what follows we use the convention that matrices are written in bold--face, vectors as columns in normal font
with an arrow above (\emph{i.e.} $\vec{Z}$) and scalar values in normal font.
By $\mathbf{I}$ we denote the identity matrix of appropriate size. For a matrix $\mathbf{M}$, $\mathbf{M}^{T}$
denotes
the transpose of $\mathbf{M}$, $\mathbf{M}^{-1}$ the inverse and $\mathbf{M}^{-T}$ the transpose of the inverse of $\mathbf{M}$.
The Hadamard product between two matrices is denoted by $\ast$ and the Kronecker product by $\otimes$.
For a matrix $\mathbf{M}$ the operation $\mathrm{vec}(\mathbf{M})$ means
the vectorization, \emph{i.e.} stacking of columns onto each other, of the
matrix $\mathbf{M}$. The notation $\mathcal{N}(\vec{\mu},\mathbf{V})$ means a
normal distribution with mean vector $\vec{\mu}$ and covariance matrix $\mathbf{V}$.
For two random variables $X$ and $Y$ the notation $X \independent Y$ means that they
are independent.

We assume that the phylogenetic tree is rooted and has a time arrow attached to it. 
We use the word time to describe the branch lengths 
allowing for different biological interpretations of it.
On the phylogenetic tree, different times connected to species require distinct notation,
\begin{itemize}
\item $t_{i}$ is the time of the $i$--th species on the phylogeny (we do not need to assume anywhere an
ultrametric tree),
\item $t_{a_{ij}}$ is time of divergence of tip species $i$ and $j$,
\item $t_{1_{i}},t_{2_{i}},\ldots,t_{m_{i}}$ 
are the consecutive times of changes of niches on the lineage ending at tip species $i$,
with $t_{0_{i}}$ denoting the time of the root and $t_{m_{i}}=t_{i}$.
\end{itemize}
See Figure 1 in Paper II for the illustration of $t_{i}$ and $t_{a_{ij}}$ on the phylogenetic tree.

\newpage
\section{The general regression framework}\label{secReg}
The linear regression model can be considered to be one of the most common tools
to study relationships between variables. It has been studied
for over two hundred years now with Gauss describing the least squares method
at the end of the eighteenth century. Due to the variety of fields and situations
where it is applied a multitude of variations and particular cases of it have
been considered. The phylogenetic comparative methods field is no exception.

We begin by reminding the reader of the multivariate linear regression model,
\begin{displaymath}
\begin{array}{rcl}
\mathbf{Y} & = &\mathbf{X}\mathbf{B} + \bm{\mathcal{E}},
\end{array}
\end{displaymath}
where $\mathbf{X}:=[\vec{X}_{1};\ldots;\vec{X}_{n}]^{T}$, $\mathbf{Y}:=[\vec{Y}_{1};\ldots;\vec{Y}_{n}]^{T}$,
$n$ is the number of observations 
and $\mathrm{vec}(\bm{\mathcal{E}}) \sim \mathcal{N}(\mathbf{0},\mathbf{V})$ where
$\mathbf{V}=\mathbf{I} \otimes \mathbf{\Sigma}_{e}$. 
We assume that each observed $\vec{Y}_{i}$ is of length $k_{Y}$ and each observed
$\vec{X}_{i}$ is of length $k_{X}$ so that the matrix $\mathbf{B}$ has $k_{X}$ rows and $k_{Y}$ columns.
The $k_{X}$ predictor variables can be either continuous (\emph{e.g.} certain trait values) or categorical
(\emph{e.g.} different environments). 
To estimate the unknown $\mathbf{B}$ and $\mathbf{\Sigma}_{e}$ parameters the least squares (LS) estimation is 
most commonly used, which is,
\begin{equation}
\begin{array}{rcl}
\hat{\mathbf{B}} & = & (\mathbf{X}^{T}\mathbf{X})^{-1}\mathbf{X}^{T}\mathbf{Y} \\
\hat{\mathbf{\Sigma}}_{e} & = & \frac{1}{n-1}\sum_{i=1}^{n}(\vec{Y}_{i}-\hat{\mathbf{B}}\vec{X}_{i})(\vec{Y}_{i}-\hat{\mathbf{B}}\vec{X}_{i})^{T}.
\end{array}\label{eqLinReg}
\end{equation}

This estimator (after \cite{NDraHSmi}) is unbiased
\begin{displaymath}
\begin{array}{c}
\E{\hat{\mathbf{B}}}=\E{(\mathbf{X}^{T}\mathbf{X})^{-1}\mathbf{X}^{T}\mathbf{Y}}=
(\mathbf{X}^{T}\mathbf{X})^{-1}\mathbf{X}^{T}\mathbf{X}\mathbf{B} = \mathbf{B}.
\end{array}
\end{displaymath}
We will assume (to avoid cumbersome notation) for the moment that $\mathbf{\Sigma}_{e}=\sigma^{2}\mathbf{I}$
and then the variance of the estimate of the $j$--th column of $\mathbf{B}$ (since $\mathbf{\Sigma}_{e}$ is diagonal
this will be the same for each column) is,
\begin{displaymath}
\begin{array}{c}
\Var{\hat{\mathbf{B}_{\cdot j}}} = 
(\mathbf{X}^{T}\mathbf{X})^{-1}\mathbf{X}^{T}\sigma^{2}\mathbf{I}\mathbf{X}(\mathbf{X}^{T}\mathbf{X})^{-1}
=\sigma^{2}(\mathbf{X}^{T}\mathbf{X})^{-1}.
\end{array}
\end{displaymath}
The confidence bound for the $i$--th element of each column $j$ of $\mathbf{B}$ from equation (\ref{eqLinReg}) is 
$\hat{\mathbf{B}}_{ij} \pm t_{\alpha/2}S\sqrt{((\mathbf{X}^{T}\mathbf{X})^{-1})_{ii}}$,
where $S^{2}$ is the usual unbiased estimate for $\sigma^{2}$
and $t_{\alpha/2}$ is the appropriate quantile of the $t$ distribution with $k_{Y}\cdot n - k_{X}\cdot k_{Y}-1$
degrees of freedom.

The above description assumed that the different observations are independent of each other. This will not always
be the case and there are many applications where the noise terms are correlated, for example time series data. 
We denote the noise covariance structure by the matrix $\mathbf{V}$ and remind the reader, after \cite{BoxJen}, of
the theory in this situation.

It is known that the na\"ive least squares estimator, 
$\hat{\mathbf{B}}$ as in equation (\ref{eqLinReg}), is still an unbiased estimator of $\mathbf{B}$ 
even with arbitrary dependencies in the noise. However the 
second moment properties will not hold in this situation.
If $\mathbf{V}$ is known up to a scalar, $\sigma^{2}$, then one can use the generalized least squares (GLS) estimator,
\begin{displaymath}
\begin{array}{rcl}
\hat{\mathbf{B}} & = & (\mathbf{X}^{T}\mathbf{V}^{-1}\mathbf{X})^{-1}\mathbf{X}^{-1}\mathbf{V}^{-1}\mathbf{Y} \\
\hat{\sigma^{2}} & = & \frac{1}{n-k_{X}\cdot k_{Y}}(\vec{Y}-\mathrm{vec}(\mathbf{X}\hat{\mathbf{B}}))^{T}\mathbf{V}^{-1}(\vec{Y}-\mathrm{vec}(\hat{\mathbf{B}}\mathbf{X})),
\end{array}
\end{displaymath}
where $\vec{Y}:=\mathrm{vec}(\mathbf{Y})$
and this will have the exact same nice properties as the LS estimator. 

Following \cite{BoxJen} (see also \cite{NDraHSmi}) we assume that our data model is $\vec{Y} = \mathbf{D}\vec{\beta} + \vec{\epsilon}$, 
where 
the response vector $\vec{Y} \in \mathbb{R}^{n\cdot k_{Y}}$ (this corresponds to $\mathrm{vec}(\mathbf{Y})$
from the independent observations case),
the design matrix $\mathbf{D} \in \mathbb{R}^{(n\cdot k_{Y})\times k_{X}}$
($\mathbf{D}=\mathbf{X}\otimes \mathbf{I}$ in the independent observations case)
is of full rank $k_{X}$, the unknown coefficient vector 
$\vec{\beta} \in \mathbb{R}^{k_{Y}\cdot k_{X}}$  ($\vec{\beta}=\mathrm{vec}(\mathbf{B}^{T})$
in the independent observations case)
and the $\vec{\epsilon}$ vector 
is distributed as $\mathcal{N}(\vec{0},\mathbf{V})$, with the covariance matrix 
$\mathbf{V} \in \mathbb{R}^{(n\cdot k_{Y}) \times (n\cdot k_{Y})}$ known up to a positive constant $\sigma^{2}$
($\mathbf{V}=\mathbf{I} \otimes \mathbf{\Sigma}_{e}$ in the independent observations case). 
The aim is to estimate the unknown $\vec{\beta}$ vector
and via a linear transformation we can restate the problem as an ordinary least squares (OLS) one. 

As $\mathbf{V}^{-1}$ is
symmetric positive definite it can be decomposed by the Cholesky decomposition as
$\mathbf{V}^{-1}=\mathbf{P}\mathbf{P}^{T}/\sigma^{2}$, where $\mathbf{P}$ is lower--triangular. Then we can write,
\begin{displaymath}
\begin{array}{rcl}
\mathbf{P}^{T}\vec{Y} & = & \mathbf{P}^{T}\mathbf{D}\vec{\beta} + \mathbf{P}^{T}\vec{\epsilon},
\end{array}
\end{displaymath}
and denoting $\vec{Y}^{\ast}:=\mathbf{P}^{T}\vec{Y}$,  $\mathbf{D}^{\ast}:=\mathbf{P}^{T}\mathbf{D}$ 
and $\vec{\epsilon}~^{\ast}:=\mathbf{P}^{T}\vec{\epsilon}$
we get
\begin{displaymath}
\begin{array}{rcl}
\vec{Y}^{\ast} & = & \mathbf{D}^{\ast}\vec{\beta} + \vec{\epsilon}~^{\ast}.
\end{array}
\end{displaymath}
As now, $\Cov{\vec{\epsilon}~^{\ast}}=\Cov{\mathbf{P}^{T}\vec{\epsilon}~}
=\mathbf{P}^{T}\mathbf{V}\mathbf{P}=\mathbf{P}^{T}\mathbf{P}^{-T}\mathbf{P}^{-1}\mathbf{P}=\mathbf{I}$,
so $\vec{\epsilon}~^{\ast}\sim \mathcal{N}(\vec{0},\mathbf{I})$
and this means that the noise terms and thereby the entries of the transformed response vector $\vec{Y}^{\ast}$
are independent.
We get that the estimate of $\vec{\beta}$ is,
\begin{displaymath}
\begin{array}{rcl}
\hat{\vec{\beta}} & = & (\mathbf{D}^{\ast^{T}}\mathbf{D}^{\ast})^{-1}\mathbf{D}^{\ast^{T}}\vec{Y}^{\ast} \\
& = & (\mathbf{D}\mathbf{P}\mathbf{P}^{T}\mathbf{X})^{-1}\mathbf{D}\mathbf{P}\mathbf{P}^{T}\vec{Y}\\
\end{array}
\end{displaymath}
which gives the formula for the generalized least squares estimator,
\begin{equation}\label{eqGLS}
\begin{array}{rcl}
\hat{\vec{\beta}} & = & (\mathbf{D}\mathbf{V}^{-1}\mathbf{D})^{-1}\mathbf{D}\mathbf{V}^{-1}\vec{Y}.
\end{array}
\end{equation}
To estimate $\sigma^{2}$,
\begin{displaymath}
\begin{array}{rcl}
\hat{\sigma^{2}} & = & \frac{1}{n-k_{X}\cdot k_{Y}}(\vec{Y}^{\ast}-\mathbf{D}^{\ast}\hat{\vec{\beta}})^{T}(\vec{Y}^{\ast}-\mathbf{D}^{\ast}\hat{\vec{\beta}})\\
& = & \frac{1}{n-k_{X}\cdot k_{Y}}(\mathbf{P}^{T}\vec{Y}-\mathbf{P}^{T}\mathbf{D}\hat{\vec{\beta}})^{T}(\mathbf{P}^{T}\mathbf{Y}-\mathbf{P}^{T}\mathbf{D}\hat{\vec{\beta}})\\
& = & \frac{1}{n-k_{X}\cdot k_{Y}}(\vec{Y}-\mathbf{D}\hat{\vec{\beta}})^{T}\mathbf{P}\mathbf{P}^{T}(\vec{Y}-\mathbf{D}\hat{\vec{\beta}})\\
& = & \frac{1}{n-k_{X}\cdot k_{Y}}(\vec{Y}-\mathbf{D}\hat{\vec{\beta}})^{T}\mathbf{V}^{-1}(\vec{Y}-\mathbf{D}\hat{\vec{\beta}})
\end{array}
\end{displaymath}
and all the properties can be derived in the same way,
\begin{displaymath}
\begin{array}{rcl}
\E{\hat{\vec{\beta}}} & = & \vec{\beta} \\
\Var{\hat{\vec{\beta}}} & = & \sigma^{2}(\mathbf{D}^{T}\mathbf{V}^{-1}\mathbf{D})^{-1}
\end{array}
\end{displaymath}
with the confidence bounds for the $i$--th element of $\hat{\vec{\beta}}$ being, \\
$\hat{\vec{\beta}}_{i} \pm t_{\alpha/2}\sqrt{\hat{\sigma^{2}}}\sqrt{((\mathbf{D}^{T}\mathbf{V}^{-1}\mathbf{D})^{-1})_{ii}}$,
where $t_{\alpha/2}$ is the appropriate point from the $t$ distribution with $n\cdot k_{Y} - k_{Y}\cdot k_{X}-1$
degrees of freedom.

In phylogenetic comparative analysis we run into exactly this issue that our data points
are dependent. If we are doing a phylogenetic regression then the noise
covariance structure $\mathbf{V}$ is non--diagonal. The covariance matrix $\mathbf{V}$ will
depend on the phylogeny but it is not obvious how. To deal with this one needs
to assume some sort of stochastic model of evolution. If for example we assume 
a Brownian motion model of evolution then $\mathbf{V}=\mathbf{T}\otimes(\mathbf{\Sigma}\mathbf{\Sigma}^{T})$
where $\mathbf{T}$ is the matrix of divergence times of species on the phylogeny,
and $\mathbf{\Sigma}$ the diffusion matrix, see also chapter \ref{secBM} and \cite{JFel85}.
Using the earlier introduced notation the $i$--th, $j$--th entry of $\mathbf{T}$ will equal
$t_{a_{ij}}$.

\newpage
\section{Presentation of Paper I}\label{secIntroPapI}
Paper I concerns the issue of correcting for measurement error 
in regression where the predictor variables and measurement errors can
be correlated in an arbitrary way. 
The bias caused by measurement error in regression is a well studied
topic in the case of independent observations. However as we show in the 
Paper I everything becomes more complicated when the predictor variables
are dependent. If we allow for the measurement errors to be dependent
between observations we get an even more complicated situation. 

In comparative studies both cases are important. The predictor 
variables are usually also species' traits evolving on the phylogeny.
The ``measurement errors'' are often due to intra--species
variability. The most common trait data in a comparative analysis are species
means from a number of observations. Attached to these means is the
variability which can be treated as ``measurement error''. As the species
are related by their phylogeny the intra--species variability should be too.
Whether in practice this will be possible
to quantify is another matter but we provide a theoretical framework
and a method to deal with this. 
Included with Paper I is a short \texttt{R} script, \texttt{GLSME.R} 
(\texttt{G}eneral \texttt{L}east \texttt{S}quares with \texttt{M}easurement \texttt{E}rror) 
that implements the described bias correction.  The code
is available for download from 
http://www.math.chalmers.se/$\sim$krzbar/GLSME

In the next section we present the description of the general model considered in Paper I.

\subsection{The measurement error bias}\label{secMEbias}
In Paper I we use the following
notation: A variable with subscript $_{o}$ will denote an observed (with error)
variable, while the subscript $_{t}$ will mean the true, unobserved value of the variable.
We consider the general linear model,
\begin{displaymath}
\begin{array}{rclc}
\vec{Y}_{t} & = & \mathbf{D}_{t}\vec{\beta} + \vec{r}_{t}, & \vec{r}_{t} \sim \mathcal{N}(\vec{0},\mathbf{V}_{t}),
\end{array}
\end{displaymath}
where $\vec{Y}_{t}$ is a vector of $n$ observations of the dependent variable, $\vec{\beta}$ is a
vector of $m$ parameters to be estimated, $\mathbf{D}_{t}$ is an $n \times m$ design matrix 
and $\vec{r}_{t}$ is a vector of $n$ noise terms with $n\times n$ covariance
matrix $\mathbf{V}_{t}$. 
Due to the correlation in the noise for estimating $\beta$ we use
the generalized least squares estimator of equation (\ref{eqGLS}).
To the above general linear model we want to introduce an error measurement model. 
We can write the model with errors in the design matrix and the response variables as,
\begin{displaymath}
\begin{array}{rcl}
\mathbf{D}_{o} & = & \mathbf{D}_{t}+ \mathbf{U} \\
\vec{Y}_{o} & = & \vec{Y}_{t}+ \vec{e}_{y},
\end{array}
\end{displaymath}
where $\mathbf{U}$ is a $n\times m$ matrix of random observation errors in the elements of $\mathbf{D}_{t}$,
and $\vec{e}_{y}$ is a vector of length $n$ of observation errors in $\vec{Y}_{t}$. Each column
of $\mathbf{U}$ is a vector of observation errors for a predictor variable. Furthermore
we assume that $\mathrm{vec}(\mathbf{D}_{t})$ and $\mathrm{vec}(\mathbf{U})$ are normal random vectors, 
\emph{i.e.} $\mathrm{vec} (\mathbf{D}_{t}) \sim \mathcal{N}(\vec{0},\mathbf{V}_{d})$ and 
$\mathrm{vec} (\mathbf{U}) \sim \mathcal{N}(\vec{0},\mathbf{V}_{u})$. 
To simplify the derivations we assume that $\mathbf{D}_{t}$ has zero--mean.
Summing up, the regression with observation error model is the following,
\begin{displaymath}
\begin{array}{rclcc}
\vec{Y}_{t} & = & \mathbf{D}_{t}\vec{\beta} + \vec{r}_{t} 
& \mathrm{vec}(\mathbf{D}_{t}) \sim \mathcal{N}(\vec{0},\mathbf{V}_{d}), & \vec{r}_{t} \sim \mathcal{N}(\vec{0},\mathbf{V}_{t}) \\
\vec{Y}_{o} & = & \vec{Y}_{t} + \vec{e}_{y} & \vec{e}_{y} \sim \mathcal{N}(\vec{0},\mathbf{V}_{e}) \\
\mathbf{D}_{o} & = & \mathbf{D}_{t} + \mathbf{U} & \mathrm{vec}(\mathbf{U}) \sim \mathcal{N}(\vec{0},\mathbf{V}_{u}) \\
\vec{Y}_{o} & = & (\mathbf{D}_{o}-\mathbf{U})\vec{\beta} + \vec{r}_{t}+\vec{e}_{y}. 
\end{array}
\end{displaymath}

We assume that all the errors are independent of each other and the other variables,
\emph{i.e.} $\vec{r}_{t} \independent \vec{e}_{y}$, $\vec{r}_{t} \independent \mathbf{U}$,
$\mathbf{U} \independent \vec{e}_{y}$, $\mathbf{U} \independent \mathbf{D}_{t}$ and 
$\vec{Y}_{t} \independent \vec{e}_{y}$. We can write the last equation as, 
$\vec{Y}_{o}  =  (\mathbf{D}_{o}-\mathbf{U})\vec{\beta} + \vec{r}$,
where $\vec{r}=\vec{r}_{t}+\vec{e}_{y}$, so
$\vec{r} \sim \mathcal{N}(\vec{0},\mathbf{V})$,
where $\mathbf{V}=\mathbf{V}_{t}+\mathbf{V}_{e}$.
In Paper I we write the noise as $\vec{r}=\vec{r}_{t}+\vec{e}_{y} - \mathbf{U}\vec{\beta}$
and consider the regression model $\vec{Y}_{o}  =  \mathbf{D}_{o}\vec{\beta} + \vec{r}$ but 
here we do not do this so that we do not have to consider iterative procedures.
We do not assume any structure (in particular diagonality) of the covariance matrices $\mathbf{V}$, $\mathbf{V}_{t}$, 
$\mathbf{V}_{e}$, $\mathbf{V}_{d}$ or $\mathbf{V}_{u}$. 
The generalized least squares estimator of $\vec{\beta}$ 
from equation (\ref{eqGLS}) can be written, as
\begin{displaymath}
\begin{array}{rcl}  
\hat{\vec{\beta}}&=&(\mathbf{D}_{o}^{T}\mathbf{V}^{-1}\mathbf{D}_{o})^{-1}\mathbf{D}_{o}^{T}\mathbf{V}^{-1}((\mathbf{D}_{o}-\mathbf{U})\vec{\beta}+\vec{r})
\\&=&
(\mathbf{D}_{o}^{T}\mathbf{V}^{-1}\mathbf{D}_{o})^{-1}\mathbf{D}_{o}^{T}\mathbf{V}^{-1}(\mathbf{D}_{t}\vec{\beta}+\vec{r}).
\end{array}
\end{displaymath}

We want to compute the expectation of this estimator conditional on the observed predictor variables $\mathbf{D}_{o}$.
We assumed $\vec{r}$ had zero mean and is independent of $\mathbf{D}_{o}$ so we have,
\begin{displaymath}
\begin{array}{rcl}
\E{\hat{\vec{\beta}} \vert \mathbf{D}_{o}} 
&=&
(\mathbf{D}_{o}^{T}\mathbf{V}^{-1}\mathbf{D}_{o})^{-1}\mathbf{D}_{o}^{T}\mathbf{V}^{-1}\E{\mathbf{D}_{t} \vert \mathbf{D}_{o}}\vec{\beta} 
 \\ &=&
(\mathbf{D}_{o}^{T}\mathbf{V}^{-1}\mathbf{D}_{o})^{-1}\mathbf{D}_{o}^{T}\mathbf{V}^{-1}\E{\mathbf{D}_{o} -\mathbf{U} \vert \mathbf{D}_{o}}\vec{\beta}
 \\ &=&
(\mathbf{I}-(\mathbf{D}_{o}^{T}\mathbf{V}^{-1}\mathbf{D}_{o})^{-1}\mathbf{D}_{o}^{T}\mathbf{V}^{-1}\E{\mathbf{U} \vert \mathbf{D}_{o}})\vec{\beta}.
\end{array}
\end{displaymath}

It remains to calculate $\E{\mathbf{U} \vert \mathbf{D}_{o}}$, 
to do this we need to work with its vectorized form, 
$\E{\mathrm{vec}(\mathbf{U}) \vert \mathbf{D}_{o}}$.
Using common facts about the multivariate normal distribution
and that $\mathbf{D}_{o}=\mathbf{D}_{t}+\mathbf{U}$ where $\mathbf{D}_{t}$, $\mathbf{U}$
have zero mean and are independent of each other we derive,
\begin{displaymath}
\begin{array}{ll}
&\E{\mathrm{vec}(\mathbf{U}) \vert \mathbf{D}_{o}} \\
=&\E{\mathrm{vec}(\mathbf{U})}+\cov{\mathrm{vec}(\mathbf{U})}{\mathrm{vec}(\mathbf{D}_{o})}
\Cov{\mathrm{vec}(\mathbf{D}_{o})}^{-1}(\mathrm{vec}(\mathbf{D}_{o})-\E{\mathrm{vec}(\mathbf{D}_{o})})\\
=&\Cov{\mathrm{vec}(\mathbf{U})}\Cov{\mathrm{vec}(\mathbf{D}_{o})}^{-1}\mathrm{vec}(\mathbf{D}_{o})\\
=&\mathbf{V}_{u}\mathbf{V}_{o}^{-1}\mathrm{vec}(\mathbf{D}_{o}),
\end{array}
\end{displaymath}
where $\mathbf{V}_{o}=\mathbf{V}_{d}+\mathbf{V}_{u}$.
(In setup of the model of \cite{THanJPieSOrzSLOUCH} $\mathbf{V}_{d}$ would be the covariance matrix resulting from the
assumed Brownian--motion of the predictors).
Therefore we have that,
\begin{equation}
\E{\hat{\vec{\beta}} \vert \mathbf{D}_{o}} =
\mathbf{K}\vec{\beta},
\label{eqBias}
\end{equation}
where the so--called reliability matrix $\mathbf{K}$ is given by
\begin{equation}
\mathbf{K} =
\mathbf{I}-(\mathbf{D}_{o}^{T}\mathbf{V}^{-1}\mathbf{D}_{o})^{-1}\mathbf{D}_{o}^{T}\mathbf{V}^{-1}
\mathrm{vec}^{-1}(\mathbf{V}_{u}\mathbf{V}_{o}^{-1}\mathrm{vec}(\mathbf{D}_{o}),
\label{eqK}
\end{equation}
where $\mathrm{vec}^{-1}(\mathbf{v})$ for a vector $\mathbf{v}$ means the inverse
of the $\mathrm{vec}$ operation with appropriate matrix size.

The above is a general formula for any covariance matrices $\mathbf{V}_{o}$ and $\mathbf{V}_{u}$.
If we impose some structure on them then it will simplify.
No assumptions are needed on the covariance matrices $\mathbf{V}_{t}$ 
and $\mathbf{V}_{e}$. If we write out the matrix multiplications elementwise we will see that 
all the predictors influence the bias of all the others. However the directions and magnitudes of these biases
depend on the entries of $\mathbf{D}_{o}$ and the covariance matrices and so they can be arbitrary.

We write below some special cases of $\mathbf{V}_{o}$ and $\mathbf{V}_{u}$ 
where $\mathbf{K}$ simplifies with the details behind them in section \ref{secPapIA}.
\\
\emph{Independent observations of predictors} where $\mathbf{\Sigma}_{d}$ is the covariance 
matrix of the true (unobserved) predictors in a given observation and $\mathbf{\Sigma}_{u}$
is the covariance matrix of the measurement errors in predictors in a given observation
\begin{equation}\label{eqMultIndepBias}
\mathbf{K} = (\mathbf{\Sigma}_{d}+\mathbf{\Sigma}_{u})^{-1}\mathbf{\Sigma}_{d}.
\end{equation}
If the predictor is one dimensional instead of $\mathbf{K}$, $\mathbf{\Sigma}_{d}$, $\mathbf{\Sigma}_{u}$
we use $\kappa$, $\sigma^{2}_{d}$ and $\sigma^{2}_{u}$ respectively and the above becomes,
\begin{equation}\label{eqSingIndepBias}
\kappa = \frac{\sigma^{2}_{d}}{\sigma^{2}_{d}+\sigma^{2}_{u}}.
\end{equation}
\\
\emph{Independent predictors} (\emph{i.e.} the predictors and their errors are independent
between each other but not between observations), now
$\mathbf{\Sigma}_{d_{i}}$ is the covariance 
matrix of the true (unobserved) $i$th predictor between observations and $\mathbf{\Sigma}_{u_{i}}$
the covariance matrix of the measurement error and $\mathbf{d}_{o_{i}}$ denotes the $i$th column of $\mathbf{D}_{o}$,
\begin{equation}\label{eqMultIndepPredBias}
\mathbf{K} = \mathbf{I}-
(\mathbf{D}_{o}^{T}\mathbf{V}^{-1}\mathbf{D}_{o})^{-1}\mathbf{D}_{o}^{T}\mathbf{V}^{-1}
\left[\mathbf{\Sigma}_{u_{1}}\mathbf{\Sigma}_{o_{1}}^{-1}\mathbf{d}_{o_{1}};
\ldots ;\mathbf{\Sigma}_{u_{m}}\mathbf{\Sigma}_{o_{m}}^{-1}\mathbf{d}_{o_{m}}
\right].
\end{equation}
Further simplifications of this case are possible if one assumes independence of errors or proportionality 
between all of the $\mathbf{\Sigma}_{u_{i}}$ matrices and these are shown in section \ref{secPapIA}.
\\
\emph{Single predictor}
\begin{equation}\label{eqSingBias}
\mathbf{K} = \mathbf{I}-(\mathbf{D}_{o}^{T}\mathbf{V}^{-1}\mathbf{D}_{o})^{-1}\mathbf{D}_{o}^{T}\mathbf{V}^{-1}\mathbf{V}_{u}\mathbf{V}_{o}^{-1}\mathbf{D}_{o}
\end{equation}

\subsection{Mean square error analysis}
In the case of a single predictor equation with independent observations 
it is well known that $\kappa \in (-1,1)$, 
see equation (\ref{eqSingIndepBias}).
Therefore the bias corrected estimator has a larger variance than the uncorrected one. This immediately
implies a trade--off if we use the mean square error, $\E{(\hat{\beta}-\beta)^{2}}$, 
to compare estimators. 

In Paper I we study how the situation looks with dependent 
observations of a single predictor. We transform the formulae for the mean square
errors of both estimators into a criterion in terms of estimable quantities,
$\sigma^{2}_{\beta}/\kappa^{2} < \sigma^{2}_{\beta}+(\kappa-1)^{2}\beta^{2}$,
where $\sigma^{2}_{\beta}$ is the 
variance of the estimator of $\beta$.

Unlike in the independent observations case $\kappa$ can take values outside
the interval $(-1,1)$. This results in a much more complicated situation
as is described in section \ref{sbsecSingPredMSE} of this thesis.

In the supplementary material to Paper I we present simulation results
that show that this criterion works well even when all of its terms are estimated from the data.

\newpage
\section{Presentation of Paper II}\label{secIntroPapII}
Paper II is devoted to a multivariate 
stochastic differential equation model of trait evolution.
With Paper II is an
\texttt{R} software package,
\texttt{mvSLOUCH} (\texttt{m}ulti\texttt{v}ariate \texttt{S}tochastic \texttt{L}inear \texttt{O}rnstein--\texttt{U}hlenbeck 
models for phylogenetic \texttt{C}omparative \texttt{H}ypothesis) which 
covers nearly all cases possible in the framework of \cite{THan97}\cite{EMarTHan96}\cite{EMarTHan97}
(with the exception of some cases where the drift matrix
is singular or does not have an eigendecomposition).
The package will be available for download from \\
http://www.math.chalmers.se/$\sim$krzbar/mvSLOUCH
with a detailed description of the input and output routines.

We start our presentation of Paper II with
a brief introduction to the 
most relevant stochastic processes in the field of phylogenetic comparative methods. 

\subsection{Brownian Motion}\label{secBM}
The first modelling approach via stochastic differential equations for 
comparative methods in a continuous trait setting is due to Felsenstein
\cite{JFel85}. It proposed that the traits evolve as a Brownian motion along the phylogeny. 
A trait $X$ is evolving as a Brownian motion if it can be described by the following 
stochastic differential equation,
\begin{displaymath}
\ud X(t) =  \sigma \ud W(t),
\end{displaymath}
where $\ud W(t)$ is white noise. 
More formally $W(t)$ is the Wiener process, it has independent increments with \mbox{ $W(t)-W(s) \sim
\mathcal{N}(0,t-s), 0 \le s \le t$}. This means that the displacement of the trait after time $t$ from its initial 
value is $X(t) \sim \mathcal{N}(X(0),\sigma^{2} t)$. This model assumes that there is absolutely no selective pressure
on trait $X$ whatsoever, so that the trait value just randomly fluctuates from its initial starting point. 

A multi--dimensional Brownian
motion model is immediate,
\begin{displaymath}
\ud \vec{X}(t) = \mathbf{\Sigma} \ud \vec{W}(t),
\end{displaymath}
where $\mathbf{\Sigma}$ will be the diffusion matrix, $\vec{W}$ will be 
the multidimensional Wiener process. All components of the random vector are normally distributed as 
$\vec{W}(t)-\vec{W}(s) \sim \mathcal{N}(\vec{0},(t-s)\mathbf{I})$ and so
$\vec{X}(t) \sim \mathcal{N}(\vec{X}(0), t\mathbf{\Sigma}\mathbf{\Sigma}^{T})$. Such a trait evolution has the property that 
it is time reversible. It does not matter whether we look at it taking $\vec{X}(0)$ (moving forward) or $\vec{X}(t)$ (moving backward) 
as the starting point. Both $\vec{X}(t) - \vec{X}(0)$ and $\vec{X}(0) - \vec{X}(t)$ will be identically distributed. From
this one can find an independent sample on the phylogenetic tree, \emph{i.e.} the contrasts between nodes such that
no pair of nodes shares a branch in the path connecting them. If the distance between two nodes  is $t_{i}-t_{j}$
then the contrast vector will be distributed as $\vec{X}(t_{i})-\vec{X}(t_{j}) \sim \mathcal{N}(\vec{0},(t_{i}-t_{j})\mathbf{\Sigma}\mathbf{\Sigma}^{T})$.
The covariance matrix of all of the data will be $\mathbf{V}=\mathbf{T}\otimes (\mathbf{\Sigma}\mathbf{\Sigma}^{T})$
where $\mathbf{T}$ is the matrix of divergence times of species on the phylogeny.
In order to estimate the ancestral state and $\mathbf{\Sigma}$ in \cite{JFel85} the independent 
contrasts algorithm was proposed which is similar to the Cholesky decomposition 
described in section \ref{secReg}.

\subsection{Ornstein--Uhlenbeck process}\label{secOU}
The lack of the possibility of adaptation in the Brownian Motion model of \cite{JFel85} was
noticed in \cite{THan97} and therefore a more complex 
Ornstein--Uhlenbeck type of framework was proposed,
\begin{equation}
\begin{array}{rcl}\label{eqOU}
\ud \vec{Z} (t) & = & -\mathbf{F}(\vec{Z}(t)-\vec{\Psi}(t))\ud t + \mathbf{\Sigma} \ud \vec{W}(t),
\end{array}
\end{equation}
termed here the Ornstein--Uhlenbeck (OU) model. The $\mathbf{F}$ matrix is called the drift matrix, $\vec{\Psi}(t)$
the drift vector and $\mathbf{\Sigma}$ the diffusion matrix. 
This stochastic differential equation is a linear equation in the narrow sense.
The general model (\ref{eqOU}) has been only partially implemented in some \texttt{R} packages and 
in this work we present a nearly complete implementation of it. We use the word nearly,
as the package does not allow for certain very specific parameter combinations 
which should not be of biological significance. 

We also consider in detail model with  a very important decomposition 
of $\mathbf{F}$ from equation (\ref{eqOU}) which results in
the multivariate Ornstein--Uhlenbeck Brownian motion (mvOUBM) model
\begin{equation}\label{eqGenModel2}
\begin{array}{c}
\ud \left[\begin{array}{c} \vec{Y} \\ \vec{X} \end{array}\right] (t) 
=  -\left[ \begin{array}{c|c} \mathbf{A} & \mathbf{B} \\ \hline \mathbf{0} & \mathbf{0} \end{array} \right]
\left(\left[\begin{array}{c} \vec{Y} \\ \vec{X} \end{array} \right] (t) -
\left[\begin{array}{c} \vec{\psi} \\ \vec{0} \end{array} \right] (t) \right)\ud t 
 \\ +
\left[\begin{array}{c|c} \mathbf{\Sigma}^{yy} & \mathbf{\Sigma}^{yx} \\ \hline \mathbf{\Sigma}^{xy} & \mathbf{\Sigma}^{xx} \end{array} \right]
\ud \mathbf{W}(t).
\end{array}
\end{equation}
This is a multivariate generalization of the OUBM model from \cite{THanJPieSOrzSLOUCH}.

The different parameters in the above model have biological interpretations
but one must be careful how to interpret them. 
The $\vec{\Psi}(t)$ function represents a deterministic optimum value for $\vec{Z}(t)$, equation (\ref{eqOU})
and $\vec{\psi}(t)$ the deterministic part of the optimum for $\vec{Y}(t)$ in equation (\ref{eqGenModel2})
given that $\mathbf{F}$ or $\mathbf{A}$ have positive real part eigenvalues respectively. 
The individual entries
in $\mathbf{F}$ and $\mathbf{A}$ can be tricky to interpret. However their
eigendecomposition is much easier. The eigenvalues, if they have positive 
real part, can be understood to control the speed of the traits' approach to their
optima while the eigenvectors indicate the path towards the optimum.
The diffusion matrix in general
represents stochastic perturbations to the traits' approaching their optimum. These
perturbations can come from many sources internal or external, \emph{e.g.}
they can represent unknown/unmeasured components of the system under study
or some random genetic changes linked to the traits.

We have described here the processes as if they were evolving on a single lineage.
In the phylogenetic comparative setting these processes are evolving
on a phylogenetic tree. At each internal node they branch off into independently
evolving components. We assume that we know the tree and include 
these branching times in the estimation procedure. The mathematics of this 
is straightforward but it is non--trivial to effectively implement the calculation
of the log--likelihood function.

\subsection{The mvSLOUCH package}
With Paper II we present an \texttt{R} package, mvSLOUCH, that implements a 
(heuristic) maximum likelihood estimation to estimate the parameters
of the stochastic differential equation (\ref{eqOU}). The package
also allows for the estimation of parameters of a special,
but very appealing, submodel of equation (\ref{eqOU}) where 
some of the variables behave as
Brownian motions, equation (\ref{eqGenModel2}). This is a multivariate generalization
of the model presented in \cite{THanJPieSOrzSLOUCH}.

The implementation of the estimation procedure requires
combining advanced linear algebra and computing techniques 
with probabilistic and statistical knowledge. This is due
to the high model dimensionality and nearly arbitrary 
dependency structure between observations. The most
interesting techniques employed are described in
Paper II, its appendix, as well as in sections \ref{secPapIIB}
and \ref{secPapIIC} of this thesis. 

The package can estimate not only the parameters but simulate the trajectories
of described stochastic differential equation models, both on an arbitrary
phylogeny and on a single lineage. 

\begin{figure}[!h]
\begin{center}
\includegraphics[width=0.6\textwidth]{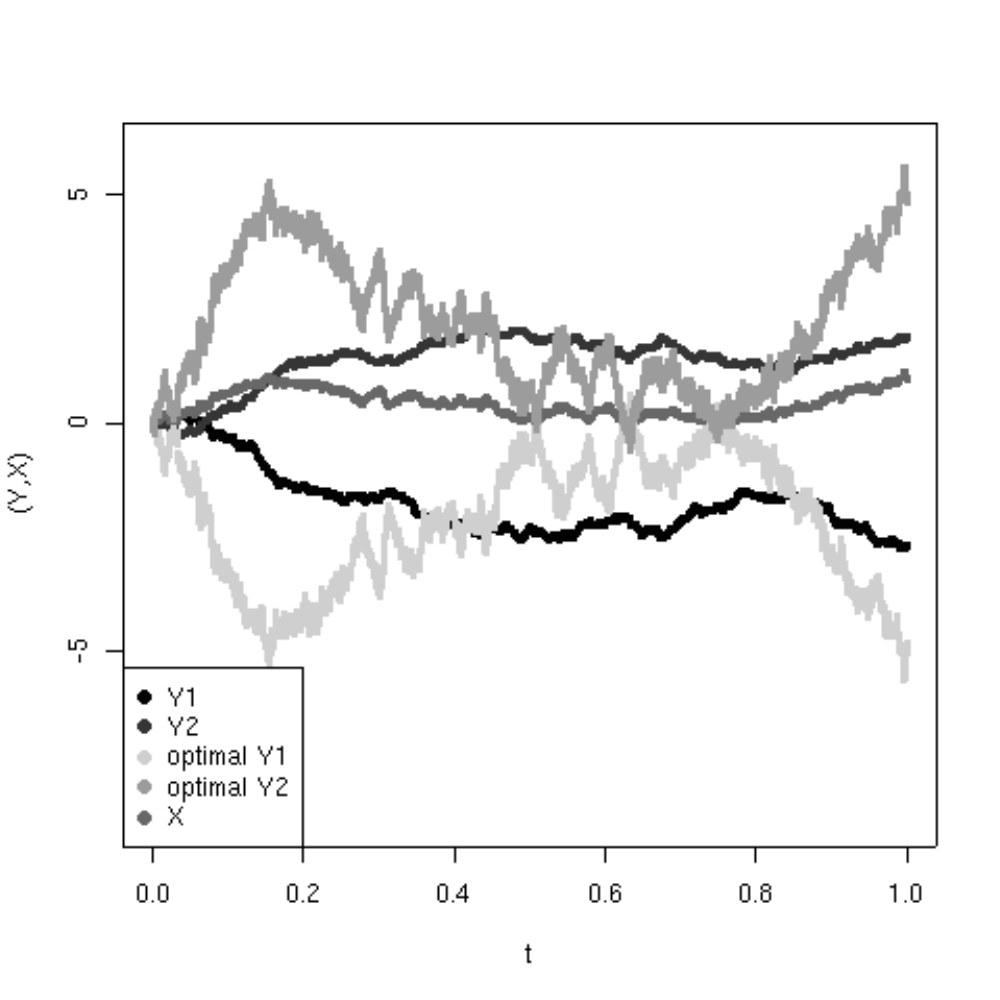}
\caption{
Trajectory and optimal values of the 
multivariate Ornstein--Uhlenbeck Brownian Motion (mvOUBM)
stochastic process.
}\label{figTrajIntro}
\end{center}
\end{figure}

In Figure \ref{figTrajIntro} we can see such a trajectory simulated 
from the model of equation (\ref{eqGenModel2}). 
The simulation reveals that small changes in the predictor process, $X$ (\emph{e.g.} environment), 
cause very
large changes in the optimal processes for the two traits $Y_{1}$ and $Y_{2}$
(see Paper II for the used nomenclature). 
This kind of behaviour could be biologically interpreted as specialization, when small 
changes of the surroundings require the organism to make large changes
in order to be best adapted to it. 
We can also see that the traits are tracking their respective optimal process, but 
with a lag and have a very gentle response to extremes in them. 
This lag means that the organism is unable to change quickly enough
to adapt and misses out on narrow (in time) optimal niches which
results in being close to optimality when the optimal process becomes toned down 
again. 
The trait
and optimal processes are obviously negatively correlated. 
Other examples of simulated trajectories are given in 
section \ref{secPapIIB}, Figure \ref{figParSim}
and in Paper II.
Through these examples one can see that these processes can exhibit all sorts of dynamics.

\subsection{Example data analysis}
In Paper II we reanalyze a data set of deer antler length, and male and female body masses
from \cite{FPlaCBonJGai} under our presented models. Our results
confirm those in \cite{FPlaCBonJGai}, that (i) 
there is positive linear relationship between the logarithms of antler length and male body mass
and (ii) the mating tactic does not influence directly the antler length
and also not the male body mass. The allometry between antler length and
male body mass is greater than $1/3$ indicating that there is more than just 
a proportional increase of antler length (one dimensional)
when body mass increases (three dimensional). We can also observe
that the estimates of the effect of breeding group size
on the antler length and male body mass are larger with the
increase in breeding group size. 

However our analysis in addition shows
that antler length and male body mass are adapting very rapidly
to changes in female body mass. We can also see that they
adapt independently of each other. There is no direct influence of
one variable on the primary optimum of the other. All dependencies
between antler length and male body mass are due to the common female
body mass predictor variable and correlations in noise pushing them 
away from their respective optimum values. The latter could be 
\emph{e.g.} due to mutations in some common genetic regulatory mechanism.
This however is pure speculation as our models show only that there
is some general ``noise'' mechanism and not what it is due to. 
This adaptation result was not observed in
the study of \cite{FPlaCBonJGai} simply because they did not consider 
a model that allowed for it. To account for phylogenetic effects 
they assumed a Brownian motion model of phenotypic trait evolution.

\clearpage

\newpage
\section{Future work}\label{secFuture}
The results of Paper I can be developed in the direction of studying the 
effects of measurement error on the estimation of covariance structures
under a Brownian motion model of evolution and possibly more complex
models afterwards.

Furthermore we have three directions of development from Paper II.
The first one concerns modelling
trait evolution and interactions among the traits and the environment.
All the models based on continuous stochastic differential equations considered so far have
assumed that the drift function is a deterministic step function. 
The regimes on the phylogeny where it is constant are presumed
to be known or preestimated. The next step in developing these models
would be to assume a random process for the drift function
coupled with the stochastic differential model. 

The randomness
can be added on a number of levels. The simplest case is to assume
two possible regime types with exponential times between regime changes 
(if the times of change are known then this would be the same as 
estimating what regime was at the root of the tree). Mathematically speaking
we have a function taking on two possible values (which are unknown) 
driven by a Poisson process and this is joined with 
a stochastic differential equation. Another possibility is that 
the drift function takes on a finite number of (unknown) values 
assuming that a Markov structure governs their changes. To connect
this with the traits, we can assume 
that the corresponding
probability transition matrix functionally depends on the current trait values.

The second direction is to address the various
statistical questions that arise from considering such a dependency structure
between different observations and aspects of the stochastic processes
used for modelling that might not have been traditionally looked at.
To the best
of our knowledge issues of parameter estimability, effective 
sample size nor the correct way of constructing confidence intervals
have not been addressed in a satisfactory manner. The 
need to capture interactions between the modelled traits
requires us to study the behaviour of the mean and covariance 
functions. Their asymptotics are well known but it remains to describe
\emph{e.g.} how their sign changes with time or how do the magnitudes
of the regression slopes evolve. 

The third idea of 
development would be to join the contents of the two papers
presented in the thesis together, to correct for bias
due to measurement error 
when estimating the parameters of the processes
evolving on the phylogeny. This might seem straightforward at first
but it is not. The measurement error model considered in Paper I
assumes that we essentially know all covariance structures 
(covariance between predictors, covariance between measurement errors and 
noise covariance structure) up to at most a scalar. The estimation
procedures implemented in this package can estimate
the covariance structure between predictors and noise.
However in Paper I it is noticed that correcting
for bias is not always a good idea as it might increase the mean square error. 
This was in a situation where the noise and predictor covariance 
structures were assumed to be known. It remains to study how
the mean square error is affected by the bias correction in a situation
where the other covariance structures are estimated.
 
We are aware that conditioning on the phylogeny is a gross simplification
and it would be ideal to combine phenotypic and molecular data in order
to jointly study the times of species diversification with the 
development of their traits. However this would be very complex
and computationally extremely demanding. At the moment
the estimation of the phylogeny with estimation of parameters of the process
modelling trait evolution is only done for the Brownian motion process \cite{JHueBRan}.
We hope to be able to develop this into more complex stochastic models.

\newpage
\section{Supplementary material A: Bias caused by measurement error; to correct or not to correct (Paper I)}\label{secPapIA}
In section \ref{secMEbias} we derived a general formula for the reliability matrix $\mathbf{K}$,
equation (\ref{eqK}) and afterwards presented formulae in situations where some kind of 
independence was assumed. In sections \ref{sbsecIndob} -- \ref{sbsecSingPred}.
we treat the subject in more detail. Section \ref{secMSEbias} answers the question
on whether it is worth to correct for the bias caused by measurement error.
Section \ref{sbsecFixed} considers the case where we have fixed effects and more
generally predictors observed without error. Finally section \ref{secUncondBias}
finds discusses conditions when it is possible to derive analytical formulae
for the unconditional (of the observed design matrix) bias caused by measurement error in predictors.

\subsection{Independent observations of predictors}\label{sbsecIndob}
The case of independent observations is widely described in the literature, see \emph{e.g.} 
\cite{MAicCRit} (where they show that in the multivariate case the bias can be in their words 
``\emph{of arbitrary sign and magnitude}'' and  
``\emph{There do not seem to be any rules of thumb that would permit one to make even
qualitative statements about the nature of this bias.}''), \cite{WFul1987}, \cite{WFul1995}
(where the effect of error in data analysis is shown on examples) or
\cite{LGle1992} (where some
statistical results are derived \emph{e.g.} mean square error results, connection to 
maximum likelihood estimates, estimability and distributional
properties). 

The situation with a single predictor is described in Paper I (see equation (12) there).
Here we will briefly describe the multipredictor case.
Let each row of $\mathbf{D}_{t}$ (the vector of predictors) be distributed
as $\mathcal{N}(0,\mathbf{\Sigma}_{d})$ and each row of $\mathbf{U}$ 
(vector of errors for each observation) be distributed
as $\mathcal{N}(0,\mathbf{\Sigma}_{u})$. Then each row of $\mathbf{D}_{o}$
is distributed as $\mathcal{N}(0,\mathbf{\Sigma}_{d}+\mathbf{\Sigma}_{u})$ and 
we have the distributions
$\mathrm{vec}(\mathbf{D}_{o}) \sim \mathcal{N}(0,(\mathbf{\Sigma}_{d}+\mathbf{\Sigma}_{u})\otimes\mathbf{I})$
and $\mathrm{vec}(\mathbf{U}) \sim \mathcal{N}(0,\mathbf{\Sigma}_{u}\otimes\mathbf{I})$ .
Putting $\mathbf{\Sigma}_{o}:=\mathbf{\Sigma}_{d}+\mathbf{\Sigma}_{u}$
we arrive at equation (\ref{eqMultIndepBias}),
\begin{displaymath}
\E{\hat{\vec{\beta}} \vert \mathbf{D}_{o}} =	
(\mathbf{I}-(\mathbf{D}_{o}^{T}\mathbf{V}^{-1}\mathbf{D}_{o})^{-1}\mathbf{D}_{o}^{T}\mathbf{V}^{-1}\mathbf{D}_{o}
\mathbf{\Sigma}_{o}^{-1}\mathbf{\Sigma}_{u})
\vec{\beta} = 
(\mathbf{I}-\mathbf{\Sigma}_{o}^{-1}\mathbf{\Sigma}_{u})\vec{\beta}
\end{displaymath}
which simplifies to $\mathbf{\Sigma}_{o}^{-1}\mathbf{\Sigma}_{d}\vec{\beta}$.
This also gives us the unconditional bias as in this case the formula does not depend on $\mathbf{D}_{o}$,
$\E{\hat{\vec{\beta}}}=\mathbf{\Sigma}_{o}^{-1}\mathbf{\Sigma}_{d}\vec{\beta}$.

We can see that the above formula is a straightforward generalization of the single predictor case. 
Some discussion about this situation can be found in \cite{JBuo}, with
detailed derivations given in \cite{LGle1992}.
There are examples (see \cite{MAicCRit}) showing that the one--dimensional
shrinkage to zero does not generalize and no qualitative statements about the bias
in the general case can be made. In particular in 
\cite{LGle1992} it is shown that if the reliability matrix ($\mathbf{\Sigma}_{o}^{-1}\mathbf{\Sigma}_{t}$) is estimated by maximum likelihood, then
the bias corrected estimates will be the maximum likelihood ones. 

In the multivariate setting we can get an upward and downward bias of arbitrary size
in a component depending on its correlation with another component 
(compare with example $2$ in \cite{MAicCRit} even for predictors that have a slope of $0$, \emph{i.e.}
they are not related to the response).
To illustrate this  we will write out the formula for the two dimensional case. Denote,
\begin{displaymath}
\begin{array}{cc}
\mathbf{\Sigma}_{u} = \left[ \begin{array}{cc} \sigma_{11} & \sigma_{12} \\ \sigma_{12} & \sigma_{22} \end{array}\right] &
\mathbf{\Sigma}_{o}^{-1} = \left[ \begin{array}{cc} p_{11} & p_{12} \\ p_{12} & p_{22} \end{array}\right]
\end{array}
\end{displaymath}
and let $\vec{\beta}=[\beta_{1}, \beta_{2}]^{T}$. We then have
\begin{displaymath}
\begin{array}{rcl}
\E{\left[\begin{array}{c} \hat{\beta}_{1} \\ \hat{\beta}_{2}\end{array}\right]} & = &
\left[\begin{array}{c} 
(1-(p_{11}\sigma_{11}+p_{12}\sigma_{12}))\beta_{1} -(p_{11}\sigma_{12}+p_{12}\sigma_{22})\beta_{2} \\ 
(1-(p_{22}\sigma_{22}+p_{12}\sigma_{12}))\beta_{2} -(p_{22}\sigma_{12}+p_{12}\sigma_{11})\beta_{2}
\end{array}\right]
\end{array}.
\end{displaymath}
So we can see that both predictors cause bias in each other but the direction of the bias depends
on the covariance between the predictors and between errors (as shown in  \cite{MAicCRit} 
and compare to a similar example on pages 109--110 of \cite{JBuo}).

\subsection{Independent predictors}
In the general case of dependent observations the formula for
$\E{\hat{\vec{\beta}} \vert \mathbf{D}_{o}}$ does not simplify so well. However if we assume that
the predictors are independent (the columns of $\mathbf{D}_{t}$ are independent)
and that the errors in different predictors are also independent (the columns of $\mathbf{U}$ are independent)
then some simplifications can be made.

Let us assume that the $i$--th column of $\mathbf{D}_{t}$ is distributed as 
$\mathcal{N}(\vec{0},\mathbf{\Sigma}_{d_{i}})$ ($\mathbf{\Sigma}_{d_{i}}$
is the covariance matrix between the $i$--th predictor
in different observations) and that the error in the $i$--th predictor
is distributed as $\mathcal{N}(\vec{0},\mathbf{\Sigma}_{u_{i}})$.
The distribution of the $i$--th column of $\mathbf{D}_{o}$
will be $\mathcal{N}(\vec{0},\mathbf{\Sigma}_{d_{i}}+\mathbf{\Sigma}_{u_{i}})$ 
and we will denote $\mathbf{\Sigma}_{o_{i}}:=\mathbf{\Sigma}_{d_{i}}+\mathbf{\Sigma}_{u_{i}}$. 
We then have that,
\begin{displaymath}
\begin{array}{c}
\mathrm{vec}(\mathbf{D}_{o}) \sim
\mathcal{N}\left(\vec{0},
\left[\begin{array}{cccc}
\mathbf{\Sigma}_{o_{1}} & \mathbf{0} & \ldots & \mathbf{0} \\
\mathbf{0} & \mathbf{\Sigma}_{o_{2}} & \ldots & \mathbf{0} \\
\vdots & \vdots &\ddots  & \vdots \\
\mathbf{0} & \mathbf{0} &\ldots &  \mathbf{\Sigma}_{o_{m}}
\end{array}\right]\right)
\\
\mathrm{vec}(\mathbf{U}) \sim
\mathcal{N}\left(\vec{0},
\left[\begin{array}{cccc}
\mathbf{\Sigma}_{u_{1}} & \mathbf{0} & \ldots &  \mathbf{0} \\
\mathbf{0} & \mathbf{\Sigma}_{u_{2}} & \ldots & \mathbf{0} \\
\vdots & \vdots &\ddots & \vdots \\
\mathbf{0} & \mathbf{0} &\ldots & \mathbf{\Sigma}_{u_{m}}
\end{array}\right]\right)
\end{array}.
\end{displaymath}
This then gives us after performing the matrix multiplication in the 
formula for the conditional expectation that 
the $i$--th column of $\E{\mathbf{U} \vert \mathbf{D}_{o}}$
is $\mathbf{\Sigma}_{u_{i}}\mathbf{\Sigma}_{o_{i}}^{-1}\mathbf{d}_{o_{i}}$,
where $\mathbf{d}_{o_{i}}$ is the $i$--th column of $\mathbf{D}_{o}$.
This gives us equation (\ref{eqMultIndepPredBias}),
\begin{displaymath}
\E{\hat{\vec{\beta}} \vert \mathbf{D}_{o}} = 
(\mathbf{I}-(\mathbf{D}_{o}^{T}\mathbf{V}^{-1}\mathbf{D}_{o})^{-1}\mathbf{D}_{o}^{T}\mathbf{V}^{-1}
\left[\mathbf{\Sigma}_{u_{1}}\mathbf{\Sigma}_{o_{1}}^{-1}\mathbf{d}_{o_{1}};
\ldots ;\mathbf{\Sigma}_{u_{m}}\mathbf{\Sigma}_{o_{m}}^{-1}\mathbf{d}_{o_{m}}
\right])
\vec{\beta}.
\end{displaymath}

If we assume further that errors in different observations are independent and 
that the covariance matrices are of the form
$\mathbf{\Sigma}_{u_{i}}=\sigma_{i}^{2}\sigma_{u}^{2}\mathbf{I}$ and
$\mathbf{\Sigma}_{d_{i}}=\sigma_{i}^{2}\mathbf{T}$,
where $\mathbf{T}$ is some symmetric positive--definite matrix (predictors evolve as independent Brownian
Motions along the phylogenetic tree),
then
\mbox{
$\mathbf{\Sigma}_{d_{i}}+\mathbf{\Sigma}_{u_{i}}=\sigma^{2}_{i}(\mathbf{T}+\sigma_{u}^{2}\mathbf{I})$,} \\
\mbox{
$\mathbf{V}_{u}=\mathrm{diag}(\sigma^{2}_{1},\ldots,\sigma^{2}_{m})\otimes \sigma^{2}_{u}\mathbf{I}$} and 
\mbox{
$\mathbf{V}_{o}=\mathrm{diag}(\sigma^{2}_{1},\ldots,\sigma^{2}_{m})\otimes (\mathbf{T}+\sigma^{2}_{u}\mathbf{I})$,} where
$\mathrm{diag}(\mathbf{v})$  is a diagonal matrix with the vector 
$\mathbf{v}$ on the diagonal. 

From 
\begin{displaymath}
\mathbf{V}_{u}\mathbf{V}_{o}^{-1}\mathrm{vec}(\mathbf{D}_{o}) =
(\mathbf{I}\otimes(\frac{1}{\sigma^{2}_{u}}\mathbf{T}+\mathbf{I})^{-1})\mathrm{vec}(\mathbf{D}_{o}),
\end{displaymath}
we get
\begin{displaymath}
\E{\mathbf{U} \vert \mathbf{D}_{o}} =
(\frac{1}{\sigma^{2}_{u}}\mathbf{T}+\mathbf{I})^{-1}\mathbf{D}_{o},
\end{displaymath}
so that 
\begin{equation}\label{eqBiasIndepPredII}
\E{\hat{\vec{\beta}} \vert \mathbf{D}_{o}} = 
(\mathbf{I}-(\mathbf{D}_{o}^{T}\mathbf{V}^{-1}\mathbf{D}_{o})^{-1}\mathbf{D}_{o}^{T}\mathbf{V}^{-1}
(\frac{1}{\sigma^{2}_{u}}\mathbf{T}+\mathbf{I})^{-1}\mathbf{D}_{o})\vec{\beta}.
\end{equation}
If we do not assume that the errors between different observations are independent,
\emph{i.e.} \mbox{
$\mathbf{\Sigma}_{u_{i}}=\sigma_{i}^{2}\mathbf{\Sigma}_{u}$,} this will change as, \\
\mbox{
$\mathbf{V}_{u}=\mathrm{diag}(\sigma^{2}_{1},\ldots,\sigma^{2}_{m})\otimes \mathbf{\Sigma}_{u}$} and 
\mbox{
$\mathbf{V}_{o}=\mathrm{diag}(\sigma^{2}_{1},\ldots,\sigma^{2}_{m})\otimes 
(\mathbf{T}+\mathbf{\Sigma}_{u})$.} 
It follows 
\begin{displaymath}
\mathbf{V}_{u}\mathbf{V}_{o}^{-1}\mathrm{vec}(\mathbf{D}_{o}) =
(\mathbf{I}\otimes(\mathbf{T}\mathbf{\Sigma}_{u}^{-1}+\mathbf{I})^{-1})\mathrm{vec}(\mathbf{D}_{o}),
\end{displaymath}
implying
\begin{displaymath}
\E{\mathbf{U} \vert \mathbf{D}_{o}} =
(\mathbf{T}\mathbf{\Sigma}_{u}^{-1}+\mathbf{I})^{-1}\mathbf{D}_{o},
\end{displaymath}
which gives 
\begin{equation}\label{eqBiasIndepPredIII}
\E{\hat{\vec{\beta}} \vert \mathbf{D}_{o}} = 
(\mathbf{I}-(\mathbf{D}_{o}^{T}\mathbf{V}^{-1}\mathbf{D}_{o})^{-1}\mathbf{D}_{o}^{T}\mathbf{V}^{-1}
(\mathbf{T}\mathbf{\Sigma}_{u}^{-1}+\mathbf{I})^{-1}\mathbf{D}_{o})\vec{\beta}.
\end{equation}

\subsection{Single predictor}\label{sbsecSingPred}
The single predictor case of the general formula (\ref{eqBias})
presents an illustrative example 
which will be treated here in detail.
The matrix $\mathbf{D}_{o}$ is a single column matrix and the generalized least squares estimator is,
\begin{displaymath}
\begin{array}{rcl}
\hat{\beta} & = & (\mathbf{D}_{o}^{T}\mathbf{V}^{-1}\mathbf{D}_{o})^{-1}\mathbf{D}_{o}^{T}\mathbf{V}^{-1}(\mathbf{D}_{t}\beta+\vec{r}_{t}+\vec{e}_{y}),
\end{array}
\end{displaymath}
and its conditional expectation (as $\vec{r}_{t}$ and $\vec{e}_{y}$ are independent of the rest) is,
\begin{displaymath}
\begin{array}{rcl}
\E{\hat{\beta} \vert \mathbf{D}_{o}}& = & (\mathbf{D}_{o}^{T}\mathbf{V}^{-1}\mathbf{D}_{o})^{-1}\mathbf{D}_{o}^{T}\mathbf{V}^{-1}\E{\mathbf{D}_{t}\vert\mathbf{D}_{o}}\beta,
\end{array}
\end{displaymath}
and using $\E{\mathbf{D}_{t}\vert \mathbf{D}_{o}}=\mathbf{D}_{o}-\E{\mathbf{U}\vert \mathbf{D}_{o}}$ 
(here as the matrix contains one column we can work with it directly there is no need to vectorize it) we get
\begin{displaymath}
\begin{array}{rcl}
\E{\hat{\beta} \vert \mathbf{D}_{o}}& = & (1-(\mathbf{D}_{o}^{T}\mathbf{V}^{-1}\mathbf{D}_{o})^{-1}\mathbf{D}_{o}^{T}\mathbf{V}^{-1}\E{\mathbf{U}\vert\mathbf{D}_{o}})\beta.
\end{array}
\end{displaymath}

The trick again is to compute $\E{\mathbf{U\vert\mathbf{D}_{o}}}$, as we assumed that all the variables 
are multivariate normal, then together the $\mathbf{U}$ and $\mathbf{D}_{o}$ will also be 
multivariate normal and so we get,
\begin{displaymath}
\begin{array}{rcl}
\E{\mathbf{U} \vert \mathbf{D}_{o}} & = & \E{\mathbf{U}} + \cov{\mathbf{U}}{\mathbf{D}_{o}}\Cov{\mathbf{D}_{o}}^{-1}(\mathbf{D}_{o}-\E{\mathbf{D}_{o}})
\end{array}
\end{displaymath}
Since we assumed that $\E{\mathbf{U}}=\E{\mathbf{D}_{o}}=\mathbf{0}$ we have,
\begin{displaymath}
\begin{array}{rcccccl}
\cov{\mathbf{U}}{\mathbf{D}_{o}} & = & \cov{\mathbf{U}}{\mathbf{D}_{t}+\mathbf{U}}& = &\Cov{\mathbf{U}}&=&\mathbf{V}_{u},
\end{array}
\end{displaymath}
which gives $\E{\mathbf{U}\vert\mathbf{D}_{o}}=\mathbf{V}_{u}\mathbf{V}_{o}^{-1}\mathbf{D}_{o}$ and from this we can write the final formula
of equation (\ref{eqSingBias}),
\begin{displaymath}
\begin{array}{rcl}
\E{\hat{\beta} \vert \mathbf{D}_{o}}& = & (1-(\mathbf{D}_{o}^{T}\mathbf{V}^{-1}\mathbf{D}_{o})^{-1}\mathbf{D}_{o}^{T}\mathbf{V}^{-1}\mathbf{V}_{u}\mathbf{V}_{o}^{-1}\mathbf{D}_{o})\beta.
\end{array}
\end{displaymath}
If $\mathbf{V}_{u}=\sigma^{2}_{u}\mathbf{I}$ and $\mathbf{V}_{o}=(\sigma^{2}_{d}+\sigma^{2}_{u})\mathbf{I}$ then we get
equation (\ref{eqSingIndepBias}),
\begin{displaymath}
\begin{array}{rcl}
\E{\hat{\beta} \vert \mathbf{D}_{o}}& = & (1-\frac{\sigma^{2}_{u}}{\sigma^{2}_{d}+\sigma^{2}_{u}})\beta = \frac{\sigma^{2}_{d}}{\sigma^{2}_{d}+\sigma^{2}_{u}}\beta =
\E{\hat{\beta}},
\end{array}
\end{displaymath}
which is consistent with the standard regression case if we estimate $\sigma^{2}_{u}$ and $\sigma_{o}^{2}$ from the 
sum of squares.

\subsection{Mean square error (MSE) analysis}\label{secMSEbias}
In the derivation of the conditional expectation of the generalized least squares estimator we used
the matrix $\mathbf{V}$. One could however be oblivious to the fact
of error in the response and use $\mathbf{V}_{t}$ instead. As we shall see here this
does not have an effect on the form of bias derivations (but will numerically change the results). 
Let us first consider a more general version of the generalized least squares estimator,
\mbox{
$\hat{\vec{\beta}} = (\mathbf{D}_{o}^{T}\mathbf{Z}\mathbf{D}_{o})^{-1}\mathbf{D}_{o}^{T}\mathbf{Z}\vec{Y}_{o}$,}
where $\mathbf{Z}$ is some symmetric, positive definite matrix. 
Then doing the same calculations
as before we get that,
\begin{displaymath}
\E{\hat{\vec{\beta}} \vert \mathbf{D}_{o}} =
\left(\mathbf{I}-(\mathbf{D}_{o}^{T}\mathbf{Z}\mathbf{D}_{o})^{-1}\mathbf{D}_{o}^{T}\mathbf{Z}
\mathrm{vec}^{-1}(\mathbf{V}_{u}\mathbf{V}_{o}^{-1}\mathrm{vec}(\mathbf{D}_{o})\right)\vec{\beta},
\end{displaymath}
so the bias correction is of the same form for any symmetric, positive definite matrix $\mathbf{Z}$, 
however the covariance matrix of $\hat{\vec{\beta}}$ can substantially change and then so will
the mean square error. 
The mean square error of an estimator is defined as,
\begin{displaymath}
\begin{array}{c}
\mathrm{MSE}[\hat{\vec{\beta}} \vert \mathbf{D}_{o}]=
\E{(\hat{\vec{\beta}}-\vec{\beta})^{T}(\hat{\vec{\beta}}-\vec{\beta})\vert \mathbf{D}_{o}} 
\end{array}
\end{displaymath}
and equals to 
$\mathrm{Tr}(\Cov{\hat{\vec{\beta}}\vert \mathbf{D}_{o}}) + \E{\hat{\vec{\beta}}-\vec{\beta}\vert \mathbf{D}_{o}}^{T}\E{\hat{\vec{\beta}}-\vec{\beta}\vert \mathbf{D}_{o}}$,
where $\mathrm{Tr}(\mathbf{M})$ is the trace of a matrix $\mathbf{M}$.

The covariance of the estimator is,
\begin{displaymath}
\begin{array}{rcl}
(\mathbf{D}^{T}_{o}\mathbf{Z}\mathbf{D}_{o})^{-1}\mathbf{D}^{T}_{o}\mathbf{Z}
\left(\Cov{\mathbf{U}\vec{\beta}\vert \mathbf{D}_{o}}+\mathbf{V}\right)
\mathbf{Z}\mathbf{D}_{o}(\mathbf{D}^{T}_{o}\mathbf{Z}\mathbf{D}_{o})^{-1}.
\end{array}
\end{displaymath}
It remains to calculate $\Cov{\mathbf{U}\vec{\beta}\vert \mathbf{D}_{o}}$, as we assumed both
$\mathbf{U}$ and $\mathbf{D}_{o}$ are multivariate normal we get this from the formula
for the conditional covariance,
\begin{displaymath}
\begin{array}{rcl}
\Cov{\mathbf{U}\vec{\beta}\vert \mathbf{D}_{o}} & = & \Cov{\mathbf{U}\vec{\beta}}-\cov{\mathbf{U}\vec{\beta}}{\mathbf{D}_{o}}\Cov{\mathbf{D}_{o}}^{-1}\cov{\mathbf{D}_{o}}{\mathbf{U}\vec{\beta}},
\end{array}
\end{displaymath}
where element $i$, $j$ of $\Cov{\mathbf{U}\vec{\beta}}$ is $\sum_{r=1}^{m}\sum_{k=1}^{m}\vec{\beta}_{k}\vec{\beta}_{r}\mathbf{V}_{u}[(r-1)n+i,(k-1)n+j]$
and element $i$, $(k-1)n+j$ of $\cov{\mathbf{U}\vec{\beta}}{\mathbf{D}_{o}}$ is $\sum_{r=1}^{m}\vec{\beta}_{r}\mathbf{V}_{u}[(r-1)n+i,(k-1)n+j]$.

\subsubsection{Mean square error for different GLS estimators}
We will consider the mean square error for the general case of some $\mathbf{Z}$ matrix
as the residual covariance structure.
To see that there is no qualitative difference when we take 
$\mathbf{Z}=\mathbf{V}^{-1}$ or $\mathbf{Z}=\mathbf{V}_{t}^{-1}$ or any other matrix  we introduce 
the general notation,
\begin{displaymath}
\begin{array}{rcl}
\mathbf{A} & := & (\mathbf{D}_{o}^{T}\mathbf{Z}^{-1}\mathbf{D}_{o})^{-1}, \\
\mathbf{B} & := & (\mathrm{vec}^{-1}(\mathbf{V}_{u}\mathbf{V}_{o}^{-1}\mathrm{vec}(\mathbf{D}_{o})))^{T}\mathbf{Z}^{-1}\mathbf{D}_{o}.
\end{array}
\end{displaymath}
We introduce a third matrix $\mathbf{C}$ whose definition will depend on the choice of $\mathbf{Z}$.
If $\mathbf{Z}=\mathbf{V}^{-1}$ then 
\begin{displaymath}
\begin{array}{rcl}
\mathbf{C} & := & \mathbf{D}_{o}^{T}\mathbf{V}^{-1}\Cov{\mathbf{U}\vec{\beta} \vert \mathbf{D}_{o}}\mathbf{V}^{-1}\mathbf{D}_{o} 
\end{array}
\end{displaymath}
and if $\mathbf{Z}=\mathbf{V}_{t}^{-1}$ then 
\begin{displaymath}
\begin{array}{rcl}
\mathbf{C} & := & \mathbf{D}_{o}^{T}\mathbf{V}_{t}^{-1}\left(\Cov{\mathbf{U}\vec{\beta} \vert \mathbf{D}_{o}}+\mathbf{V}_{e}\right)\mathbf{V}_{t}^{-1}\mathbf{D}_{o} .
\end{array}
\end{displaymath}
We can notice immediately that $\mathbf{A}$ and $\mathbf{C}$ are symmetric positive definite matrices
and recognize $\mathbf{A}$ as the estimated covariance of the estimate of $\vec{\beta}$.

In the case of the uncorrected estimator, 
\begin{equation}
\begin{array}{c}
\mathrm{MSE}[\hat{\vec{\beta}} \vert \mathbf{D}_{o}] =
\mathrm{Tr}\left(\mathbf{A}+\mathbf{A}
\mathbf{C}\mathbf{A}\right) 
+ 
\vec{\beta}^{T}\left(
\mathbf{B}\mathbf{A}\mathbf{A}\mathbf{B}^{T}\right)
\vec{\beta}
\end{array}\label{eqMSEuncorr}
\end{equation}
and for the corrected estimator the mean square error becomes,
\begin{equation}
\begin{array}{l}
\mathrm{MSE}[\hat{\vec{\beta}}_{corr} \vert \mathbf{D}_{o}] =
\mathrm{Tr}\left(
\left(\mathbf{I}-\mathbf{A}
\mathbf{B}\right)^{-1}
\left(\mathbf{A}+\mathbf{A}\mathbf{C}\mathbf{A}
\right)
\left(\mathbf{I}-
\mathbf{A}\mathbf{B}
\right)^{-T}\right).
\end{array}\label{eqMSEcorr}
\end{equation}
We notice that in both cases the formula for the mean square error depends on $\vec{\beta}$,
in the second formula only through $\mathbf{C}$.

\subsubsection{Mean square error for a single predictor}\label{sbsecSingPredMSE}
Turning to the
single predictor case 
one can find more visible conditions for 
the corrected estimator having a smaller mean square error. 
Our notation for the matrix components $\mathbf{A}$, $\mathbf{B}$ and $\mathbf{C}$ 
of the mean square error formulae simplify to,
\begin{displaymath}
\begin{array}{l}
a  :=  (\mathbf{D}_{o}^{T}\mathbf{Z}^{-1}\mathbf{D}_{o})^{-1}>0,\\
b  :=  \mathbf{D}_{o}^{T}\mathbf{Z}^{-1}_{o}\mathbf{V}_{u}\mathbf{V}^{-1}\mathbf{D}_{o}=\mathbf{D}_{o}^{T}\mathbf{V}^{-1}\mathbf{V}_{u}\mathbf{V}^{-1}_{o}\mathbf{D}_{o}.
\end{array}
\end{displaymath}
If $\mathbf{Z}=\mathbf{V}^{-1}$ then 
\begin{displaymath}
\begin{array}{l}
c  :=  \mathbf{D}_{o}^{T}\mathbf{V}^{-1}\Cov{\mathbf{U}\vert \mathbf{D}_{o}}\mathbf{V}^{-1}\mathbf{D}_{o}>0~~~~~~~~~~~~~
\end{array}
\end{displaymath}
and if $\mathbf{Z}=\mathbf{V}_{t}^{-1}$ then 
\begin{displaymath}
\begin{array}{l}
c  :=  \mathbf{D}_{o}^{T}\mathbf{V}_{t}^{-1}\left(\Cov{\mathbf{U}\vert \mathbf{D}_{o}}+\mathbf{V}_{e}\right)\mathbf{V}_{t}^{-1}\mathbf{D}_{o}>0.~~
\end{array}
\end{displaymath}

\begin{figure}[h!]
\begin{minipage}[b]{0.5\linewidth}
\centering
\includegraphics[width=0.6\textwidth]{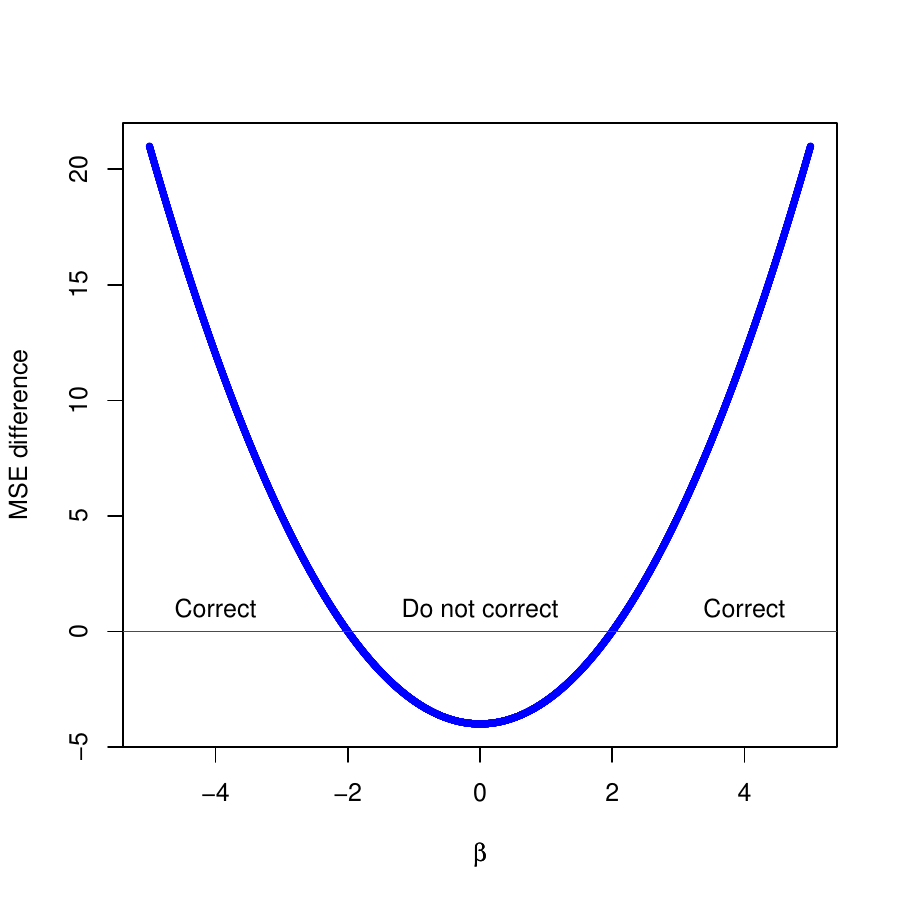} \\
\includegraphics[width=0.6\textwidth]{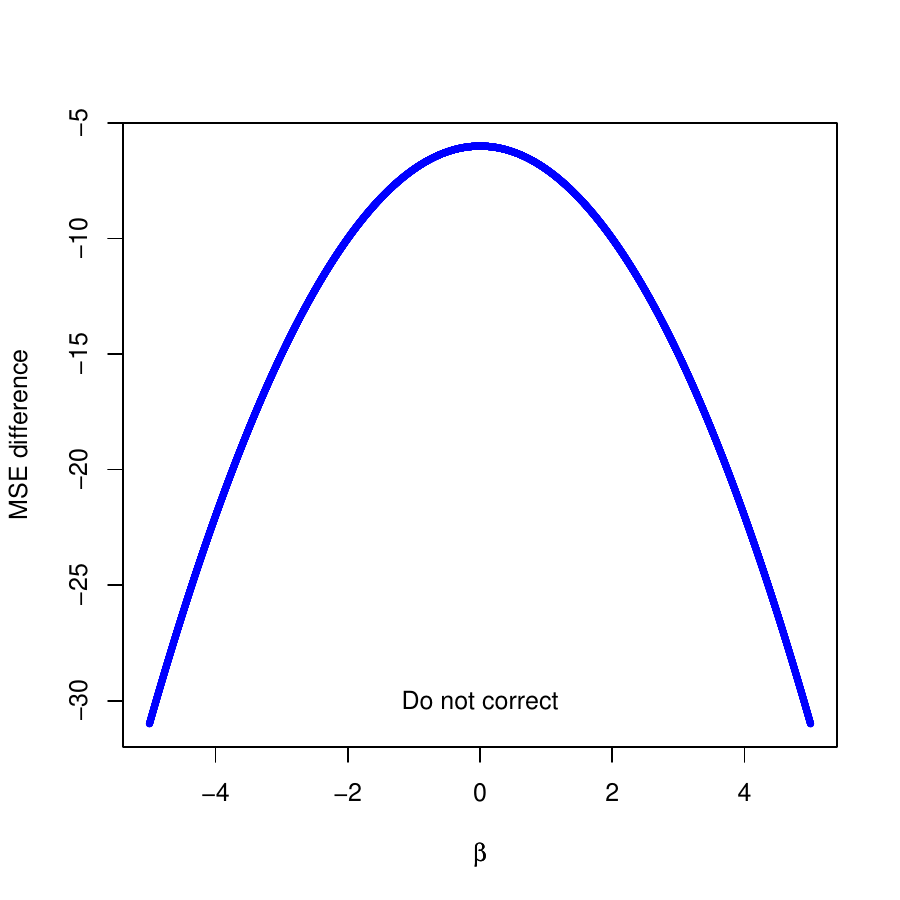}
\end{minipage}
\begin{minipage}[b]{0.5\linewidth}
\caption{Depending whether the second degree polynomial is positive or negative 
when \mbox{
$0<ab<2$} for a given $\beta$ it might be beneficial to correct for the bias or not, top:
sign of coefficient in front of $\beta^{2}$ positive, bottom: negative.}\label{figMSEreg}
\end{minipage}
\end{figure}

The corrected estimator has a smaller mean square error if the following condition, derived 
from combining the univariate versions of equations (\ref{eqMSEuncorr}) 
and (\ref{eqMSEcorr}), holds,
\begin{equation}
\begin{array}{rcl}
a^{2}b^{2}\beta^{2}+(a+a^{2}c\beta^{2})(1-\frac{1}{(1-ab)^{2}}) &>& 0.
\end{array}\label{eq1predMSEdiff}
\end{equation}
Notice that $(1-ab)$ is the bias coefficient in equation (\ref{eqSingBias}).
We can immediately see that if $b=0$ (this will be when we have \emph{e.g.}
no measurement error) then both estimators have equal mean square errors. 
The other conclusion is that if $ab \in (-\infty,0)\cup [2,\infty)$ then
the corrected estimator is better which implies 
that for all $b<0$ it is better to correct. This is because for these values
the variance of the corrected estimator is smaller than that of the uncorrected one. 
When $ab \in(0,2)$ we have a trade--off
between the bias and variance. The condition can be reformulated as,
\begin{displaymath}
\begin{array}{rcl}
a^{2}(b^{2}+c\frac{(1-ab)^{2}-1}{(1-ab)^{2}})\beta^{2}+a(1-\frac{1}{(1-ab)^{2}}) & > & 0,
\end{array}
\end{displaymath}
and depending on the sign of $a^{2}(b^{2}+c\frac{(1-ab)^{2}-1}{(1-ab)^{2}})$ (when $0<ab<2$),
we will have one of the two situations illustrated in Figure \ref{figMSEreg}. Notice that the
formula is symmetric around $0$ in $\beta$.
We can see that if $0<ab<2$ the magnitude of $c$ will determine whether it
is worth correcting or not for a given region of $\beta$s. If $c$ is large
then the coefficient will be negative and the second degree polynomial will
always be negative so it is not worth correcting. If $c$ is small
then it will be worth to correct for large $\beta$s while not for small ones.
We can see that if $0<ab<2$ then for $\beta=0$ the biased estimator
is better.

We can rewrite the mean square error difference condition (\ref{eq1predMSEdiff}) in terms
of the bias coefficient $(1-ab)$ which we will label as $\kappa_{x}:=1-ab$. The condition
then becomes,
\begin{equation}
\begin{array}{rcl}
(\kappa_{x}-1)(\kappa_{x}^{3}-\kappa_{x}^{2} + (\frac{a}{\beta^{2}}+a^{2}c)\kappa_{x} + \frac{a}{\beta^{2}}+a^{2}c) & > & 0
\end{array}\label{eq1predMSEk}
\end{equation}
which can be also written as,
\begin{displaymath}
\begin{array}{rcl}
(\kappa_{x}-1)((a^{2}c(\kappa_{x}+1)+\kappa_{x}^{2}(\kappa_{x}-1))\beta^{2}+a(\kappa_{x}+1)) & > & 0.
\end{array}
\end{displaymath}

\begin{figure}[h!]
\begin{center}
\includegraphics[width=0.4\textwidth]{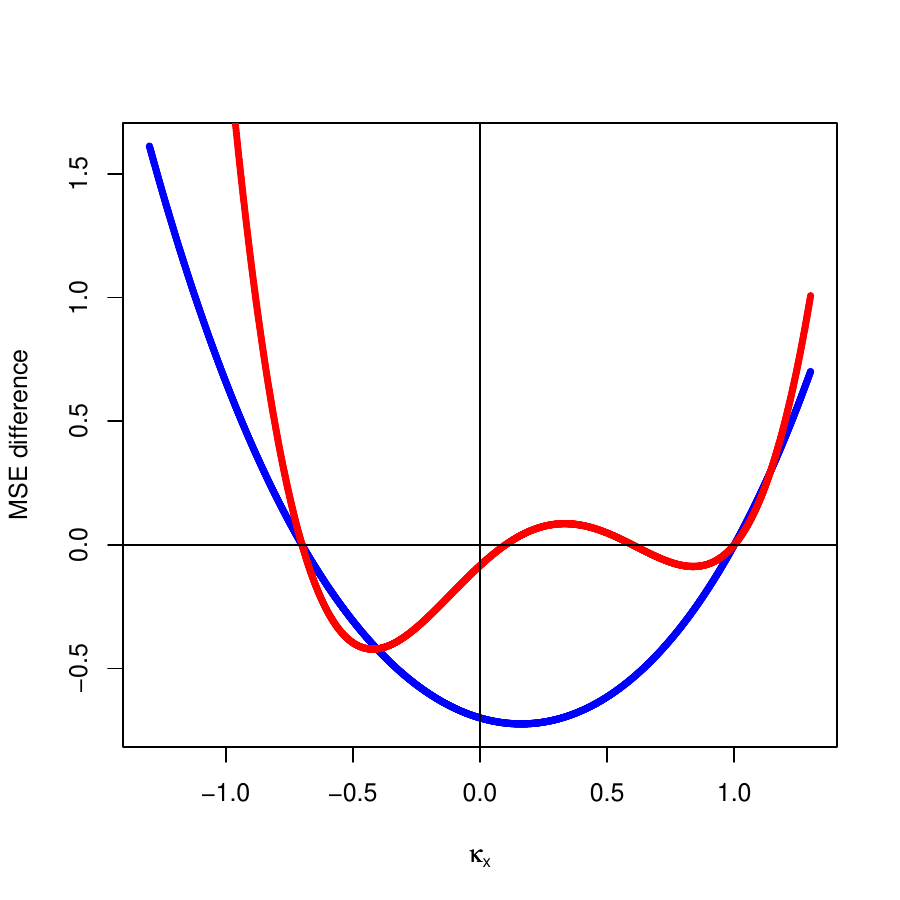}
\caption{The two possibilities how the mean square error difference will be depending on the parameters.}\label{figMSEregkappax}
\end{center}
\end{figure}

We know from previous calculations when $0<ab<2$ then there is a potential trade -- off between
the variance and the bias. In terms of $\kappa_{x}$ this is $-1 < \kappa_{x} < 1$.
We can also see that when $\kappa_{x}=0$ then it is always better not to correct and
when $\kappa_{x}=-1$ it is always better to correct. This means that equation (\ref{eq1predMSEk})
has to have a root between $-1$ and $0$ (another root is $\kappa_{x}=1$). It remains
to study whether it has any other real roots. If we have that 
$a^{2}c > \kappa_{x}^{2}(1-\kappa_{x})/(\kappa_{x}+1)$ then for $\kappa_{x} \in (0,1)$ the function will always be negative.
This will happen if $a^{2} c > 0.09017$ as $\kappa_{x}^{2}(1-\kappa_{x})/(\kappa_{x}+1)$ when $\kappa_{x} \in (0,1)$
attains a maximum at $\frac{\sqrt{5}-1}{2}$ equal to just below $0.09017$. 
As we know that we have to have a root between $(-1,0)$ let us assume that it is $-\kappa_{0}$ 
(so $\kappa_{0}>0$). Then we can write the condition polynomial as,
$(\kappa_{x}-1)(\kappa_{x}+\kappa_{0})(\kappa_{x}^{2}-(1+\kappa_{0})\kappa_{x}+(a/\beta^{2}+a^{2}c)/\kappa_{0})$.
If it has two more real roots they both have to be positive 
(and therefore between $0$ and $1$ as $1$ will not be a root of the equation), as they will be,
\begin{displaymath}
\begin{array}{rcl}
\kappa_{x_{1,2}} & = & \frac{1}{2}\left((1+\kappa_{0})\pm\sqrt{(1+\kappa_{0})^{2}-4\frac{\frac{a}{\beta^{2}}+a^{2}c}{\kappa_{0}}}\right),
\end{array}
\end{displaymath}
The two possibilities for how the mean square error difference can look like
are presented symbolically in Figure \ref{figMSEregkappax}.

The above result can be extended further for other values of the $\mathbf{Z}$ matrix. The mean square error
difference can be written in a general form as,
\begin{displaymath}
\begin{array}{rcl}
(\kappa_{x}-1)\left(\kappa_{x}^{3}-\kappa_{x}^{2}+\kappa_{x}\frac{\Var{\hat{\beta}\vert \mathbf{D}_{o}}}{\beta_{2}^{2}}+\frac{\Var{\hat{\beta}\vert \mathbf{D}_{o}}}{\beta_{2}^{2}}\right)
\end{array}
\end{displaymath}
and as $\Var{\hat{\beta}\vert \mathbf{D}_{o}}/\beta_{2}^{2}>0$ all of the previous discussion
holds with just $a/\beta^{2}+a^{2}c$ changed to $\Var{\hat{\beta}\vert \mathbf{D}_{o}}/\beta_{2}^{2}$.
The condition for which it will be better never to correct when $\kappa_{x} \in (0,1)$
changes to \\
\mbox{
$\Var{\hat{\beta}\vert \mathbf{D}_{o}}/\beta_{2}^{2} > \kappa_{x}^{2}(1-\kappa_{x})/(\kappa_{x}+1)$} 
and by the previous discussion this will be if 
\mbox{
$\Var{\hat{\beta}\vert \mathbf{D}_{o}}/\beta_{2}^{2} > 0.09017$.}

If we make the simplifying assumption that $\mathbf{V}_{u}\mathbf{V}_{o}^{-1}=\kappa \mathbf{I}$,
then 
\mbox{
$b=\kappa/a$} and the mean square error difference becomes,
\begin{displaymath}
\begin{array}{rcl}
(\kappa^{2}-\frac{a^{2}c\kappa(2-\kappa)}{(1-\kappa)^{2}})\beta^{2}-\frac{a\kappa(2-\kappa)}{(1-\kappa)^{2}} & > & 0,
\end{array}
\end{displaymath}
we can immediately see that if the true slope is $0$ then the biased estimator is 
superior in terms of mean square error (compare to the discussed later result of \cite{LGle1992}).
Because $0<\kappa<1$ the free term is 
always negative, so there will always be a region of small $\beta$s where the correction increases the mean square error.
Therefore depending on the sign of the coefficient in front of $\beta^{2}$ it will be better sometimes
to correct and sometimes not to correct. If it is negative then the biased estimator
will be always better, if it is positive then for small $\beta$s it will be better not
to correct, while for large ones it will be better. 

The coefficient in front of $\beta^{2}$ can be negative if $a$ (the estimate of the estimator's variance
in the case of no measurement error) or $c$ is large. This gives us a rule of thumb, if our
estimator has large variance it is better not to correct. However if we compare this 
to the situation, discussed previously, where our observations are dependent then this rule of thumb does not hold,
the dependencies confound it.

The multipredictor case with independent observations is studied in detail in \cite{LGle1992}, here
we just state the mean square error condition in our formulation.
In the multipredictor case if we assume that $\mathbf{V}_{u} = \mathbf{\Sigma}_{u} \otimes \mathbf{I}$
and $\mathbf{V}_{o} = (\mathbf{\Sigma}_{d}+\mathbf{\Sigma}_{u}) \otimes \mathbf{I}$ (our observations of predictors are independent)
and denoting,
\begin{displaymath}
\begin{array}{rcl}
\mathbf{A} & = & (\mathbf{D}_{o}^{T} \mathbf{V}^{-1}\mathbf{D}_{o})^{-1} \\
\mathbf{B} & = & \mathbf{\Sigma}_{u} - \mathbf{\Sigma}_{u}(\mathbf{\Sigma}_{d}+\mathbf{\Sigma}_{u})^{-1}\mathbf{\Sigma}_{u} \\
\mathbf{C} & = & \mathbf{D}_{o}^{T} \mathbf{V}^{-2} \mathbf{D}_{o} \\
\mathbf{G} & = & (\mathbf{I}-(\mathbf{\Sigma}_{d}+\mathbf{\Sigma}_{u})^{-1}\mathbf{\Sigma}_{u})^{-1},
\end{array}
\end{displaymath}
we get that the difference between the mean square error of the uncorrected and
corrected estimator is,
\begin{displaymath}
\begin{array}{l}
\mathrm{MSE}[\hat{\vec{\beta}} \vert \mathbf{D}_{o}] - \mathrm{MSE}[\hat{\vec{\beta}}_{corr} \vert \mathbf{D}_{o}] 
\\=
\mathrm{Tr}\left(\mathbf{A}+\vec{\beta}^{T}\mathbf{B}\vec{\beta}\mathbf{A}\mathbf{C}\mathbf{A}-
\mathbf{G}(\mathbf{A}+\vec{\beta}^{T}\mathbf{B}\vec{\beta}\mathbf{A}\mathbf{C}\mathbf{A})\mathbf{G}^{T}\right)
\\+
\vec{\beta}^{T}\mathbf{\Sigma}_{u}(\mathbf{\Sigma}_{u}+\mathbf{\Sigma}_{d})^{-2}\mathbf{\Sigma}_{u}\vec{\beta}.
\end{array}
\end{displaymath}
The mean square error of the situation with independent observations is considered in \cite{LGle1992}. The author 
states the result that the trace of the covariance matrix of the corrected estimator
is greater than the trace of the uncorrected one and so there is a trade--off between the
bias and variance in the mean square error. Notice that from the single predictor case discussion we can
see that this result does not generalize to the case of dependent predictors. 
It is shown that if the sample size is large, the bias overwhelms
the variance so then it is better to correct. Also the mean square error of estimating $\mathbf{T}^{-1}\vec{\beta}$ is 
considered, where 
\mbox{
$\mathbf{T}\mathbf{J}\mathbf{T^{-1}}=(\mathbf{I}-(\mathbf{\Sigma}_{d}+\mathbf{\Sigma}_{u})^{-1}\mathbf{\Sigma}_{u})$,}
where $\mathbf{J}$ is a diagonal eigenvalue matrix. If the true slope is zero then it is shown that
the biased estimator of $\mathbf{T}^{-1}\vec{\beta}$ has smaller risk
otherwise a condition is given which estimator to choose.
In \cite{WFul1995} the regression with a single predictor, intercept and independent observations
is considered. The mean square error is shortly studied in relation to sample size, as the main topic of 
applications were survey studies. 

\subsection{Effect on intercept, fixed effects}\label{sbsecFixed}
The previous section considered the situation where $\E{\mathbf{D}_{t}}=\E{\mathbf{D}_{o}}=\mathbf{0}$, this in particular
implies that there is no intercept in the model which is of course not realistic. Here we consider a more general 
case, where the observed design matrix $\mathbf{D}_{o}$ consists of fixed effects (\emph{e.g.} an intercept),
random effects observed without error and random effects observed with error.

We first consider the case where all random effects are observed with error.
Let $\mathbf{D}_{t}=\left[\mathbf{F} ;\mathbf{X}_{t} \right]$, where $\mathbf{F}$
is meant to denote the fixed effect part of the matrix and $\mathbf{X}_{t}$
the random part with measurement errors. Then 
$\mathbf{D}_{o}=\left[\mathbf{F} ;\mathbf{X}_{t} \right] + \left[\mathbf{0} ;\mathbf{U} \right]
=\left[\mathbf{F} ;\mathbf{X}_{o} \right]$ and the model is,
\begin{displaymath}
\begin{array}{rcl}
\vec{Y}_{o} & = & \mathbf{F}\vec{\beta}_{F} + (\mathbf{X}_{o}-\mathbf{U})\vec{\beta}_{X} + \vec{r},
\end{array}
\end{displaymath}
we denote by $\vec{\beta}=[\vec{\beta}_{F}^{T};\vec{\beta}_{X}^{T}]^{T}$.
We need to calculate $\E{\mathrm{vec}(\left[\mathbf{0} ;\mathbf{U} \right])\vert \mathbf{D}_{o}}$.
Below we divide the matrices into blocks of appropriate sizes (relating to fixed and random effects),
\begin{displaymath}
\begin{array}{c}
\E{\mathrm{vec}(\left[\mathbf{0};\mathbf{U}\right])\vert \mathbf{D}_{o}} = 
\left[\begin{array}{c}\mathbf{0} \\ \cov{\mathrm{vec}(\mathbf{U})}{\mathrm{vec}(\mathbf{X}_{o})}\Cov{\mathrm{vec}(\mathbf{X}_{o})}^{-1}\mathrm{vec}(\mathbf{X}_{o}) \end{array}\right].
\end{array}
\end{displaymath}
Using equation (\ref{eqBias}), we write 
$\E{\hat{\vec{\beta}}\vert \mathbf{D}_{o}}$ as
\begin{displaymath}
\begin{array}{l}
~~~\left(\mathbf{I}-(\left[\mathbf{F} ; \mathbf{X}_{o}\right]^{T}\mathbf{V}^{-1}\left[\mathbf{F} ; \mathbf{X}_{o}\right])^{-1}\left[\mathbf{F} ; \mathbf{X}_{o}\right]^{T}
\mathbf{V}^{-1}\left[\begin{array}{c}\mathbf{0} \\\hline \E{\mathbf{U} \vert \mathbf{D}_{o}} \end{array}\right]\right)\vec{\beta} 
\\=
\left(\mathbf{I} -
\left[\begin{array}{c|c} \mathbf{F}^{T}\mathbf{V}^{-1}\mathbf{F}& \mathbf{F}^{T}\mathbf{V}^{-1}\mathbf{X}_{o}
\\ \hline \vec{X}^{T}_{o}\mathbf{V}^{-1}\mathbf{F}& \vec{X}^{T}_{o}\mathbf{V}^{-1}\mathbf{X}_{o} \end{array}\right]^{-1}
\left[\begin{array}{c|c} \mathbf{0} &  \mathbf{F}^{T}\mathbf{V}^{-1}\E{\mathbf{U}\vert \mathbf{D}_{o}} \\\hline \mathbf{0} & \vec{X}^{T}_{o}\mathbf{V}^{-1}\E{\mathbf{U}\vert \mathbf{D}_{o}}\end{array}\right]
\right)\vec{\beta} 
\\=
\left[\begin{array}{c|c}
\mathbf{I} & -\mathbf{A}\mathbf{F}^{T}\mathbf{V}^{-1}\E{\mathbf{U} \vert \mathbf{D}_{o}}-\mathbf{B}\mathbf{X}_{o}^{T}\mathbf{V}^{-1}\E{\mathbf{U} \vert \mathbf{D}_{o}} \\ \hline
\mathbf{0} & -\mathbf{C}\mathbf{F}^{T}\mathbf{V}^{-1}\E{\mathbf{U} \vert \mathbf{D}_{o}}-\mathbf{G}\mathbf{X}_{o}^{T}\mathbf{V}^{-1}\E{\mathbf{U} \vert \mathbf{D}_{o}}
\end{array}\right]\vec{\beta},
\end{array}
\end{displaymath}
where 
\begin{displaymath}
\begin{array}{rcl}
\mathbf{A} & = &\left(\mathbf{F}^{T}\mathbf{V}^{-1}\mathbf{F}-\mathbf{F}^{T}\mathbf{V}^{-1}\mathbf{X}_{o}(\mathbf{X}_{o}^{T}\mathbf{V}^{-1}\mathbf{X}_{o})^{-1}\mathbf{X}_{o}^{T}\mathbf{V}^{-1}\mathbf{F}\right)^{-1}, \\
\mathbf{B} &= &-\left(\mathbf{F}^{T}\mathbf{V}^{-1}\mathbf{F}-\mathbf{F}^{T}\mathbf{V}^{-1}\mathbf{X}_{o}(\mathbf{X}_{o}^{T}\mathbf{V}^{-1}\mathbf{X}_{o})^{-1}\mathbf{X}_{o}^{T}\mathbf{V}^{-1}\mathbf{F}\right)^{-1}
 \\ && \times
\mathbf{F}^{T}\mathbf{V}^{-1}\mathbf{X}_{o}(\mathbf{X}_{o}^{T}\mathbf{V}^{-1}\mathbf{X}_{o})^{-1},\\
\mathbf{C} &= &-\left(\mathbf{X}_{o}^{T}\mathbf{V}^{-1}\mathbf{X}_{o}-\mathbf{X}_{o}^{T}\mathbf{V}^{-1}\mathbf{F}(\mathbf{F}^{T}\mathbf{V}^{-1}\mathbf{F})^{-1}\mathbf{F}^{T}\mathbf{V}^{-1}\mathbf{X}_{o}\right)^{-1}
 \\ && \times
\mathbf{X}_{o}^{T}\mathbf{V}^{-1}\mathbf{F}(\mathbf{F}^{T}\mathbf{V}^{-1}\mathbf{F})^{-1}, \\
\mathbf{G}& =& \left(\mathbf{X}_{o}^{T}\mathbf{V}^{-1}\mathbf{X}_{o}-\mathbf{X}_{o}^{T}\mathbf{V}^{-1}\mathbf{F}(\mathbf{F}^{T}\mathbf{V}^{-1}\mathbf{F})^{-1}\mathbf{F}^{T}\mathbf{V}^{-1}\mathbf{X}_{o}\right)^{-1},
\end{array}
\end{displaymath}
coming from the formula for the inverse of a block matrix.
So the final formula for $\E{\hat{\vec{\beta}}\vert \mathbf{D}_{o}}$ is,
\begin{equation}
\begin{array}{rcl}
\left[
\begin{array}{c}
\vec{\beta}_{F}-\left(\mathbf{A}\mathbf{F}^{T}\mathbf{V}^{-1}\E{\mathbf{U} \vert \mathbf{D}_{o}}+\mathbf{B}\mathbf{X}_{o}^{T}\mathbf{V}^{-1}\E{\mathbf{U} \vert \mathbf{D}_{o}}\right)\vec{\beta}_{X}
\\ 
\left(\mathbf{I}-(\mathbf{C}\mathbf{F}^{T}\mathbf{V}^{-1}\E{\mathbf{U} \vert \mathbf{D}_{o}}+\mathbf{G}\mathbf{X}_{o}^{T}\mathbf{V}^{-1}\E{\mathbf{U} \vert \mathbf{D}_{o}})\right)\vec{\beta}_{X}
\end{array}
\right].
\end{array}\label{eqFnFbias}
\end{equation}
We can see that there is an influence on the estimate of the fixed effects from the measurement error
and the fixed effects coefficient does not have any influence on the bias of the random effects estimates
(compare with example $7$ in \cite{MAicCRit}). Equation (\ref{eqFnFbias})
additionally shows how to correct for the bias in the fixed effects. We can ask what assumptions are needed for
formula (\ref{eqFnFbias}) to simplify. 

First we will consider whether it is possible to transform this equation so that the
bias of the coefficients of the random effects is of the same functional form as the one
in equation (\ref{eqBias}). It can be shown that in general these biases will not be equal
but if we consider the following transformation of the random part of the design matrix
$\mathbf{X}_{o}^{\ast}:=\mathbf{X}_{o}-\mathbf{F}(\mathbf{F}^{T}\mathbf{V}^{-1}\mathbf{F})^{-1}\mathbf{F}^{T}\mathbf{V}^{-1}\mathbf{X}_{o}$
(notice that $\E{\mathbf{X}_{o}^{\ast}}=\mathbf{0}$ which is crucial for the derivations) then, 
$\mathbf{F}^{T}\mathbf{V}^{-1}\mathbf{X}_{o}^{\ast}=\mathbf{0}$ which gives
that 
$\E{[\hat{\vec{\beta}}_{F}^{\ast^{T}}~\hat{\vec{\beta}}_{X}^{T}]^{T}\vert \mathbf{D}_{o}}$ 
is,
\begin{displaymath}
\begin{array}{rcl}
\left[
\begin{array}{c}
\vec{\beta}_{F}^{\ast}-(\mathbf{F}^{T}\mathbf{V}^{-1}\mathbf{F})^{-1}\mathbf{F}^{T}\mathbf{V}^{-1}\E{\mathbf{U} \vert \mathbf{D}_{o}}\vec{\beta}_{X}
\\ 
(\mathbf{I}-(\mathbf{X}_{o}^{\ast T}\mathbf{V}^{-1}\mathbf{X}_{o}^{\ast})^{-1}\mathbf{X}_{o}^{\ast T}\mathbf{V}^{-1}\E{\mathbf{U} \vert \mathbf{D}_{o}})\vec{\beta}_{X}
\end{array}
\right],
\end{array}
\end{displaymath}
where $\vec{\beta}_{F}^{\ast}:=\vec{\beta}_{F}+(\mathbf{F}^{T}\mathbf{V}^{-1}\mathbf{F})^{-1}\mathbf{F}^{T}\mathbf{V}^{-1}\mathbf{X}_{o}\vec{\beta}_{X}$.

In section \ref{secUncondBias} we will look at what assumptions are needed for the bias to have
a simplified formula. If we follow the same chain of reasoning for the fixed
effects to be unbiased we require that $\E{\mathbf{U} \vert \mathbf{D}_{o}}=\mathbf{X}_{o}\mathbf{K}$
for some matrix $\mathbf{K}$ (the situation when this will happen was discussed in section \ref{sbsecIndob}).
If this is the case then we will have the fixed effects unbiased and the reliability matrix for the
random effects,
\begin{displaymath}
\begin{array}{rcl}
\E{\hat{\vec{\beta}}\vert \mathbf{D}_{o}} & = & \left[\begin{array}{c} \vec{\beta}_{F} \\ (\mathbf{I}-\mathbf{K})\vec{\beta}_{X}\end{array} \right].
\end{array}
\end{displaymath}
\subsubsection{Fixed effects, random effects measured with and without error}
The above derivations did not consider the situation that there could be random effects measured without error. 
We now denote 
$\mathbf{D}_{t}=[\mathbf{F};\mathbf{X}_{c};\mathbf{X}_{t}]$ and $\mathbf{D}_{o}=\mathbf{D}_{t}+[\mathbf{0};\mathbf{0};\mathbf{U}]$,
where $\mathbf{X}_{c}$ are the random effects measured without error. The random effects covariance
matrix is $\mathbf{V}_{o}=\left[\begin{array}{cc} \mathbf{V}_{c} & \mathbf{V}_{cx} \\ \mathbf{V}_{cx}^{T} & \mathbf{V}_{x} \end{array} \right]$
(where $\mathbf{V}_{x}=\Cov{\mathrm{vec}(\mathbf{X}_{t})}+\mathbf{V}_{u}$)
and the regression model is,
\begin{displaymath}
\begin{array}{rcl}
\vec{Y}_{o} & = & \mathbf{F}\vec{\beta}_{F}+\mathbf{X}_{c}\vec{\beta}_{c}+(\mathbf{X}_{o} - \mathbf{U}) \vec{\beta}_{X}+\vec{r}.
\end{array}
\end{displaymath}
As before we need $\E{\mathbf{U}\vert \mathbf{D}_{o}}$, which can be calculated as,
\begin{displaymath}
\begin{array}{rcl}
\mathrm{vec}^{-1}(\mathbf{V}_{u}\Cov{\mathrm{vec}(\mathbf{X}_{o}) \vert \mathbf{X}_{c}}^{-1}(\mathrm{vec}(\mathbf{X}_{o})-\mathbf{V}_{cx}\mathbf{V}_{c}^{-1}\mathrm{vec}(\mathbf{X}_{c}))),
\end{array}
\end{displaymath}
and this results in the following formula for the bias, 
\begin{equation}
\begin{array}{rcl}
\E{\hat{\vec{\beta}} \vert \mathbf{D}_{o} }=
\left[\begin{array}{c} 
\vec{\beta}_{F} - (\mathbf{A}\mathbf{F}^{T}+\mathbf{B}\mathbf{X}_{c}^{T}+\mathbf{C}\mathbf{X}_{o}^{T})\mathbf{V}^{-1}\E{\mathbf{U} \vert \mathbf{D}_{o}}\vec{\beta}_{X} \\
\vec{\beta}_{c} - (\mathbf{G}\mathbf{F}^{T}+\mathbf{H}\mathbf{X}_{c}^{T}+\mathbf{L}\mathbf{X}_{o}^{T})\mathbf{V}^{-1}\E{\mathbf{U} \vert \mathbf{D}_{o}}\vec{\beta}_{X} \\
(\mathbf{I} -(\mathbf{P}\mathbf{F}^{T}+\mathbf{R}\mathbf{X}_{c}^{T}+\mathbf{Q}\mathbf{X}_{o}^{T})\mathbf{V}^{-1}\E{\mathbf{U} \vert \mathbf{D}_{o}})\vec{\beta}_{X}
\end{array} \right],
\end{array}\label{eqFnFnMnFE}
\end{equation}
where
\begin{displaymath}
\begin{array}{rcl}
\left[\begin{array}{ccc} 
\mathbf{F}^{T}\mathbf{V}^{-1}\mathbf{F} & \mathbf{F}^{T}\mathbf{V}^{-1}\mathbf{X}_{c} & \mathbf{F}^{T}\mathbf{V}^{-1}\mathbf{X}_{o} \\
\mathbf{X}_{c}^{T}\mathbf{V}^{-1}\mathbf{F} & \mathbf{X}_{c}^{T}\mathbf{V}^{-1}\mathbf{X}_{c} & \mathbf{X}_{c}^{T}\mathbf{V}^{-1}\mathbf{X}_{o} \\
\mathbf{X}_{o}^{T}\mathbf{V}^{-1}\mathbf{F} & \mathbf{X}_{o}^{T}\mathbf{V}^{-1}\mathbf{X}_{c} & \mathbf{X}_{o}^{T}\mathbf{V}^{-1}\mathbf{X}_{o}
\end{array} \right]^{-1}& =
& \left[\begin{array}{ccc}\mathbf{A} & \mathbf{B} & \mathbf{C} \\ \mathbf{G} & \mathbf{H} & \mathbf{L} \\ \mathbf{P} & \mathbf{R} & \mathbf{Q} \end{array} \right].
\end{array}
\end{displaymath}
We can see the same property that the coefficients of the correctly measured effects do not influence the bias.
It is only the coefficient
corresponding to the predictors measured with error that causes a bias in all the other coefficients
and this bias can have any size and sign. One can compare this to the examples on pages 109 -- 113 in
\cite{JBuo} and we have the same conclusion that ``\emph{measurement error can bias the coefficients that are 
measured without error}'' \cite{JBuo}.

We will do the same kind of formula transformation as in the previous section.
We perform the following transformation of the random part of the design matrix
$\mathbf{X}_{c}^{\ast}:=\mathbf{X}_{c}-\mathbf{F}(\mathbf{F}^{T}\mathbf{V}^{-1}\mathbf{F})^{-1}\mathbf{F}^{T}\mathbf{V}^{-1}\mathbf{X}_{c}$,
$\mathbf{X}_{o}^{\ast}:=\mathbf{X}_{o}-\mathbf{F}(\mathbf{F}^{T}\mathbf{V}^{-1}\mathbf{F})^{-1}\mathbf{F}^{T}\mathbf{V}^{-1}\mathbf{X}_{o}$
(notice that $\E{\mathbf{X}_{c}^{\ast}}=\mathbf{0}$, $\E{\mathbf{X}_{o}^{\ast}}=\mathbf{0}$ still
which again is crucial for the derivations) then, 
$\mathbf{F}^{T}\mathbf{V}^{-1}\mathbf{X}_{c}^{\ast}=\mathbf{0}$ and
$\mathbf{F}^{T}\mathbf{V}^{-1}\mathbf{X}_{o}^{\ast}=\mathbf{0}$ which give
that 
$\E{[\hat{\vec{\beta}}_{F}^{\ast^{T}}~\hat{\vec{\beta}}_{c}^{T}~\hat{\vec{\beta}}_{X}^{T}]^{T}\vert \mathbf{D}_{o}}$
equals,
\begin{displaymath}
\begin{array}{rcl}
\left[
\begin{array}{c}
\vec{\beta}_{F}^{\ast}-(\mathbf{F}^{T}\mathbf{V}^{-1}\mathbf{F})^{-1}\mathbf{F}^{T}\mathbf{V}^{-1}\E{\mathbf{U} \vert \mathbf{D}_{o}}\vec{\beta}_{X}
\\ 
\vec{\beta}_{c}-(\mathbf{A}\mathbf{X}_{c}^{\ast T}+\mathbf{B}\mathbf{X}_{o}^{\ast T})\mathbf{V}^{-1}\E{\mathbf{U} \vert \mathbf{D}_{o}}\vec{\beta}_{X}
\\
(\mathbf{I}-\mathbf{C}\mathbf{X}_{c}^{\ast T}+\mathbf{G}\mathbf{X}_{o}^{\ast T})\mathbf{V}^{-1}\E{\mathbf{U} \vert \mathbf{D}_{o}})\vec{\beta}_{X}
\end{array}
\right],
\end{array}
\end{displaymath}
where $\vec{\beta}_{F}^{\ast}:=\vec{\beta}_{F}+(\mathbf{F}^{T}\mathbf{V}^{-1}\mathbf{F})^{-1}\mathbf{F}^{T}\mathbf{V}^{-1}(\mathbf{X}_{c}\vec{\beta}_{c}+\mathbf{X}_{o}\vec{\beta}_{X})$
and 
\begin{displaymath}
\left[\begin{array}{cc}
\mathbf{X}_{c}^{\ast T}\mathbf{V}^{-1}\mathbf{X}_{c}^{\ast} & \mathbf{X}_{c}^{\ast T}\mathbf{V}^{-1}\mathbf{X}_{o}^{\ast} \\
\mathbf{X}_{o}^{\ast T}\mathbf{V}^{-1}\mathbf{X}_{c}^{\ast} & \mathbf{X}_{o}^{\ast T}\mathbf{V}^{-1}\mathbf{X}_{o}^{\ast}
\end{array}\right]^{-1} =
\left[\begin{array}{cc} \mathbf{A} & \mathbf{B} \\ \mathbf{C} & \mathbf{G} \end{array}\right].
\end{displaymath}

\subsubsection{Single predictor}\label{subsecSingpredZeroMean}
The single predictor with intercept model can be written as,
\begin{displaymath}
\begin{array}{rcl}
\vec{Y}_{o} & = & \vec{1}\beta_{1} + (\vec{X}_{o} - \mathbf{U}) \beta_{2} + \vec{r}, 
\end{array}
\end{displaymath}
where $\vec{1}$ denotes of vector of 1s of length $n$. Since
$\E{\mathbf{U}\vert \mathbf{D}_{o}}= \mathbf{V}_{u}\mathbf{V}_{o}^{-1}\vec{X}_{o}$ (where 
\mbox{
$\Cov{\vec{X}_{o}}=\mathbf{V}_{o}$),}
we have, 
\begin{displaymath}
\begin{small}
\begin{array}{rcl}
\E{[\hat{\beta}_{1}~ \hat{\beta}_{2}]^{T} \vert \mathbf{D}_{o}} & = &
\left[\begin{array}{c} \beta_{1} - \kappa_{1}\beta_{2} \\ (1-\kappa_{2})\beta_{2} \end{array}\right],
\end{array}
\end{small}
\end{displaymath}
where
\begin{displaymath}
\begin{array}{rcl}
\kappa_{1} & = & \frac{(\vec{X}_{o}^{T}\mathbf{V}^{-1}\vec{X}_{o})(\vec{1}^{T}\mathbf{V}^{-1}\mathbf{V}_{u}\mathbf{V}_{o}^{-1}\vec{X}_{o})-(\vec{X}_{o}^{T}\mathbf{V}^{-1}\mathbf{V}_{u}\mathbf{V}_{o}^{-1}\vec{X}_{o})(\vec{1}^{T}\mathbf{V}^{-1}\vec{X}_{o})}
{(\vec{X}_{o}^{T}\mathbf{V}^{-1}\vec{X}_{o})(\vec{1}^{T}\mathbf{V}^{-1}\vec{1})-(\vec{1}^{T}\mathbf{V}^{-1}\vec{X}_{o})^{2}}, \\
\kappa_{2} & = & \frac{(\vec{X}_{o}^{T}\mathbf{V}^{-1}\mathbf{V}_{u}\mathbf{V}_{o}^{-1}\vec{X}_{o})(\vec{1}^{T}\mathbf{V}^{-1}\vec{1})-(\vec{X}_{o}^{T}\mathbf{V}^{-1}\vec{1})(\vec{1}^{T}\mathbf{V}^{-1}\mathbf{V}_{u}\mathbf{V}_{o}^{-1}\vec{X}_{o})}
{(\vec{X}_{o}^{T}\mathbf{V}^{-1}\vec{X}_{o})(\vec{1}^{T}\mathbf{V}^{-1}\vec{1})-(\vec{1}^{T}\mathbf{V}^{-1}\vec{X}_{o})^{2}}.
\end{array}
\end{displaymath}

We can ask as in the general case how does the coefficient in front of the slope, $\beta_{2}$, 
$1-\kappa_{2}$
compare to the corresponding one in
equation (\ref{eqSingBias}). In can be shown that in the general dependent case 
(unless $\mathbf{V}_{u}=\kappa \mathbf{V}_{o}$)
they are not equal (assuming that $\vec{X}_{o}=\mathbf{D}_{o}$).
This is another difference caused by dependence between predictors. 
However one can as before transform the bias
coefficient in front of $\beta_{2}$ to be of the same functional form as the one in equation (\ref{eqSingBias}).
If we transform $\vec{X}_{o}$ to
\mbox{
$\vec{X}_{o}^{\ast}:=\vec{X}_{o}-(\vec{1}^{T}\mathbf{V}^{-1}\vec{X}_{o})/(\vec{1}^{T}\mathbf{V}^{-1}\vec{1})\vec{1}$}
then we still have a zero mean design matrix as $\E{\vec{X}_{o}^{\ast}}=\mathbf{0}$ and \\
\mbox{
$\vec{1}^{T}\mathbf{V}^{-1}\vec{X}_{o}^{\ast}=0$.}
The model becomes,
\begin{displaymath}
\begin{array}{rcl}
\vec{Y}_{o} & = & \vec{1}(\beta_{1}+\frac{\vec{1}^{T}\mathbf{V}^{-1}\vec{X}_{o}}{\vec{1}^{T}\mathbf{V}^{-1}\vec{1}}) + (\vec{X}_{o}^{\ast} - \mathbf{U}) \beta_{2} + \vec{r}.
\end{array}
\end{displaymath}
Defining $\beta_{1}^{\ast}:=\beta_{1}+(\vec{1}^{T}\mathbf{V}^{-1}\vec{X}_{o})/(\vec{1}^{T}\mathbf{V}^{-1}\vec{1})$,
we get,
\begin{displaymath}
\begin{small}
\begin{array}{rcl}
\E{\left[\begin{array}{c} \hat{\beta}_{1}^{\ast} \\ \hat{\beta}_{2} \end{array}\right] \vert \mathbf{D}_{o}} & = & 
\left[
\begin{array}{c}
\beta_{1}^{\ast} - (\vec{1}^{T}\mathbf{V}^{-1}\vec{1})^{-1}\vec{1}^{T}\mathbf{V}^{-1}\mathbf{V}_{u}\mathbf{V}_{o}^{-1}\vec{X}_{o}\beta_{2} 
\\
(1-(\vec{X}_{o}^{\ast T}\mathbf{V}^{-1}\vec{X}_{o}^{\ast})^{-1}\vec{X}_{o}^{\ast T}\mathbf{V}^{-1}\mathbf{V}_{u}\mathbf{V}_{o}^{-1}\vec{X}_{o})\beta_{2}
\end{array}
\right].
\end{array}
\end{small}
\end{displaymath}
Notice that in the above formula we have both $\vec{X}_{o}$ and $\vec{X}_{o}^{\ast}$.

If we do not have an intercept but some general fixed effects vector then we write the model as,
\begin{displaymath}
\begin{array}{rcl}
\vec{Y}_{o} & = & \mathbf{F}\vec{\beta}_{F} + (\vec{X}_{o} - \mathbf{U})\beta_{2} + \vec{r},
\end{array}
\end{displaymath}
$\E{\mathbf{U}\vert \mathbf{D}_{o}}$ does not change, as this only depends on the random effects and the 
formula ,
\begin{displaymath}
\begin{small}
\begin{array}{rcl}
\E{[ \hat{\vec{\beta}}_{F}^{T}~ \hat{\beta}_{2}]^{T} \vert \mathbf{D}_{o}} & =&
\left[\begin{array}{c} \vec{\beta}_{F} - \kappa_{1}\beta_{2} \\ (1-\kappa_{2})\beta_{2} \end{array}\right],
\end{array}
\end{small}
\end{displaymath}
with 
\begin{displaymath}
\begin{array}{rcl}
\kappa_{1} & =& \frac{(\vec{X}_{o}^{T}\mathbf{V}^{-1}\vec{X}_{o})(\mathbf{F}^{T}\mathbf{V}^{-1}\mathbf{V}_{u}\mathbf{V}_{o}^{-1}\vec{X}_{o})-(\vec{X}_{o}^{T}\mathbf{V}^{-1}\mathbf{V}_{u}\mathbf{V}_{o}^{-1}\vec{X}_{o})(\mathbf{F}^{T}\mathbf{V}^{-1}\vec{X}_{o})}
{(\vec{X}_{o}^{T}\mathbf{V}^{-1}\vec{X}_{o})(\mathbf{F}^{T}\mathbf{V}^{-1}\mathbf{F})-(\mathbf{F}^{T}\mathbf{V}^{-1}\vec{X}_{o})^{2}}, \\
\kappa_{2} & = & \frac{(\vec{X}_{o}^{T}\mathbf{V}^{-1}\mathbf{V}_{u}\mathbf{V}_{o}^{-1}\vec{X}_{o})(\mathbf{F}^{T}\mathbf{V}^{-1}\mathbf{F})-(\vec{X}_{o}^{T}\mathbf{V}^{-1}\mathbf{F})(\mathbf{F}^{T}\mathbf{V}^{-1}\mathbf{V}_{u}\mathbf{V}_{o}^{-1}\vec{X}_{o})}
{(\vec{X}_{o}^{T}\mathbf{V}^{-1}\vec{X}_{o})(\mathbf{F}^{T}\mathbf{V}^{-1}\mathbf{F})-(\mathbf{F}^{T}\mathbf{V}^{-1}\vec{X}_{o})^{2}}.
\end{array}
\end{displaymath}

We can do the same trick as with the intercept and instead of $\vec{X}_{o}$ consider,
$\vec{X}_{o}^{\ast}:=\vec{X}_{o}-(\mathbf{F}^{T}\mathbf{V}^{-1}\vec{X}_{o})/(\mathbf{F}^{T}\mathbf{V}^{-1}\vec{1})\vec{1}$
This has the effect that $\mathbf{F}^{T}\mathbf{V}^{-1}\vec{X}_{o}^{\ast}=0$ but still 
$\E{\vec{X}_{o}^{\ast}}=\mathbf{0}$. The formula then becomes,
\begin{displaymath}
\begin{small}
\begin{array}{rcl}
\E{\left[\begin{array}{c} \hat{\vec{\beta}}_{F}^{\ast} \\ \hat{\beta}_{2} \end{array}\right] \vert \mathbf{D}_{o}} & = & 
\left[
\begin{array}{c}
\vec{\beta}_{F}^{\ast} - (\mathbf{F}^{T}\mathbf{V}^{-1}\mathbf{F})^{-1}\mathbf{F}^{T}\mathbf{V}^{-1}\mathbf{V}_{u}\mathbf{V}_{o}^{-1}\vec{X}_{o}\beta_{2} 
\\
(1-(\vec{X}_{o}^{\ast T}\mathbf{V}^{-1}\vec{X}_{o}^{\ast})^{-1}\vec{X}_{o}^{\ast T}\mathbf{V}^{-1}\mathbf{V}_{u}\mathbf{V}_{o}^{-1}\vec{X}_{o}\beta_{2}
\end{array}
\right],
\end{array}
\end{small}
\end{displaymath}
where $\vec{\beta}_{F}^{\ast}:=\vec{\beta}_{F}+(\mathbf{F}^{T}\mathbf{V}^{-1}\vec{X}_{o})/(\mathbf{F}^{T}\mathbf{V}^{-1}\vec{1})\vec{1}$
and so the slope's coefficient is of the same form as that in equation (\ref{eqSingBias}).
Notice again that in the last formula we have both $\vec{X}_{o}$ and $\vec{X}_{o}^{\ast}$.

If $\mathbf{F}$ is not a vector but a fixed effects matrix (\emph{i.e.} $\vec{\beta}_{F}$ is a vector and not a scalar) then 
again $\E{\mathbf{U}\vert \mathbf{D}_{o}}$ does not change
and the formula becomes,
\begin{displaymath}
\begin{array}{rcl}
\E{\hat{\vec{\beta}}\vert \mathbf{D}_{o}} & = &
\left[
\begin{array}{c}
\vec{\beta}_{F}-\left(\mathbf{A}\mathbf{F}^{T}\mathbf{V}^{-1}\mathbf{V}_{u}\mathbf{V}_{o}^{-1}\vec{X}_{o}+\mathbf{B}\vec{X}_{o}^{T}\mathbf{V}^{-1}\mathbf{V}_{u}\mathbf{V}_{o}^{-1}\vec{X}_{o}\right)\beta_{2}
\\ 
\left(1-(\mathbf{C}\mathbf{F}^{T}\mathbf{V}^{-1}\mathbf{V}_{u}\mathbf{V}_{o}^{-1}\vec{X}_{o}+\mathbf{G}\vec{X}_{o}^{T}\mathbf{V}^{-1}\mathbf{V}_{u}\mathbf{V}_{o}^{-1}\vec{X}_{o})\right)\beta_{2}
\end{array}
\right]
\\&=& 
\left[\begin{array}{c} \beta_{1} - \mathbf{K}_{1}\beta_{2} \\ (1-\kappa_{2})\beta_{2} \end{array}\right],
\end{array}
\end{displaymath}
where, as previously
\begin{displaymath}
\begin{array}{rcl}
\mathbf{A} & =  & \left(\mathbf{F}^{T}\mathbf{V}^{-1}\mathbf{F}-\mathbf{F}^{T}\mathbf{V}^{-1}\vec{X}_{o}(\vec{X}_{o}^{T}\mathbf{V}^{-1}\vec{X}_{o})^{-1}\vec{X}_{o}^{T}\mathbf{V}^{-1}\mathbf{F}\right)^{-1}, \\
\mathbf{B} & = & -\left(\mathbf{F}^{T}\mathbf{V}^{-1}\mathbf{F}-\mathbf{F}^{T}\mathbf{V}^{-1}\vec{X}_{o}(\vec{X}_{o}^{T}\mathbf{V}^{-1}\vec{X}_{o})^{-1}\vec{X}_{o}^{T}\mathbf{V}^{-1}\mathbf{F}\right)^{-1}
\\ && \times
\mathbf{F}^{T}\mathbf{V}^{-1}\vec{X}_{o}(\vec{X}_{o}^{T}\mathbf{V}^{-1}\vec{X}_{o})^{-1},\\
\mathbf{C} & = & -\left(\vec{X}_{o}^{T}\mathbf{V}^{-1}\vec{X}_{o}-\vec{X}_{o}^{T}\mathbf{V}^{-1}\mathbf{F}(\mathbf{F}^{T}\mathbf{V}^{-1}\mathbf{F})^{-1}\mathbf{F}^{T}\mathbf{V}^{-1}\vec{X}_{o}\right)^{-1}
\\ && \times
\vec{X}_{o}^{T}\mathbf{V}^{-1}\mathbf{F}(\mathbf{F}^{T}\mathbf{V}^{-1}\mathbf{F})^{-1},\\
\mathbf{G} & = & \left(\vec{X}_{o}^{T}\mathbf{V}^{-1}\vec{X}_{o}-\vec{X}_{o}^{T}\mathbf{V}^{-1}\mathbf{F}(\mathbf{F}^{T}\mathbf{V}^{-1}\mathbf{F})^{-1}\mathbf{F}^{T}\mathbf{V}^{-1}\vec{X}_{o}\right)^{-1},
\end{array}
\end{displaymath}

We can find a simplification of the formula as before.
We notice that after the study of the single predictor case it might be 
tempting to also consider
$\vec{X}_{o}^{\ast}:=\vec{X}_{o}-\mathbf{J}(\mathbf{F}^{T}\mathbf{V}^{-1}\mathbf{J})^{-1}\mathbf{F}^{T}\mathbf{V}^{-1}\vec{X}_{o}$,
where $\mathbf{J}$ is a matrix of ones, with number of rows equaling the number of observations
and columns equaling the number of columns of $\mathbf{F}$. However one cannot use this, as 
$\mathbf{F}^{T}\mathbf{V}^{-1}\mathbf{J}$ is singular. We define as in the general case
\mbox{
$\vec{X}_{o}^{\ast}:=\vec{X}_{o}-\mathbf{F}(\mathbf{F}^{T}\mathbf{V}^{-1}\mathbf{F})^{-1}\mathbf{F}^{T}\mathbf{V}^{-1}\vec{X}_{o}$}
and as before $\mathbf{F}^{T}\mathbf{V}^{-1}\vec{X}_{o}^{\ast}=\mathbf{0}$.
This results in,
\begin{displaymath}
\begin{array}{rcl}
\mathbf{A} &=& \left(\mathbf{F}^{T}\mathbf{V}^{-1}\mathbf{F}\right)^{-1},\\
\mathbf{B} &=& \mathbf{0}$, $\mathbf{C} = \mathbf{0},\\
\mathbf{G} &=& \left(\vec{X}_{o}^{T}\mathbf{V}^{-1}\vec{X}_{o}\right)^{-1}
\end{array}
\end{displaymath}
and the formula becomes,
\begin{displaymath}
\begin{array}{rcl}
\E{\left[\begin{array}{c}\hat{\vec{\beta}}_{F}^{\ast} \\ \beta_{2} \end{array} \right]\vert \mathbf{D}_{o}} & = &
\left[
\begin{array}{c}
\vec{\beta}_{F}^{\ast}-(\mathbf{F}^{T}\mathbf{V}^{-1}\mathbf{F})^{-1}\mathbf{F}^{T}\mathbf{V}^{-1}\mathbf{V}_{u}\mathbf{V}_{o}^{-1}\vec{X}_{o}\beta_{2} \\
\left(1-(\vec{X}_{o}^{\ast T}\mathbf{V}^{-1}\vec{X}_{o}^{\ast})^{-1}\vec{X}_{o}^{\ast T}\mathbf{V}^{-1}\mathbf{V}_{u}\mathbf{V}_{o}^{-1}\vec{X}_{o})\right)\beta_{2}
\end{array}
\right],
\end{array}
\end{displaymath}
where $\vec{\beta}_{F}^{\ast}:=\beta_{F}+(\mathbf{F}^{T}\mathbf{V}^{-1}\mathbf{F})^{-1}\mathbf{F}^{T}\mathbf{V}^{-1}\vec{X}_{o}$.
This simplification depends on $\E{\vec{X}_{o}}=\mathbf{0}$.

\subsubsection{Mean square error analysis}
If we consider the mean square error of both the intercept and slope then we get the following conditions
using the results and notation from section \ref{sbsecSingPred}.
The corrected estimator of the slope is better if,
\begin{displaymath}
\begin{array}{rcl}
(\kappa_{x}-1)(\kappa_{x}^{3}-\kappa_{x}^{2} + (\frac{a}{\beta^{2}}+a^{2}c)\kappa_{x} + \frac{a}{\beta^{2}}+a^{2}c) & > & 0
\end{array}
\end{displaymath}
which is the same condition, equation (\ref{eq1predMSEk}), as in the no intercept case.
Thus all the results from the no intercept case can be carried over to the case when we have an intercept.
The condition for the corrected estimator of the intercept to have a smaller mean square error is that
\begin{displaymath}
\begin{array}{cl}
&(\vec{1}^{T}\mathbf{V}^{-1}\vec{1})^{-2}
(\vec{1}^{T}\mathbf{V}^{-1}\mathbf{V}_{u}\mathbf{V}_{o}^{-1}\vec{X}_{o})^{2}[(1-a^{2}c)\beta_{2}^{2} -a]
\\-& 
2\frac{(\vec{1}^{T}\mathbf{V}^{-1}\vec{1})^{-2}}{a}
\vec{1}^{T}\mathbf{V}^{-1}\mathbf{V}_{u}\mathbf{V}_{o}^{-1}\vec{X}_{o}
\vec{1}^{T}\mathbf{V}^{-1}
\Cov{\mathbf{U} \vert \vec{X}_{o}}\mathbf{V}^{-1}\vec{X}_{o}^{\ast}\beta_{2}
\end{array}
\end{displaymath}
is positive.
As we can analyze the estimators of the intercept and slope separately it might in some situations
pay--off to use for one parameter the corrected estimator and for the other the uncorrected one.
 
In all of the discussion we assumed that we know all the covariance matrices. In practice this might 
of course not be the case. The only way to know the measurement error covariance matrix
is to make multiple measurements. In comparative analysis studies this is natural as 
for each species we have a mean from a number of individuals. This will cause 
$\mathbf{V}_{e}$ and $\mathbf{V}_{u}$ to be diagonal or of the form $\mathbf{\Sigma} \otimes \mathbf{I}$ 
(in the multivariate case) and we can use this in the estimation procedure.

\subsection{Unconditional bias of GLS estimator with measurement error in predictors}\label{secUncondBias}
The main problem with calculating the bias of the generalized least squares estimator is that one needs
to be able to calculate the expectation of 
\begin{displaymath}
(\mathbf{D}_{o}^{T}\mathbf{V}^{-1}\mathbf{D}_{o})^{-1}
\mathbf{D}_{o}^{T}\mathbf{V}^{-1}\mathbf{V}_{u}\mathbf{V}_{o}^{-1}\mathbf{D}_{o}
\end{displaymath}
for arbitrary covariance matrices $\mathbf{V}$, $\mathbf{V}_{u}$ and $\mathbf{V}_{o}$.
This is difficult even in the case of one predictor as 
the matrix product $\mathbf{V}^{-1}\mathbf{V}_{u}\mathbf{V}_{o}^{-1}$
seems to have few useful for us properties. 
We will now consider the very simple case when $\beta$ is one dimensional, \emph{i.e.}
$\mathbf{D}_{o}$ is of dimension $n \times 1$ which means that we can write the 
expectation of the generalized least squares estimator as,
\begin{displaymath}
\begin{array}{rcl}
\E{\hat{\beta}} & = & \beta - \E{\frac{\mathbf{D}_{o}^{T}\mathbf{V}^{-1}\mathbf{V}_{u}\mathbf{V}_{o}^{-1}
\mathbf{D}_{o}}
{\mathbf{D}_{o}^{T}\mathbf{V}^{-1}\mathbf{D}_{o}}}\beta
\end{array}
\end{displaymath}
and we need to consider,
\begin{displaymath}
\begin{array}{c}
\E{\frac{\mathbf{D}_{o}^{T}\mathbf{V}^{-1}\mathbf{V}_{u}\mathbf{V}_{o}^{-1}\mathbf{D}_{o}}
{\mathbf{D}_{o}^{T}\mathbf{V}^{-1}\mathbf{D}_{o}}}
,
\end{array}
\end{displaymath}
where $\mathbf{D}_{o}\sim \mathcal{N}(\vec{0},\mathbf{V}_{o})$.
It can be easily shown that the matrix
\begin{displaymath}
\mathbf{V}^{-1}\mathbf{V}_{u}\mathbf{V}_{o}^{-1} = \mathbf{V}^{-1}\mathbf{V}_{u}(\mathbf{V}_{d}+\mathbf{V}_{u})^{-1}
=\mathbf{V}^{-1}(\mathbf{V}_{u}^{-1}\mathbf{V}_{d}+\mathbf{I})^{-1}
\end{displaymath}
has positive real eigenvalues. 
\begin{Lemma}\label{lemProdPD}
If $\mathbf{B}$ and $\mathbf{C}$ are symmetric positive definite matrices then $\mathbf{A}=\mathbf{B}\mathbf{C}$
has positive real eigenvalues.
\end{Lemma}
\textbf{Proof}
$\mathbf{C}=\mathbf{B}^{-1}\mathbf{A}$ is by assumption symmetric positive definite (and so Hermitian)
so for any complex $\mathbf{z}$, $\mathbf{z}^{T}\mathbf{B}^{-1}\mathbf{A}\mathbf{z} \in \mathbb{R}_{+}$.
Assume that $\mathbf{A}$ has a complex negative real part eigenvalue $\lambda=(-a)+\mathrm{i}b$ ($a>0$)
and let $\mathbf{z}$ be its (complex) eigenvector. Then
\begin{displaymath}
\begin{array}{ccl}
\mathbb{R}_{+} \in \mathbf{z}^{T}\mathbf{C}\mathbf{z} & = &
\mathbf{z}^{T}\mathbf{B}^{-1}\mathbf{A}\mathbf{z} = 
\mathbf{z}^{T}\mathbf{B}^{-1}\lambda\mathbf{z} = 
\mathbf{z}^{T}\mathbf{B}^{-1}(-a + \mathrm{i}b)\mathbf{z}
\\&=&
-a \mathbf{z}^{T}\mathbf{B}^{-1}\mathbf{z} + \mathrm{i}b \mathbf{z}^{T}\mathbf{B}^{-1}\mathbf{z},
\end{array}
\end{displaymath}
which is complex with negative real part as $\mathbf{z}^{T}\mathbf{B}^{-1}\mathbf{z}\in \mathbb{R}_{+}$
a contradiction. \begin{flushright} \emph{Q.E.D.} \end{flushright}

Results on the products of symmetric positive definite matrices
can be found in \emph{e.g.} \cite{CBal1967I}, \cite{CBal1968II}, \cite{CBal1968III}, 
\cite{CBal1970IV} or \cite{PWu1988}.
However positive real eigenvalues are not sufficient for the matrix to be positive definite 
(see \emph{e.g.} \cite{CJoh70})
and
we cannot have symmetricity guaranteed as there is a product of symmetric matrices. 
The properties of positive definiteness and symmetricity of 
$\mathbf{V}^{-1}\mathbf{V}_{u}\mathbf{V}_{o}^{-1}$ are needed as results for distributions 
of ratios of quadratic forms assume this.
Let us therefore assume that $\mathbf{V}^{-1}\mathbf{V}_{u}\mathbf{V}_{o}^{-1}$ is symmetric positive definite
and define 
\begin{displaymath}
\begin{array}{rcl}
\mathbf{Q} & = & (\mathbf{V}^{-1}\mathbf{V}_{u}\mathbf{V}_{o}^{-1}+(\mathbf{V}^{-1}\mathbf{V}_{u}\mathbf{V}_{o}^{-1})^{T})/2 \\
\mathbf{P} & = & \mathbf{V}^{-1},
\end{array}
\end{displaymath}
and then the expectation we are interested in equals 
\begin{displaymath}
\E{(\mathbf{D}_{o}^{T}\mathbf{Q}\mathbf{D}_{o})/(\mathbf{D}_{o}^{T}\mathbf{P}\mathbf{D}_{o})},
\end{displaymath}
where both $\mathbf{Q}$ and $\mathbf{P}$ are symmetric positive definite. 
Results on expectations of such ratios can be found in \emph{e.g.} \cite{JGur53}, \cite{MSmi89}, \cite{MSmi93}, \cite{MSmi96}
but they are rather complex and in most cases (even if $\mathbf{V}_{u}=\sigma^{2}\mathbf{I}$)
we cannot expect $\mathbf{V}^{-1}\mathbf{V}_{u}\mathbf{V}_{o}^{-1}$ to be symmetric positive definite.

We can also ask what conditions on $\mathbf{V}_{u}$ and $\mathbf{V}_{o}$ 
are necessary and sufficient for the bias to be
of the form $\E{\hat{\beta}\vert \mathbf{D}_{o}}=(\mathbf{I}-\mathbf{K})\vec{\beta}$ 
(a linear combination of the true parameters)
for some matrix $\mathbf{K}$ independent of $\mathbf{D}_{o}$ (in section \ref{sbsecSingPred} we showed the condition for the
single predictor case). From equation \ref{eqBias} we can see that we need 
\mbox{
$\mathrm{vec}^{-1}(\mathbf{V}_{u}\mathbf{V}_{o}^{-1}\mathrm{vec}(\mathbf{D}_{o}))=\mathbf{D}_{o}\mathbf{K}$}
for some appropriate $m\times m$ matrix $\mathbf{K}$
(as the inverse of a matrix is unique). This will be achieved for those covariance matrix 
$\mathbf{V}_{u}$, $\mathbf{V}_{o}$
pairs for which there exists an appropriate $m\times m$ matrix $\mathbf{K}$ such that 
\mbox{
$\mathbf{V}_{u}\mathbf{V}_{o}^{-1}=\mathbf{K}^{T}\otimes\mathbf{I}$.} 
As $\mathbf{V}_{o}=\mathbf{V}_{u}+\mathbf{V}_{d}$ this gives
$\mathbf{V}_{d}=\mathbf{V}_{u}(\mathbf{I} - \mathbf{K}^{T} \otimes \mathbf{I})$ which simplifies to
\begin{displaymath}
\mathbf{V}_{u}=\mathbf{V}_{d}((\mathbf{K}^{T} + \mathbf{I})^{-1} \otimes \mathbf{I})=((\mathbf{K}^{-T}-\mathbf{I})^{-1}\otimes \mathbf{I})\mathbf{V}_{d}.
\end{displaymath}

The condition $\mathbf{V}_{u}\mathbf{V}_{o}^{-1}=\mathbf{K}^{T}\otimes\mathbf{I}$ can be understood in the following
manner. The matrices $\mathbf{V}_{u}$, $\mathbf{V}^{-1}_{o}$ have full rank so their 
rows of $\mathbf{V}_{u}$ and columns of  $\mathbf{V}^{-1}_{o}$ generate the vector space $\mathbb{R}^{nm}$.
Each subset of the rows/columns generates some subspace, in particular consider the following subspaces,
\begin{displaymath}
\begin{array}{c}
\mathbf{\Psi}_{1} = \mathrm{span}\{\mathbf{V}_{u}[(r-1)n+1,], r=1,\ldots,m\} \\ 
\vdots \\ 
\mathbf{\Psi}_{h} = \mathrm{span}\{\mathbf{V}_{u}[(r-1)n+k,], r=1,\ldots,m\} \\ 
\vdots \\ 
\mathbf{\Psi}_{n} = \mathrm{span}\{\mathbf{V}_{u}[(r-1)n+n,], r=1,\ldots,m\} \\ 
\end{array}
\end{displaymath}
and
\begin{displaymath}
\begin{array}{c}
\mathbf{\Phi}_{1}=  \mathrm{span}\{\mathbf{V}_{o}^{-1}[,(r-1)n+1], r=1,\ldots,m \} \\
\vdots \\
\mathbf{\Phi}_{h}=  \mathrm{span}\{\mathbf{V}_{o}^{-1}[,(r-1)n+k], r=1,\ldots,m \} \\
\vdots \\
\mathbf{\Phi}_{n}=  \mathrm{span}\{\mathbf{V}_{o}^{-1}[,(r-1)n+n], r=1,\ldots,m \}, 
\end{array}
\end{displaymath}
where for a matrix $\mathbf{M}$, $\mathbf{M}[i,]$ means the $i$th row of $\mathbf{M}$ and $\mathbf{M}[,j]$ means the $j$th column of $\mathbf{M}$.
The first condition linking the matrices $\mathbf{V}_{u}$ and $\mathbf{V}_{o}$ is that for each
$h=1,\ldots,n$ $\mathbf{\Psi}_{h} =\mathbf{\Phi}_{h} $ and for each $i \neq j$
$\mathbf{\Psi}_{i}$ and $\mathbf{\Phi}_{j}$ are orthogonal. 

The second condition is that there
exists a matrix $\mathbf{K}=[k_{ij}]$ of size $m \times m$ such that for each pair $i$ and $j$ we have
for each $h=1,\ldots,n$ that the dot product of the $i$th vector of the $h$th $\Psi$ base
and the $j$th vector of the $h$th $\Phi$ base is $k_{ji}$ (so we are also assuming there is an ordering 
of base vectors). As the relationship between $\mathbf{V}_{d}$ and $\mathbf{V}_{u}$ is of the same form with the Kronecker 
product, the same interpretation is valid for linking $\mathbf{V}_{d}$ and $\mathbf{V}_{u}$.
We used the wording that $\mathbf{K}$ must be appropriate. This is so that the
matrices $\mathbf{V}_{d}$ and $\mathbf{V}_{u}$ will be covariance matrices. A necessary
condition is in lemma \ref{lemProdPD}, $\mathbf{K}$ must have positive real eigenvalues. 
The family of permissible $\mathbf{K}$ matrices in the case of independent observations
is described in \cite{MAicCRit}.

We can further ask what are the necessary and sufficient conditions on $\mathbf{V}_{u}$ and $\mathbf{V}_{o}$ 
for the bias to be of the form, \\
$\E{\hat{\beta} \vert \mathbf{D}_{o}}= (\mathbf{D}_{o}^{T}\mathbf{V}^{-1}\mathbf{D}_{o})^{-1}\mathbf{D}_{o}^{T}\mathbf{V}^{-1}\mathbf{H}\mathbf{D}_{o}\mathbf{K}$
for some appropriate matrices $\mathbf{H} \in \mathbb{R}^{n \times n}$ and $\mathbf{K} \in \mathbb{R}^{m \times m}$.
The needed relation is 
\mbox{
$\mathbf{V}_{u}\mathbf{V}_{o}^{-1} = \mathbf{K}^{T} \otimes \mathbf{H}$} which
using that $\mathbf{V}_{o}=\mathbf{V}_{u}+\mathbf{V}_{d}$ this becomes,
\begin{displaymath}
\mathbf{V}_{u}=\mathbf{V}_{d}(\mathbf{I}-\mathbf{K}^{T} \otimes \mathbf{H})^{-1}=(\mathbf{K}^{-T} \otimes \mathbf{H}^{-1}-\mathbf{I})^{-1}\mathbf{V}_{d}.
\end{displaymath}

If we look at the bias in the single predictor case, from  equation (\ref{eqSingBias}) we can see that for the bias to be of the ``classical'' form $\E{\hat{\beta}\vert \mathbf{D}_{o}}=(1-\kappa)\beta$
(where $\kappa$ does not depend on $\mathbf{D}_{o}$),
we need that (as $\mathbf{D}_{o}$ can take on any (random) values) $\mathbf{V}^{-1}\mathbf{V}_{u}\mathbf{V}_{o}^{-1}=\kappa \mathbf{V}^{-1}$
or $\mathbf{V}_{u}=\kappa \mathbf{V}_{o}$. Since $\mathbf{V}_{o}=\mathbf{V}_{d}+\mathbf{V}_{u}$, it follows 
\begin{displaymath}\begin{array}{rcl}\mathbf{V}_{u} & = & \frac{\kappa}{1-\kappa} \mathbf{V}_{d},\end{array}\end{displaymath}
which means that the covariance of the errors in the predictors has to be proportional to the covariance between the predictors.
Notice that for $\mathbf{V}_{u}$, $\mathbf{V}_{d}$ to be covariance matrices we need $0<\kappa<1$ and since $\kappa/(1-\kappa)$ is bijective on $(0,1)\rightarrow \mathbb{R}_{+}$,
in this setting the estimate of $\beta$ will always be biased downwards. Notice that the multivariate formulae
are a generalization of this.

\newpage
\section{Supplementary material B: Stochastic differential equations for the Orn\-stein--Uhlen\-beck model (Paper II)}\label{secPapIIB} 

\subsection{Convergence of stochastic process}
One of the most important properties of the considered processes is whether the processes
converge or not. Convergence of a process can be interpreted as adaptation. Below I review the definitions
(see \emph{e.g.}  \cite{GGriDSti} or \cite{DWil}) and the most basic convergence theorems.

\begin{definition}[a.s. convergence]
A stochastic process $\{X(t)\}$ converges \emph{almost surely} to a random variable $X$ if
$P(\{\omega : X(t,\omega) \rightarrow X(\omega) \})=1$.
\end{definition}

\begin{definition}[convergence in probability]
A stochastic process $\{X(t)\}$ converges \emph{in probability} to a random variable $X$ if
for any $\epsilon > 0$ $P(\{\omega : \vert X(t,\omega) - X(\omega) \vert > \epsilon \})\rightarrow 0$
as $t\rightarrow \infty$.
\end{definition}

\begin{definition}[$L^{p}$ convergence]
A stochastic process $\{X(t)\}$ converges in $L^{p}$ to a random variable $X$ if
$\E{\vert X(t) \vert^{p}} < \infty$ for all $t$ and \\
\mbox{
$\E{\vert X(t) - X\vert^{p}} \rightarrow 0$}
as $t\rightarrow \infty$.
\end{definition}

\begin{definition}[weak convergence]
A stochastic process $\{X(t)\}$ converges \emph{weakly} (or \emph{in distribution}) to a random variable $X$ if
the distribution functions $F_{X(t)}(x)$ converge to the distribution function of $X$, $F_{X}(x)$
at every point $x$ of continuity of $F_{X}$.
\end{definition}

The convergences are related as almost sure convergence implies convergence in probability
and convergence in probability implies weak convergence. 
$L^{p}$ convergence implies convergence in probability. The implications do not hold the
other way in general.

\begin{theo}[\cite{AWen}]\label{thCov1}
For every $t$ let $\E{\vert\zeta_{t}\vert^{2}}< \infty$. It is sufficient and necessary for
$\lim \zeta_{t}$ when $t\rightarrow t_{0}$ (or $t\rightarrow \infty$) to exist in $L^{2}$ that
$\E{\zeta_{t}\zeta_{s}}$ when $t,s \rightarrow t_{0}$ (or $t,s \rightarrow \infty$)
exists and is finite.
\end{theo}
 
\begin{theo}[\cite{AWen}]\label{thCov2}
It is necessary and sufficient for the limit of $\zeta_{t}$ to exist in probability for 
$t \rightarrow \infty$ that the two dimensional distributions
$\Psi_{\zeta_{t}\zeta_{s}}$ converge weakly as $t,s \rightarrow \infty$. 
\end{theo}
 
\begin{theo}[L\'evy-Cram\'er continuity theorem \cite{GGriDSti} p.190]
The following are equivalent,
\begin{itemize}
\item $F_{n}$ converges weakly to $F$
\item If $\phi_{F_{n}}(t)$ and $\phi_{F}(t)$ are respectively the characteristic functions of $F_{n}$ and $F$
and for every $t\in \mathbb{R}$ we have $\lim_{n \rightarrow \infty} \phi_{F_{n}}(t) = \phi_{F}(t)$.
\end{itemize}
\emph{i.e.} pointwise convergence of characteristic functions is equivalent to weak convergence.
\end{theo}
 
In the case when we have normal distributions they are characterized by their mean value and covariance
structure. Therefore if we have convergence of mean vectors and covariance matrices this will 
result in convergence of their characteristic functions 
($\phi_{X}(\vec{t})=\exp(i\vec{\mu}\vec{t}-\vec{t}^{T}\Sigma \vec{t}/2)$ \cite{DWil})
implying weak convergence. When we have a stochastic process with continuous time
instead of discrete then if the mean vector and covariance converges this will imply convergence
for every subsequence. In turn this will give us for every subsequence pointwise convergence of characteristic
functions and so for every subsequence weak convergence of measures which 
by definition gives us weak convergence ($\lim_{s\rightarrow t} F_{s}(x)=F_{t}(x)$).

\subsection{Matrix exponential and related integral}
In most of the models to be able to calculate the covariance structure we need to be able 
to calculate the matrix integral function 
\begin{equation}\label{eqIntExp}
\int_{0}^{t} e^{\mathbf{F}v} \mathbf{\Psi} e^{\mathbf{F}^{T}v} \ud v,
\end{equation}
for a matrix $\mathbf{\Psi}$. Given an 
eigendecomposition  $\mathbf{F}=\mathbf{P}\mathbf{\Lambda}\mathbf{P}^{-1}$,
where $\mathbf{\Lambda}$ is the diagonal matrix of eigenvalues and $\mathbf{P}$ is the invertible matrix of eigenvectors,
we have
\begin{displaymath}
\begin{array}{rcl}
e^{-t\mathbf{F}} & = & \mathbf{P} \left[ \begin{array}{ccc} e^{-t\lambda_{1}} & \ldots & 0 \\ \vdots & \ddots & \vdots \\ \ldots & 0 & e^{-t\lambda_{k_{Y}}} \end{array} \right] \mathbf{P}^{-1} \\
\end{array}
\end{displaymath} 
and therefore
\begin{displaymath}
\begin{array}{rcl}
\int_{0}^{t}e^{v\mathbf{F}}\mathbf{\Psi} e^{v\mathbf{F}^{T}}\ud v & = & 
\mathbf{P} 
\int_{0}^{t} \left[ e^{v(\lambda_{i}+\lambda_{j})} \right]_{1\le i,j \le k} \ast \mathbf{\Upsilon}
\ud v \mathbf{P}^{T} 
\end{array}
\end{displaymath} 
where $\mathbf{\Upsilon}:=\mathbf{P}^{-1}\mathbf{\Psi} \mathbf{P}^{-T}$. 
Suppose there are exactly $l\ge 0$ zero--valued eigenvalues and 
$(\lambda_{1},\ldots,\lambda_{k_{Y}})=(\lambda_{1},\ldots,\lambda_{k_{Y}-l},0,\ldots,0)$.
Then (\ref{eqIntExp}) can be obtained as
\begin{displaymath}
{\tiny
\begin{array}{rcl}
\mathbf{P} 
\left[
\begin{array}{c|c}
\left[ \frac{1}{\lambda_{i}+\lambda_{j}}(e^{t(\lambda_{i}+\lambda_{j})}-1) \cdot \Upsilon_{ij}\right]_{1\le i,j \le k-l} 
&
\left[ \frac{1}{\lambda_{i}}(e^{t\lambda_{i}}-1) \cdot \Upsilon_{ij}\right]_{\substack{1 \le i \le k-l \\(k-l+1)\le j \le k}} 
\\ \hline
\left[ \frac{1}{\lambda_{j}}(e^{t\lambda_{j}}-1) \cdot \Upsilon_{ij}\right]_{\substack{(k-l+1) \le i \le k-l \\1\le j \le k-l}}
&
t\left[\mathbf{\Upsilon}_{ij}\right]_{(k-l+1)\le i,j \le k} 
\end{array}
\right]
\mathbf{P}^{T}.
\end{array}
}
\end{displaymath} 
This might not be the most effective way of calculating this integral as in hinges on the precision
of calculating $\mathbf{F}$'s eigendecomposition. The papers
\cite{CMolCVanLoan}, \cite{CVanLoan77}, \cite{CVanLoan78} consider calculating the matrix exponential and related problems.

\subsection{Ornstein--Uhlenbeck model}
In \cite{MButAKinOUCH} a special version of the Ornstein--Uhlenbeck model, equation (\ref{eqOU}), is presented (with software), 
where $\vec{\Psi}(t)$ is a step function over the phylogeny and $\mathbf{F}$ is a symmetric positive
definite matrix (so a positive scalar in the one dimensional case). This restriction on $\mathbf{F}$ results in ease of interpretability
and simple parametrization for numerical optimization of the 
likelihood function (see section \ref{secPapIIC}). However this restriction is not necessary
and using the eigenvalue decomposition of $\mathbf{F}=\mathbf{P}\mathbf{\Lambda}\mathbf{P}^{-1}$ 
($\mathbf{P}\mathbf{\Lambda}\mathbf{P}^{T}$ for $\mathbf{F}$ symmetric positive definite) 
we can work with arbitrary $\mathbf{F}$ matrices. 

In Paper II we present a package that
in fact allows for an arbitrary invertible drift matrix using its eigendecomposition.
The traits evolve according to equation (\ref{eqOU}) whose solution is
(see \cite{LEva} or \cite{CGar}),
\begin{equation}\label{eqSolOU}
\vec{Z}(t)=e^{-\mathbf{F}t}\vec{Z}(0) + \int_{0}^{t} e^{-\mathbf{F}(t-v)}\mathbf{F}\vec{\Psi} \ud v + \int_{0}^{t} e^{-\mathbf{F}(t-v)}\mathbf{\Sigma} \ud \mathbf{W}(v).
\end{equation}
Using that $\frac{\ud}{\ud t}e^{\mathbf{F}t} = e^{\mathbf{F}t}\mathbf{F}=\mathbf{F}e^{\mathbf{F}t}$ 
and assuming $\vec{\Psi}(t)\equiv \vec{\Psi}$ is constant this can be written as,
\begin{displaymath}
\begin{array}{rcl}
\vec{Z}(t)&=&e^{-\mathbf{F}t}\vec{Z}(0) + e^{-\mathbf{F}t}(e^{\mathbf{F}t}-\mathbf{I})\vec{\Psi} + \int_{0}^{t} e^{-\mathbf{F}(t-v)}\mathbf{\Sigma} \ud \mathbf{W}(v) 
 \\&=&
\vec{\Psi}+e^{-\mathbf{F}t}(\vec{Z}(0) -\vec{\Psi}) + \int_{0}^{t} e^{-\mathbf{F}(t-v)}\mathbf{\Sigma} \ud \mathbf{W}(v),
\end{array}
\end{displaymath}
If $\vec{\Psi}(t)$ is a step function 
such that on the intervals $(t_{j-1},t_{j}]$, $\vec{\Psi}(t)$ takes
values $\vec{\Psi}_{j}$, then
\begin{displaymath}
\vec{Z}(t)=e^{-\mathbf{F}t}\left(\vec{Z}(0) +\sum_{j=1}^{m}(e^{t_{j}\mathbf{F}}-e^{t_{j-1}\mathbf{F}})\vec{\Psi}_{j}\right) + \int_{0}^{t} e^{-\mathbf{F}(t-v)}\mathbf{\Sigma} \ud \mathbf{W}(v).
\end{displaymath}
As this is a multivariate normal process we need to calculate the mean and covariance functions. They will
be, conditional on the initial value $\vec{Z}(0)$,
\begin{displaymath}
\begin{array}{rcl}
\E{\vec{Z}(t)}  & = &  e^{-\mathbf{F}t}\left(\vec{Z}(0) +\sum_{j=1}^{m}(e^{t_{j}\mathbf{F}}-e^{t_{j-1}\mathbf{F}})\vec{\Psi}_{j}\right) \\
\end{array}
\end{displaymath}
and
\begin{equation}
\begin{array}{rcl}
\Cov{\vec{Z}(t)} & = & e^{-\mathbf{F}t} \left( \int_{0}^{t}e^{\mathbf{F}v}\mathbf{\Sigma}\mathbf{\Sigma}^{T}e^{\mathbf{F}^{T}v}\ud v \right) e^{-\mathbf{A}^{T}t}.
\end{array}\label{eqCovOU}
\end{equation}
The second moment of the process can be calculated to be
\begin{displaymath}
\mathbf{P}\left(\left[ \frac{1}{\lambda_{i}+\lambda_{j}}(1-e^{-(\lambda_{i}+\lambda_{j})t})\right]_{1 \le i,j \le k} 
\ast (\mathbf{P}^{-1}\mathbf{\Sigma}\mathbf{\Sigma}^{T}\mathbf{P}^{T}) \right) \mathbf{P}^{T},
\end{displaymath}
where $k$ denotes the dimension of $\mathbf{F}$.

We are also interested in the distribution of the data on the phylogeny. To do this we need
to calculate $\cov{\vec{Z}_{i}}{\vec{Z}_{j}}$ for all species $i$ and 
$j$. We use the identity from \cite{EMarTHan96} 
\begin{displaymath}
\begin{array}{rcl}
\cov{\vec{Z}_{i}}{\vec{Z}_{j}} & = &
\cov{\E{\vec{Z}_{i}\vert \vec{Z}_{a_{ij}}}}{\E{\vec{Z}_{j}\vert \vec{Z}_{a_{ij}}}},
\end{array}
\end{displaymath}
where $\vec{Z}_{a_{ij}}$ is the value of the trait vector of the most recent common ancestor
of species $i$ and $j$. This will result in, 
\begin{equation}\label{eqOUcovSpec}
\begin{array}{rcl}
\cov{\vec{Z}_{i}}{\vec{Z}_{j}} & = &
e^{-\mathbf{F}(t_{i}-t_{a_{ij}})}\Cov{\vec{Z}_{a}}e^{-\mathbf{F}^{T}(t_{j}-t_{a_{ij}})}.
\end{array}
\end{equation}

Depending on whether $\mathbf{F}$ is has positive real part eigenvalues or not we will get different
asymptotic properties of the process.

\subsubsection{Asymptotic properties of unitrait Ornstein--Uhlenbeck model}
Consider the single trait case we have, assuming that $\Psi(t)\equiv \Psi$ is 
constant and let $\alpha>0$.
\begin{displaymath}
\begin{array}{rcl}
\ud Z(t) & = & -\alpha(Z(t)-\Psi)\ud t +\sigma\ud W \\
Z(t) & = & \Psi+(Z(0)-\Psi)e^{-\alpha t}+\sigma \int_{0}^{t} e^{-\alpha(t-v)}\ud W(v).
\end{array}
\end{displaymath}
We see that
\begin{displaymath}
\begin{array}{rcl}
\E{Z(t)} & = & \Psi+(Z(0)-\Psi)e^{-\alpha t} \rightarrow \Psi \\
\Var{Z(t)} & = & \frac{\sigma^{2}}{2\alpha}(1-e^{-2\alpha t}) \rightarrow \frac{\sigma^{2}}{2 \alpha}.
\end{array}
\end{displaymath}
It follows that $Z(t)$ converges weakly to a normal random variable $\mathcal{N}(\Psi,\frac{\sigma^{2}}{2 \alpha})$.
We can ask whether it will converge in probability.
Observe that 
\begin{displaymath}
\begin{array}{l}
\cov{Z(t)}{Z(s)}=
\frac{\sigma^{2}}{2 \alpha}(e^{-\alpha\vert t-s \vert}-e^{-\alpha(t+s)}) 
=
\frac{\sigma^{2}}{2 \alpha}e^{-\alpha\vert t-s \vert} + o(1),
\end{array}
\end{displaymath}
by direct calculation or using equation (\ref{eqOUcovSpec}).
According to theorem \ref{thCov2}
we do not have convergence for $t,s \rightarrow \infty$ as the limit depends on $\vert t -s \vert$ 
so we will be able to choose subsequences that will give different limits. Therefore the process
does not converge in probability nor almost surely, only weakly.
An intuition what this means can be that the process can have spikes but these will last
for very short periods of time.

The limiting distribution is the stationary distribution of the process, this can be seen if we assume 
that $Z(0) \sim \mathcal{N}(\Psi,\frac{\sigma^{2}}{2 \alpha})$. We then have
\begin{displaymath}
\begin{array}{rcl}
\E{Z(t)} & = & \Psi+(\E{Z(0)}-\Psi)e^{-\alpha t} = \Psi+(\Psi-\Psi)e^{-\alpha t} = \Psi \\
\Var{Z(t)} & = & e^{-2\alpha t}\Var{Z(0)} +\frac{\sigma^{2}}{2\alpha}(1-e^{-2\alpha t}) 
\\ &=& 
e^{-2\alpha t}\frac{\sigma^{2}}{2 \alpha} +\frac{\sigma^{2}}{2\alpha}(1-e^{-2\alpha t}) = 
\frac{\sigma^{2}}{2 \alpha}.
\end{array}
\end{displaymath}
If $\alpha \le 0$ then no stationary distribution exists, nor do we have weak convergence (and 
if $\alpha=0$ the process becomes a Brownian motion).

\subsubsection{Asymptotic properties of multitrait Ornstein--Uhlenbeck model}
The distribution of the vector of traits $\vec{Z}$ at time $t$ is always multivariate normal. We can ask what does this tend to
as $t\rightarrow \infty$. The distribution of $\vec{Z}$ at $t=\infty$ (assuming $\mathbf{F}$ has positive real part eigenvalues)
will be multivariate normal 
(as the characteristic functions will converge). The mean value of $\vec{Z}$ will tend to (assuming
$\vec{\Psi}(t)\equiv \vec{\Psi}$ is constant from some time point),
\begin{displaymath}
\lim_{t \rightarrow \infty} \E{\vec{Z}(t)}= \vec{\Psi}
\end{displaymath}
and the covariance structure
\begin{displaymath}
\begin{array}{rcl}
\lim\limits_{t \rightarrow \infty} \Cov{\vec{Z}(t)} &  = & \int_{0}^{\infty}e^{-\mathbf{F}(t-v)} \mathbf{\Sigma}\mathbf{\Sigma}^{T}e^{-\mathbf{F}^{T}(t-v)} \ud v \\
& =&\mathbf{P} \left(\left[ \frac{1}{\lambda_{i} + \lambda_{j}}\right]_{1\le i,j \le k_{Y}} \ast (\mathbf{P}^{-1}\mathbf{\Sigma}\mathbf{\Sigma}^{T}\mathbf{P}^{-T}) \right) \mathbf{P}^{T}.
\end{array}
\end{displaymath}
This gives us that the process describing the species' evolution on the tree 
tends, in distribution, to a multivariate
normal with the above mean and covariance structure. What is obvious from equation (\ref{eqOUcovSpec}) is that the covariance (and in turn correlation) 
between two traits in different species will be tending to $0$ as $t\rightarrow \infty$ (as the covariance at $t_{a_{ij}}$ is fixed). 

\subsection{Ornstein--Uhlenbeck Brownian Motion model}
In \cite{THanJPieSOrzSLOUCH} (with further developments in \cite{ALabJPieTHan}, \cite{JPieKBarTHanKVoj}) a very special type of the Ornstein--Uhlenbeck model (introduced in \cite{EMarTHan97}) was presented. It assumed
that the drift matrix has only non--zero values on the first row and the diffusion matrix is diagonal, this implies that the variables 
``assigned'' to the other rows evolve marginally
as independent Brownian motions. We denote by $y(t)$ the first variable and the remaining ones by $x_{1}(t),\ldots,x_{k_{x}}(t)$ 
which results in the equation,
\begin{displaymath}
\begin{array}{rcl}
\ud y_{\phantom{1}}(t) & = & -\alpha[y(t) - b_{0} - \sum_{l=1}^{k_{x}}b_{l}x_{l}(t)]\ud t + \sigma_{y} \ud W_{y}, \\
\ud x_{l}(t) & = & \sigma_{x_{l}} \ud W_{x_{l}} ,~~~~  1 \le l \le k_{x},
\end{array}
\end{displaymath}
Due to the layering structure this model is termed the Ornstein--Uhlenbeck Brownian Motion (OUBM) model.
In this particular case the solution equation (\ref{eqSolOU}) takes the form,
\begin{displaymath}
\begin{array}{rclr}
y(t) & = & e^{-\alpha t}y(0) + (1-e^{-\alpha t}) (b_{0}+\sum_{l=1}^{k_{x}} b_{l}x_{l}(0))  
\\ && +
\int_{0}^{t} e^{-\alpha(t-s)}\sigma_{y} \ud W_{y}(s) + \sum_{l=1}^{k_{x}} \int_{0}^{t}(1-e^{-\alpha(t-s)})b_{l}\sigma_{x_{l}}\ud W_{x_{l}}(s), \\
x_{l}(t) & = & x_{l}(0) + \int_{0}^{t} \sigma_{x_{l}} \ud W_{x_{l}}(s), ~~~~ 1 \le l\le k_{x}.
\end{array}
\end{displaymath}
The corresponding moments are,
\begin{displaymath}
\begin{array}{rcl}
\E{y(t)} & = & e^{-\alpha t}y(0) + (1-e^{-\alpha t}) (b_{0}+\sum_{l=1}^{k_{x}} b_{l}x_{l}(0)) \\
\E{x_{l}(t)} & = & x_{l}(0) ~~~~ 1 \le l \le k_{x} \\
\Var{y(t)} & = & \frac{1-e^{-2\alpha t}}{2 \alpha}\sigma_{y}^{2} + \left(t+\frac{1-e^{-2\alpha t}}{2\alpha}-\frac{2}{\alpha}(1-e^{-\alpha t})\right)\sum_{l=1}^{k_{x}}\sigma_{x_{l}}^{2}b_{l}^{2} \\
\Var{x_{l}(t)} & = & \sigma_{x_{l}}^{2}t ~~~~ 1 \le l \le k_{x} \\
\cov{y}{x_{l}(t)} & = & (t-\frac{1-e^{-\alpha t}}{\alpha})\sigma_{x_{l}}^{2}b_{l} ~~~~ 1 \le l \le k_{x} \\  
\cov{x_{r}}{x_{l}(t)} & = & 0 ~~~~ 1 \le r,l \le k_{x}, r\neq l.   
\end{array}
\end{displaymath}
We calculate the correlation structure between species $i$ and $j$ stemming from the phylogeny, 
\begin{displaymath}
\begin{array}{rcl}
\cov{y^{i}_{\phantom{1}}}{y^{j}_{\phantom{1}}} & = &  e^{-\alpha(t_{i}+ t_{j} - 2t_{a_{ij}})}
\left(\frac{1-e^{-2\alpha t_{a_{ij}}}}{2\alpha}(\sigma_{y}^{2}+\sum_{l=1}^{k_{x}}\sigma_{x_{l}}^{2}b_{l}^{2})
+\right.\\ &&\left.+
(t_{a_{ij}}-\frac{2(1-e^{-\alpha t_{a_{ij}}})}{\alpha})\sum_{l=1}^{k_{x}}\sigma_{x_{l}}^{2}b_{l}^{2}\right) \\
\cov{y^{i}_{\phantom{1}}}{x^{j}_{l}}  &=&  e^{-\alpha (t_{i}-t_{a_{ij}})}(t_{a_{ij}}-\frac{1-e^{-\alpha t_{a_{ij}}}}{\alpha})\sigma_{x_{l}}^{2}b_{l}
\\ &&
+(1-e^{-\alpha (t_{i}-t_{a_{ij}})})t_{a_{ij}}\sigma_{x_{l}}^{2}b_{l}\\
\cov{x^{i}_{r}}{x^{j}_{l}}  &=&  0 ~~~~~~ r \neq l \\
\cov{x^{i}_{l}}{x^{j}_{l}}  &= & t_{a_{ij}} \sigma_{x_{l}}^{2}.
\end{array}
\end{displaymath}
We can immediately see that, as long as not all $b_{l}$ are $0$ there will be no stationary 
distribution and no convergence of the process even in distribution
as the Brownian motion type variables causes the variance to escape to infinity at a linear rate. 

\subsection{Multivariate Ornstein--Uhlenbeck Brownian Motion model}
In Paper II we generalize the above Ornstein--Uhlenbeck Brownian Motion model to the situation where $y$ can be a vector $\vec{Y}(t)$ of length $k_{Y}$ and the 
$x$ variables can evolve in a correlated fashion. 
We term it multivariate Ornstein--Uhlenbeck Brownian Motion (mvOUBM).
The stochastic differential equation formulation for this model is in 
equation (\ref{eqGenModel2}) in section \ref{secOU} and we called it
the multivariate Ornstein--Uhlenbeck Brownian Motion (mvOUBM). This model has a 
very special decomposition of the drift matrix,
\begin{displaymath}
\begin{array}{rcl}
\mathbf{F} & = & \left[\begin{array}{c|c} \mathbf{A} & \mathbf{B} \\\hline \mathbf{0} & \mathbf{0} \end{array} \right].
\end{array}
\end{displaymath}
The easiness of computing any moments of the model hinges directly on the easiness of 
computing
\begin{displaymath}
\begin{array}{rcl}
e^{-\mathbf{F}t} & = &
\sum_{n=0}^{\infty}
\frac{(-t)^{n}}{n!}\mathbf{F}^{n}.
\end{array}
\end{displaymath}
We have
\begin{displaymath}
\begin{array}{rcl}
\left[\begin{array}{c|c} \mathbf{A} & \mathbf{B} \\ \hline \mathbf{0} & \mathbf{0}\end{array}\right]^{n} & = &
\left\{\begin{array}{cr}\mathbf{I} & n=0\\
\left[\begin{array}{c|c} \mathbf{A}^{n} & \mathbf{0} \\ \hline \mathbf{0} & \mathbf{0}\end{array}\right]  +
\left[\begin{array}{c|c} \mathbf{A}^{n-1} & \mathbf{0} \\ \hline \mathbf{0} & \mathbf{0}\end{array}\right] 
\left[\begin{array}{c|c} \mathbf{0} & \mathbf{B} \\ \hline \mathbf{0} & \mathbf{0}\end{array}\right] &  
n \neq 0 \end{array}\right .
\end{array}
\end{displaymath}
which can be easily shown by induction. 
Assuming $\mathbf{A}^{-1}$ exists 
we get 
\begin{displaymath}
\begin{array}{rcl}
e^{-\mathbf{F}t}&=&\mathbf{I} + 
\sum_{n=1}^{\infty}\frac{(-t)^{n}}{n!}\left(
\left[\begin{array}{c|c} \mathbf{A}^{n} & \mathbf{0} \\ \hline \mathbf{0} & \mathbf{0}\end{array}\right]  
+ 
\left[\begin{array}{c|c} \mathbf{A}^{n-1} & \mathbf{0} \\ \hline \mathbf{0} & \mathbf{0}\end{array}\right] 
\left[\begin{array}{c|c} \mathbf{0} & \mathbf{B} \\ \hline \mathbf{0} & \mathbf{0}\end{array}\right]\right) \\
&=&\mathbf{I} + 
\left[\begin{array}{c|c} \sum_{n=1}^{\infty}\frac{(-t)^{n}}{n!}\mathbf{A}^{n} & \mathbf{0} \\ \hline \mathbf{0} & \mathbf{0}\end{array}\right]  
\\&&+
\left(
\mathbf{I} + \sum_{n=1}^{\infty}\frac{(-t)^{n}}{n!}\left[\begin{array}{c|c} \mathbf{A}^{n} & \mathbf{0} \\ \hline \mathbf{0} & \mathbf{0}\end{array}\right] - \mathbf{I}
\right)
\left[\begin{array}{c|c} \mathbf{0} & \mathbf{A}^{-1}\mathbf{B} \\ \hline \mathbf{0} & \mathbf{0}\end{array}\right] 
\\ &= & 
\left[\begin{array}{c|c}e^{-t\mathbf{A}} & \mathbf{0} \\ \hline \mathbf{0} & \mathbf{I}\end{array}\right]  
+
\left( \left[\begin{array}{c|c}e^{-t\mathbf{A}} & \mathbf{0} \\ \hline \mathbf{0} & \mathbf{I}\end{array}\right]  - \mathbf{I}  \right)
\left[\begin{array}{c|c} \mathbf{0} & \mathbf{A}^{-1}\mathbf{B} \\ \hline \mathbf{0} & \mathbf{0}\end{array}\right] 
 \\ &=& 
\left[\begin{array}{c|c}e^{-t\mathbf{A}} & \mathbf{0} \\ \hline \mathbf{0} & \mathbf{I}\end{array}\right]  \left[\begin{array}{c|c} \mathbf{I} & \mathbf{A}^{-1}\mathbf{B} \\ \hline \mathbf{0} & \mathbf{I}\end{array}\right]
-\left[\begin{array}{c|c} \mathbf{0} & \mathbf{A}^{-1}\mathbf{B} \\ \hline \mathbf{0} & \mathbf{0}\end{array}\right] 
\\ &= &
\ePotegObl{-t}.
\end{array}
\end{displaymath}
If we further assume that the ``optimum'' vector $\vec{\psi}(t)$ is a step function we get that
$\E{[\vec{Y}^{T}~ \vec{X}^{T}]^{T}}(t)$ equals,
\begin{displaymath}
\begin{array}{cl}
& \ePotegObl{-t} \left[ \begin{array}{c} \vec{Y} \\ \vec{X} \end{array}\right] (0) 
\\ +&
\ePotegObl{-t} \sum_{j=1}^{m} \int_{t_{j-1}}^{t_{j}} e^{\mathbf{F}v}\mathbf{F}\ud v
\left[\begin{array}{c} \vec{\psi}_{j} \\ \vec{0} \end{array}\right]  
\\ = &
\ePotegObl{-t} \left[ \begin{array}{c} \vec{Y} \\ \vec{X} \end{array} \right] (0) 
\\ + &
\ePotegObl{-t} \sum_{j=1}^{m}\left( e^{\mathbf{F}t_{j}}
-e^{\mathbf{F}t_{j-1}}\right)
\left[ \begin{array}{c} \vec{\psi}_{j} \\ \vec{0} \end{array} \right]  
 \\ = &
\ePotegObl{-t}\left[\begin{array}{c} \vec{Y} \\ \vec{X} \end{array} \right] (0) 
 \\ + &
\ePotegObl{-t} \sum_{j=1}^{m} \left(\ePotegObl{t_{j}} 
\right.\\-&\left. 
\ePotegObl{t_{j-1}}\right)
\left[\begin{array}{c} \vec{\psi}_{j} \\ \vec{0}\end{array}\right]
 \\= &
\left[ \begin{array}{c}
-\mathbf{A}^{-1}\mathbf{B}\vec{X}(0) + e^{-t\mathbf{A}}(\vec{Y}(0)+\sum_{j=1}^{m}(e^{t_{j}\mathbf{A}}-e^{t_{j-1}\mathbf{A}})\vec{\psi}_{j}+\mathbf{A}^{-1}\mathbf{B}\vec{X}(0)) \\ \vec{X}(0) \end{array} \right]
\end{array}
\end{displaymath} 
Let us denote 
$\left[\begin{array}{c|c}\mathbf{\Sigma}^{yy} & \mathbf{\Sigma}^{yx} \\ \hline\mathbf{\Sigma}^{xy} & \mathbf{\Sigma}^{xx} \end{array}\right]
\left[\begin{array}{c|c}\mathbf{\Sigma}^{yy} & \mathbf{\Sigma}^{yx} \\ \hline \mathbf{\Sigma}^{xy} & \mathbf{\Sigma}^{xx} \end{array} \right]^{T}=:
\left[\begin{array}{c|c} \mathbf{\Sigma}_{11} & \mathbf{\Sigma}_{12} \\ \hline  \mathbf{\Sigma}_{21} & \mathbf{\Sigma}_{22} \end{array}\right]$, 
and by equation (\ref{eqCovOU}) we get the covariance equaling,
\begin{displaymath}
\begin{array}{cl}
 = &\ePotegObl{-t}
 \\  \times&\int_{0}^{t} \left[ \begin{array}{c|c}
e^{v\mathbf{A}}\mathbf{\Sigma}_{11} + (e^{v\mathbf{A}}-\mathbf{I})\mathbf{A}^{-1}\mathbf{B}\mathbf{\Sigma}_{21} & e^{v\mathbf{A}}\mathbf{\Sigma}_{12}+(e^{v\mathbf{A}}-\mathbf{I})\mathbf{A}^{-1}\mathbf{B}\mathbf{\Sigma}_{22} \\ \hline\mathbf{\Sigma}_{21} & \mathbf{\Sigma}_{22} \end{array} \right]
 \\ \times&
\ePotegOblT{v} \ud v \ePotegOblT{-t}. \end{array}
\end{displaymath}
This then becomes
\begin{displaymath}
\begin{array}{rcl}
\Cov{\vec{Y}(t)} & = &
\int_{0}^{t}e^{-\mathbf{A}v}\mathbf{\Sigma}_{11}e^{-\mathbf{A}^{T}v} \ud v  
 \\ && +
\int_{0}^{t}e^{-\mathbf{A}v}\mathbf{A}^{-1}\mathbf{B}\mathbf{\Sigma}_{21}e^{-\mathbf{A}^{T}v} \ud v  
 \\ && + 
\int_{0}^{t}e^{-\mathbf{A}v}\mathbf{\Sigma}_{12}\mathbf{B}^{T}\mathbf{A}^{-T}e^{-\mathbf{A}^{T}v} \ud v 
 \\ && +
\int_{0}^{t}e^{-\mathbf{A}v}\mathbf{A}^{-1}\mathbf{B}\mathbf{\Sigma}_{22}\mathbf{B}^{T}\mathbf{A}^{-T}e^{-\mathbf{A}^{T}v} \ud v 
 \\ && -
\mathbf{A}^{-1}\mathbf{B}(\mathbf{\Sigma}_{21}+\mathbf{\Sigma}_{22}\mathbf{B}^{T}\mathbf{A}^{-T})
\mathbf{A}^{-T}(\mathbf{I}-e^{-\mathbf{A}^{T}t})
 \\ && -
(\mathbf{I}-e^{-\mathbf{A}t})\mathbf{A}^{-1}(\mathbf{\Sigma}_{12}
+\mathbf{A}^{-1}\mathbf{B}\mathbf{\Sigma}_{22})\mathbf{B}^{T}\mathbf{A}^{-T} 
 \\ && +
t\mathbf{A}^{-1}\mathbf{B}\mathbf{\Sigma}_{22}\mathbf{B}^{T}\mathbf{A}^{-T} \\
\cov{\vec{Y}(t)}{\vec{X}(t)} & = &
(\mathbf{I}-e^{-\mathbf{A}t})\mathbf{A}^{-1}
(\mathbf{\Sigma}_{12}+\mathbf{A}^{-1}\mathbf{B}\mathbf{\Sigma}_{22})
 \\&& -
t\mathbf{A}^{-1}\mathbf{B}t\mathbf{\Sigma}_{22} \\
\Cov{\vec{X}(t)} & = & t\mathbf{\Sigma}_{22}.
\end{array}
\end{displaymath}

We calculate the covariance structure coming from the phylogeny between species $i$ and $j$,
\begin{displaymath}
\begin{array}{l}
\cov{\vec{Y}_{i}}{\vec{Y}_{j}}  =  
e^{-\mathbf{A}(t_{i}-t_{a_{ij}})}\left(
\int_{0}^{t_{a_{ij}}}e^{v\mathbf{A}}\mathbf{\Sigma}_{11}e^{v\mathbf{A}^{T}} \ud v  
\right. \\ + \left.
\int_{0}^{t_{a_{ij}}}e^{v\mathbf{A}}\mathbf{A}^{-1}\mathbf{B}\mathbf{\Sigma}_{21}e^{v\mathbf{A}^{T}} \ud v 
+ \int_{0}^{t_{a_{ij}}}e^{v\mathbf{A}}\mathbf{\Sigma}_{12}\mathbf{B}^{T}\mathbf{A}^{-T}e^{v\mathbf{A}^{T}} \ud v 
\right.  \\  + \left.
\int_{0}^{t_{a_{ij}}}e^{v\mathbf{A}}\mathbf{A}^{-1}\mathbf{B}\mathbf{\Sigma}_{22}\mathbf{B}^{T}\mathbf{A}^{-T}e^{v\mathbf{A}^{T}} \ud v 
\right)e^{-\mathbf{A}^{T}(t_{j}-t_{a_{ij}})} 
 \\  -
e^{-\mathbf{A}(t_{i}-t_{a_{ij}})}(\mathbf{I}-e^{-t_{a_{ij}}\mathbf{A}})\mathbf{A}^{-1}(\mathbf{\Sigma}_{12}+\mathbf{A}^{-1}\mathbf{B}\mathbf{\Sigma}_{22})\mathbf{B}^{T}\mathbf{A}^{-T} 
 \\  +
\mathbf{A}^{-1}\mathbf{B}(\mathbf{\Sigma}_{21}+\mathbf{\Sigma}_{22}\mathbf{B}^{T}\mathbf{A}^{-T})\mathbf{A}^{-T}(\mathbf{I}-e^{-t_{a_{ij}}\mathbf{A}^{T}})e^{-\mathbf{A}^{T}(t_{j}-t_{a_{ij}})}
\\+
t_{a_{ij}}\mathbf{A}^{-1}\mathbf{B}\mathbf{\Sigma}_{22}\mathbf{B}^{T}\mathbf{A}^{-T}
\\
\cov{\vec{Y}_{i}}{\vec{X}_{j}}  =  e^{-\mathbf{A}(t_{i}-t_{a_{ij}})}(\mathbf{I}-e^{-t_{a_{ij}}\mathbf{A}})(\mathbf{\Sigma}_{12}+\mathbf{A}^{-1}\mathbf{B}\mathbf{\Sigma}_{22})-t_{a_{ij}}\mathbf{A}^{-1}\mathbf{B}\mathbf{\Sigma}_{22} \\
\cov{\vec{X}_{i}}{\vec{X}_{j}}  =  t_{a_{ij}} \mathbf{\Sigma}_{22} .
\end{array}
\end{displaymath}

\begin{figure}[!h]
~~A~~~~~~~~~~~~~~~~~~~~~~~~~~~~~~B~~~~~~~~~~~~~~~~~~~~~~~~~~~~~C\\
\includegraphics[width=0.3\textwidth]{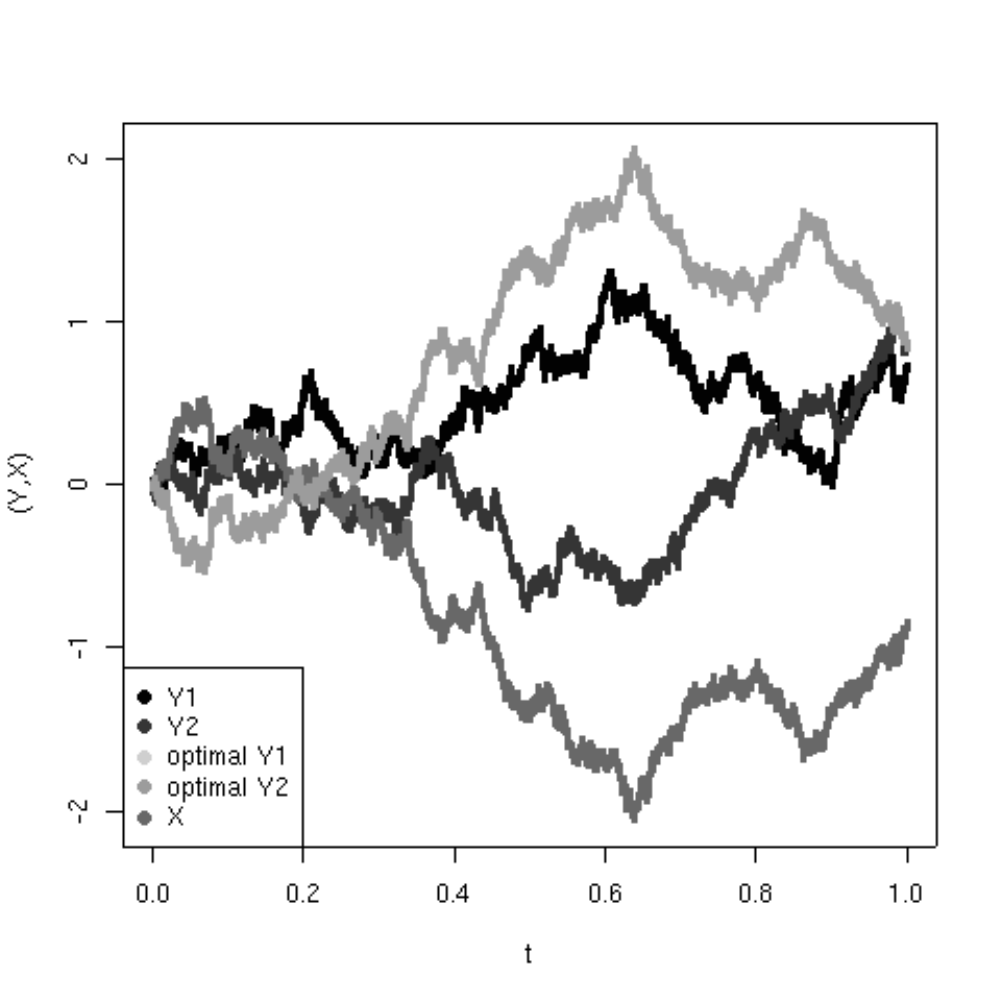}
\includegraphics[width=0.3\textwidth]{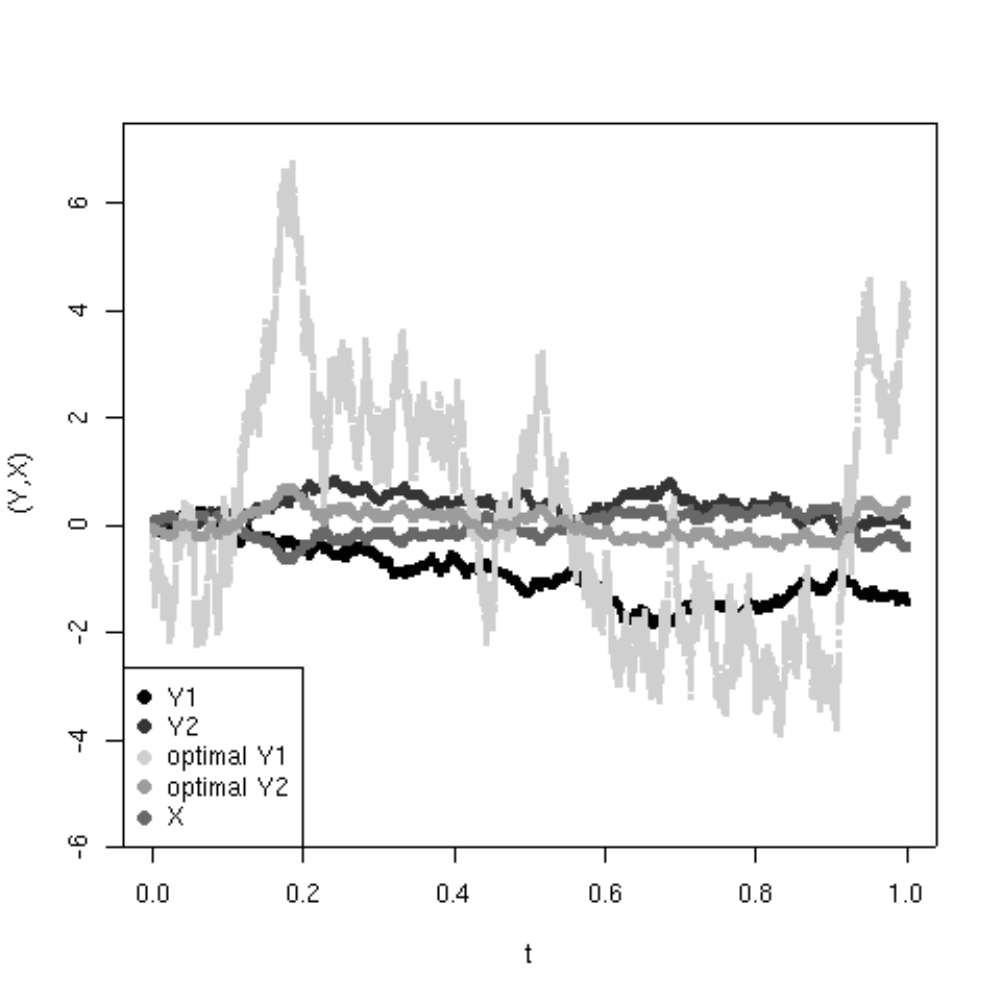}
\includegraphics[width=0.3\textwidth]{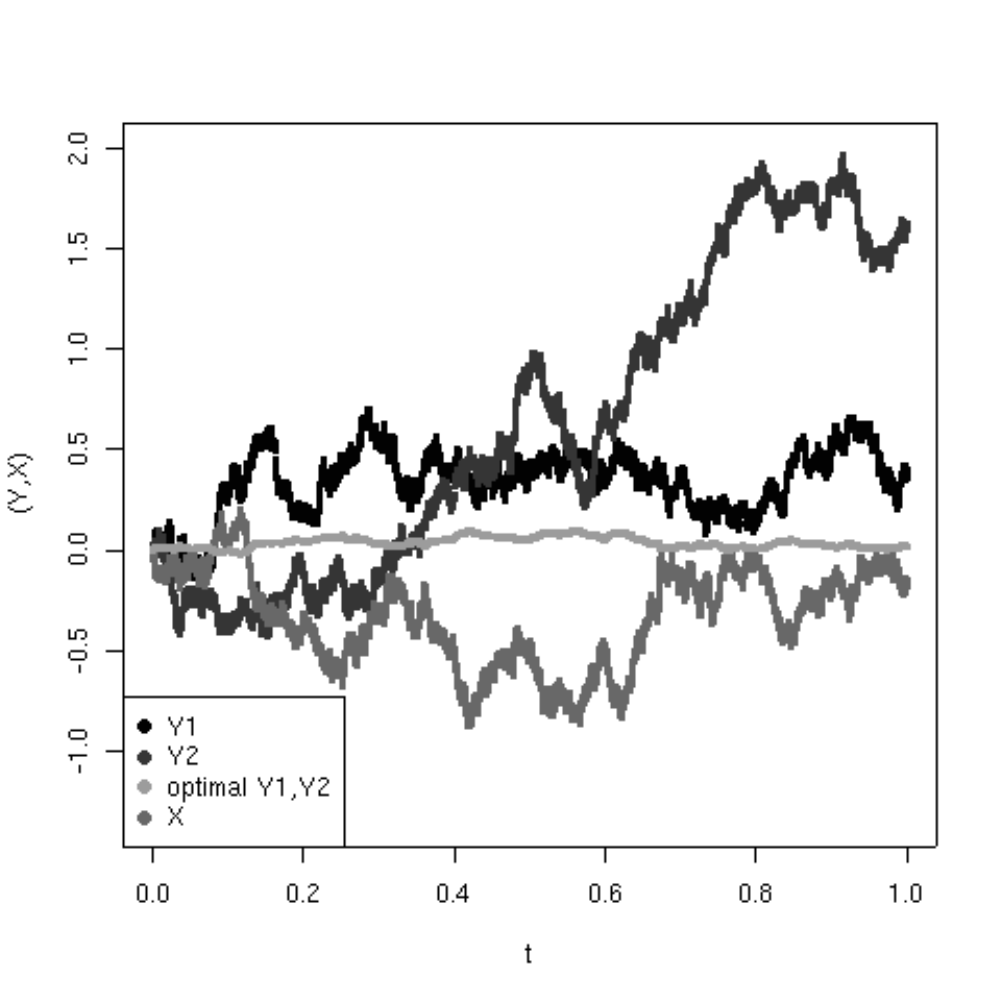}\\
~~D~~~~~~~~~~~~~~~~~~~~~~~~~~~~~~E~~~~~~~~~~~~~~~~~~~~~~~~~~~~~F\\
\includegraphics[width=0.3\textwidth]{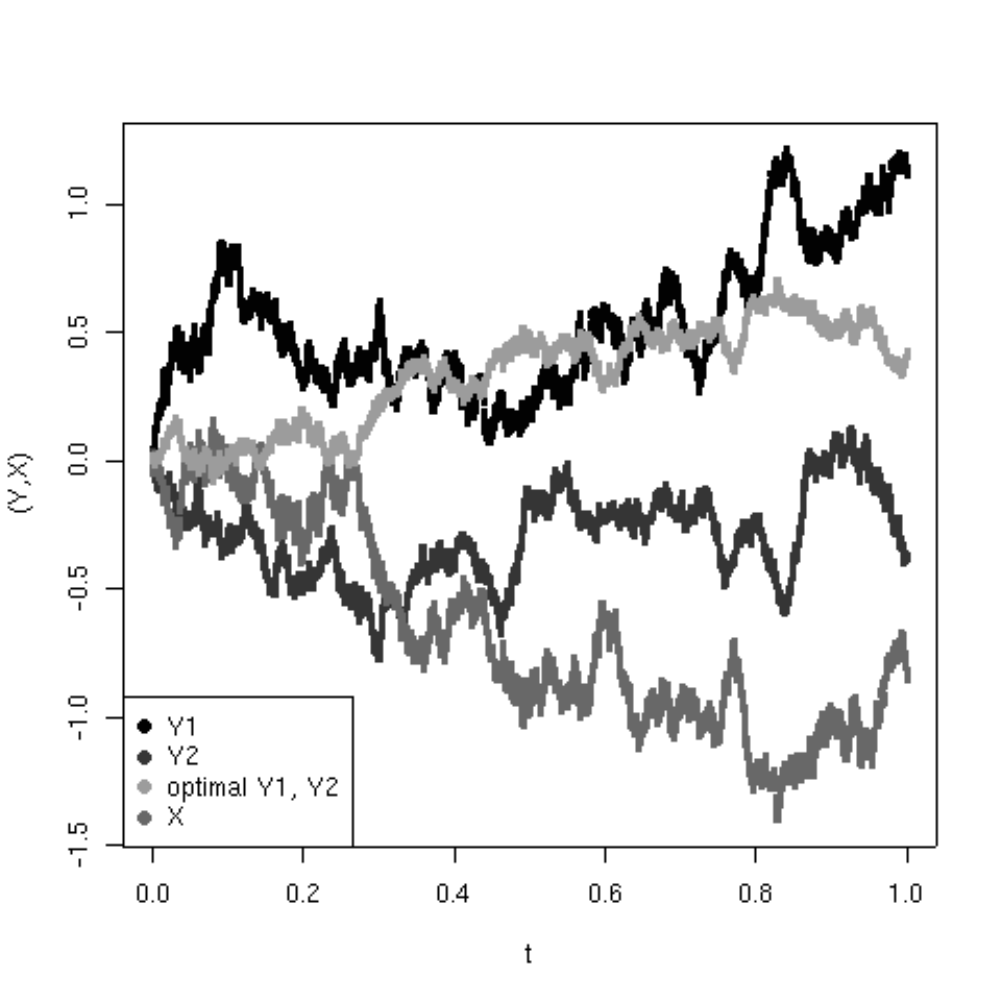}
\includegraphics[width=0.3\textwidth]{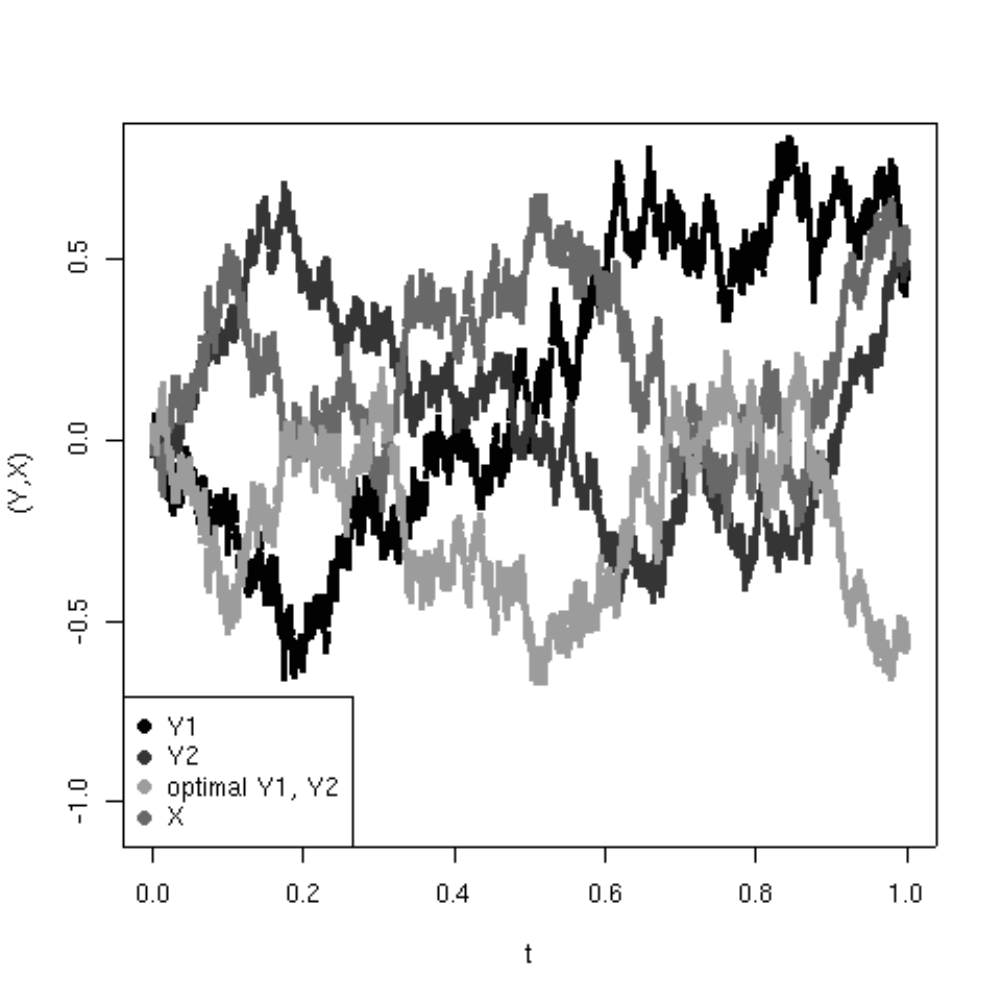}
\includegraphics[width=0.3\textwidth]{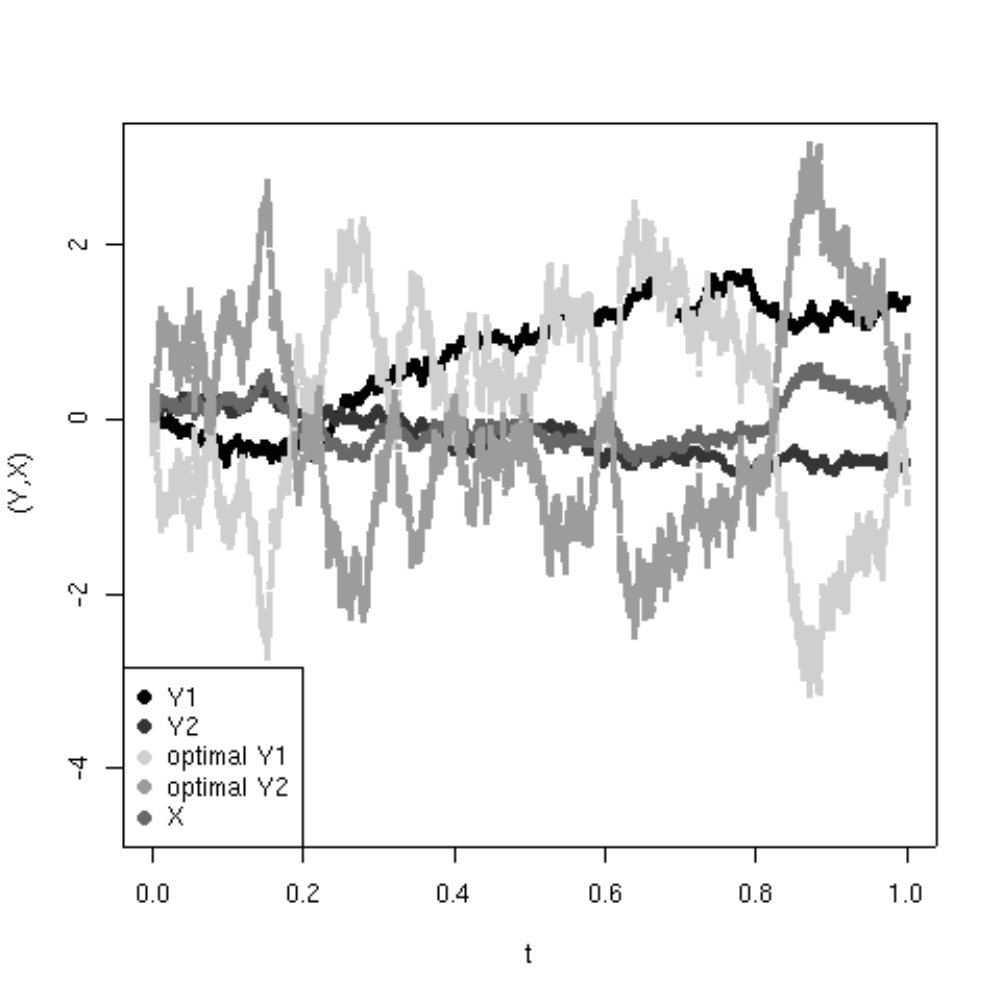}
\caption[]{
Simulations for different parameters \newline
A: $\mathbf{A}=[\begin{array}{cc} 1 & 0 \\ 0 & 1\end{array}]$ $\mathbf{B}=[\begin{array}{c} 1 \\ 1\end{array}]$ $\mathbf{\Sigma}^{yy}=[\begin{array}{cc} 1 & -0.5\\ 0 & 1\end{array}]$ \newline
B: $\mathbf{A}=[\begin{array}{cc} 0.1 & -0.001 \\ -0.01 & 1\end{array}]$ $\mathbf{B}=[\begin{array}{c} 1 \\ 1\end{array}]$ $\mathbf{\Sigma}^{yy}=[\begin{array}{cc} 1 & -0.5\\ 0 & 1\end{array}]$  \newline
C: $\mathbf{A}=[\begin{array}{cc} 0.01 & -0.001 \\ -0.001 & 0.1\end{array}]$ $\mathbf{B}=[\begin{array}{c} 0.001 \\ 0.001\end{array}]$ $\mathbf{\Sigma}^{yy}=[\begin{array}{cc} 1 & 0\\ 0 & 1\end{array}]$  \newline
D: $\mathbf{A}=[\begin{array}{cc} 1 & 0.999 \\ 0.999 & 1\end{array}]$ $\mathbf{B}=[\begin{array}{c} 1 \\ 1\end{array}]$ $\mathbf{\Sigma}^{yy}=[\begin{array}{cc} 0.894 & -0.447\\ 0 & 0.775\end{array}]$  \newline
E: $\mathbf{A}=[\begin{array}{cc} 1 & -0.999 \\ -0.999 & 1\end{array}]$ $\mathbf{B}=[\begin{array}{c} 0.001 \\ 0.001\end{array}]$ $\mathbf{\Sigma}^{yy}=[\begin{array}{cc} 0.894 & -0.447\\ 0 & 0.775\end{array}]$  \newline
F: $\mathbf{A}=[\begin{array}{cc} 1 & 0.999 \\ 0.999 & 1\end{array}]$ $\mathbf{B}=[\begin{array}{c} 0.005 \\ -0.005\end{array}]$ $\mathbf{\Sigma}^{yy}=[\begin{array}{cc} 0.894 & -0.447\\ 0 & 0.775\end{array}]$  \newline 
Other simulation examples can be found in Paper II.
}\label{figParSim}
\end{figure}

In Figure \ref{figParSim} we present a number of simulations of the trajectories of the multivariate 
Ornstein--Uhlenbeck Brownian Motion model for different parameter
values. More examples of simulations can be found in the Paper II.
In all of these cases $\mathbf{\Sigma}^{xx}=1$, $\vec{Y}(0)=\vec{0}$, $\vec{\psi}=\vec{0}$ and
$\vec{X}(0)=0$.

\subsubsection{Stationary regime of the mvOUBM model}
All of the discussion about the asymptotic properties of the multivariate Ornstein--Uhlenbeck Brownian Motion model naturally also apply to the univariate one. 
In particular, there is no stationary regime
as the covariance of the process escapes to infinity linearly in time (due to the ``$\vec{X}$'' variables evolving marginally as Brownian motion).
However, given that $\mathbf{A}$ has positive real part eigenvalues ,
the process $\vec{Y}-\mathbf{A}^{-1}\mathbf{B}\vec{X}-\vec{\psi}$ (or $y-\vec{b}\vec{X}-\psi$ in the univariate case)

tends (in distribution) to a mean $\vec{0}$ normal random variable with covariance 
\begin{displaymath}
\begin{array}{c}
\mathbf{P}\left(
[\frac{1}{\lambda_{i} + \lambda_{j}}]_{1\le i,j \le k_{Y}} \ast \left(
\mathbf{P}^{-1}\left(
\mathbf{\Sigma}_{11} + \mathbf{\Sigma}_{12}\mathbf{B}^{T}\mathbf{A}^{-T}
+\mathbf{A}^{-1}\mathbf{B}\mathbf{\Sigma}_{21} 
 \right.\right.\right. \\ \left.\left.\left. +
\mathbf{A}^{-1}\mathbf{B}\mathbf{\Sigma}_{22}\mathbf{B}^{T}\mathbf{A}^{-T}
\right)\mathbf{P}^{-T}
 \right)
\right)
\mathbf{P}^{T}.
\end{array}
\end{displaymath}
Unlike in the Ornstein--Uhlenbeck type model the covariance between traits in different
species after divergence does not disappear but tends to a constant, due to the covariance remaining
from the driving Brownian motion, 
\begin{displaymath}
\begin{array}{rcl}
\cov{\vec{Y}_{i}}{\vec{Y}_{j}}(t) & \rightarrow &  t_{a_{ij}}\mathbf{A}^{-1}\mathbf{B}\mathbf{\Sigma}_{22}\mathbf{B}^{T}\mathbf{A}^{-T}, \\
\cov{\vec{Y}_{i}}{\vec{X}_{j}}(t) & \rightarrow & -t_{a_{ij}}\mathbf{A}^{-1}\mathbf{B}\mathbf{\Sigma}_{22}, \\
\cov{\vec{X}_{i}}{\vec{X}_{j}}(t) & = & t_{a_{ij}} \mathbf{\Sigma}_{22} .
\end{array}
\end{displaymath}

\subsection{Relations between the models}
We can discuss e a hierarchy amongst the models.
In all of the presented cases the distribution of the observed data is multivariate normal.
The simplest model is independent observations of predictors with equal mean $\vec{\mu}$ and (co)variance
$\mathbf{\Sigma}$. The distribution of the data is then multivariate normal 
$\mathcal{N}(\mathbf{1}\otimes \vec{\mu},\mathbf{I} \otimes \mathbf{\Sigma})$. 

We can see that this is 
a special case of the Brownian motion model with $\mathbf{T}=\mathbf{I}$ as the distribution of the
data under the Brownian motion model is
$\mathcal{N}(\mathbf{1}\otimes \vec{X}(0),\mathbf{T} \otimes \mathbf{\Sigma})$, where
$\mathbf{T}$ is defined by the phylogeny as the matrix of times of the most recent common ancestor 
for each pair of species. 

In turn the Brownian motion model can be considered as a submodel of the
multivariate Ornstein--Uhlenbeck Brownian Motion
model. If in the multivariate Ornstein--Uhlenbeck Brownian Motion 
model we will have $\mathbf{A}=\mathbf{0}$ and
$\mathbf{B}=\mathbf{0}$ then the stochastic differential equation describing the process will be
\begin{displaymath}
\begin{array}{rcl}
\left[\begin{array}{c} \ud \vec{Y}(t) \\ \ud \vec{X}(t) \end{array} \right] & = &
\mathbf{0} \ud t + 
\left[\begin{array}{cc}  \mathbf{\Sigma}^{yy}& \mathbf{\Sigma}^{yx} \\ \mathbf{\Sigma}^{xy}& \mathbf{\Sigma}^{xx}\end{array}\right] \ud \mathbf{W}(t)
\end{array}
\end{displaymath}
and the data distribution is
$\mathcal{N}(\mathbf{1}\otimes [\vec{Y}(0)~\vec{X}(0)]^{T},\mathbf{T} \otimes \left[\begin{array}{cc}  \mathbf{\Sigma}^{yy}& \mathbf{\Sigma}^{yx} \\ \mathbf{\Sigma}^{xy}& \mathbf{\Sigma}^{xx}\end{array}\right]$.
Notice that we have to derive this from the stochastic differential equation formulation and not from the covariance structure
of the process as to calculate it we assumed that $\mathbf{A}$ is invertible.

The multivariate Ornstein--Uhlenbeck Brownian Motion 
model is in turn a special case of the generalized (to arbitrary drift matrix)
Ornstein--Uhlenbeck model, by setting the appropriate ``bottom'' (if we order the variables) rows to zero. 
This model is in turn a special case of the general model presented in \cite{EMarTHan97}
where one does not assume that $\mathbf{F}$ needs to be invertible.

Then the most general class is the general linear model where we assume any data covariance
matrix $\mathbf{V}$ and some superstructure (on the one from the previous models) of the 
mean values.

One can also consider relation between the models by looking at the different classes
of $\mathbf{F}$ (or respectively $\mathbf{A}$). The simplest one is when we assume the matrix has a single value 
on its diagonal. Then we can consider any diagonal matrix. If we assumed that the diagonal values have to be positive
then the next more general class is the class of symmetric positive definite matrices, then the
class of matrices with positive eigenvalues and then the general family of invertible matrices. 
This sort of hierarchy of models allows one to see which model fits better to the data
via \emph{e.g} the log--likelihood ratio test. There are of course other model selection
criteria like the AIC, AIC$_{c}$, BIC \emph{e.t.c.} One can connect biological hypothesis
with the different models (or matrix classes) and the model comparison
criteria tell us how each hypothesis is favoured. 

If we define by $k$ the number of parameters in the model, by $n$ the number
of observed values (number of species multiplied by number of variables minus the number of missing values) 
and by $\mathcal{L}$ the log--likelihood then the different information criteria
will have the following formulae,
\begin{displaymath}
\begin{array}{rcl}
\mathrm{AIC} & = & 2 \cdot \mathcal{L} + 2\cdot k \\
\mathrm{AIC}_{c} & = & 2 \cdot \mathcal{L} + 2\cdot k + \frac{2\cdot k \cdot (k+1)}{n-k-1} = \mathrm{AIC} + \frac{2\cdot k \cdot (k+1)}{n-k-1}\\
\mathrm{BIC} & = & 2 \cdot \mathcal{L} + k \cdot \log n.
\end{array}
\end{displaymath}

\newpage
\section{Supplementary material C: Matrix pa\-ra\-met\-riz\-ations used in \texttt{mvSLOUCH} package (Paper II)}\label{secPapIIC} 
To be able to work with estimating the parameters we have to be able to move around the space of decomposable matrices which
entails the parametrization of the $\mathbf{F}$/$\mathbf{A}$ matrix. There are two possibilities that all the eigenvalues are real or that 
some are complex. Since
the matrix has $k_{Y}^{2}$ elements we aim to parametrize it by at most $k_{Y}^{2}$ variables.
Its worth noting the paper \cite{JPinDBat} as it considers parametrizations of 
covariance matrices (symmetric, positive--definite real).
\subsection{Real eigenvalues case}
Since we assumed that none of eigenvalues are $0$, $\mathbf{P}$ is invertible and  we can do a unique 
(if the elements of $\mathbf{R}$'s diagonal are forced to be positive) QR decomposition 
of $\mathbf{P}$ \cite{RHorCJoh} (page 112). We have that $\mathbf{P}=\mathbf{Q}\mathbf{R}$ where $\mathbf{Q}$ is an orthogonal matrix and $\mathbf{R}$ is upper triangular.
Since the \texttt{R} software returns each eigenvector to be of unit length ($<x,x>=x^{T}x$) it was noticed that it returns the
$\mathbf{R}$ matrix as such that its columns are also of unit length.
\begin{Lemma}
If $\mathbf{P}$ is invertible and such that its columns are of unit length ($<x,x>=x^{T}x=1$) and if 
in its QR decomposition $\mathbf{P}=\mathbf{Q}\mathbf{R}$, $\mathbf{R}$'s diagonal elements are forced to be positive then $\mathbf{R}$'s columns
will also be of unit length.
\end{Lemma}
\textbf{Proof} 
Denote the elements of the matrices as, $\mathbf{P}_{ij}=p_{ij}$, $\mathbf{Q}_{ik}=q_{ik}$ and
$\mathbf{R}_{kj}=r_{kj}$, notice that from the properties of QR decomposition,
$\sum_{i=1}^{k_{Y}}q_{ik}q_{il}=1$ iff $l=k$, otherwise $0$ and for $k>j$ $r_{kj}=0$. 
Now consider the length of the $j$--th column of $\mathbf{P}$,
\begin{displaymath}
\begin{array}{l}
1=\mathbf{P}_{\cdot j}^{T}\mathbf{P}_{\cdot j}=\sum_{i=1}^{k_{Y}}p_{ij}^{2}=\sum_{i=1}^{k_{Y}}(\sum_{k=1}^{j}q_{ik}r_{kj})^{2}
\\=
\sum_{i=1}^{k_{Y}}\sum_{k_{1}=1}^{j}\sum_{k_{2}=1}^{j}q_{ik_{1}}q_{ik_{2}}r_{k_{1}j}r_{k_{2}j}
\\=
\sum_{k_{1}=1}^{j}\sum_{k_{2}=1}^{j}r_{k_{1}j}r_{k_{2}j}\sum_{i=1}^{k_{Y}}q_{ik_{1}}q_{ik_{2}}=
\sum_{k=1}^{j}r_{kj}^{2}=\mathbf{R}_{\cdot j}^{T}\mathbf{R}_{\cdot j},
\end{array}
\end{displaymath}
which shows that all columns of $\mathbf{R}$ are also of unit length.
\begin{flushright} \emph{Q.E.D.} \end{flushright}
We also know that an orthogonal matrix $\mathbf{Q}$ of size $k_{Y} \times k_{Y}$ can be represented by $k_{Y}(k_{Y}-1)/2$ Givens 
rotations \cite{TAndIOlkLUnd} (another approach to parametrization is in \cite{GSte}). 
Each Givens rotation can be represented by one value \cite{GGolCvanLoan}, since
we assume that $c^{2}+s^{2}=1$ (see \cite{GGolCvanLoan} for meaning of notation), and the position pair $(i,j)$ in $\mathbf{P}$ which it is meant to zero. We therefore 
have a representation by $n$ eigenvalues, $k_{Y}(k_{Y}-1)/2$ numbers representing the Givens rotations of $\mathbf{Q}$
and $k_{Y}(k_{Y}-1)/2$ values from above $\mathbf{R}$'s diagonal (since we can calculate the diagonal values as each
column is of unit length). When generating $\mathbf{R}$ its $j$th column has to be generated from the $j$ dimensional unit
sphere with the diagonal value being positive. 
And so we have a parametrization with $k_{Y}^{2}$ parameters.

\subsection{Complex eigenvalues case}
In the complex case the situation is not so simple as the QR decomposition as above would require $2k_{Y}^{2}$
parameters if applied directly. We have to try to take advantage of two facts; that if a complex number is
an eigenvalue then its conjugate will also be (so we have $k_{Y}$ parameters describing the eigenvalues)
and that if we have a complex eigenvector its conjugate will also be an eigenvector (and so $\mathbf{P}$ 
will have 
$k_{Y}^{2}$ 
distinct elements). A problem with the QR approach is that the \texttt{R} implementation \texttt{qr()}
does not work perfectly in the complex case. 
When $\mathbf{P}^{T}$ was decomposed by \texttt{R} and then the product $\mathbf{Q}^{T}\mathbf{R}^{T}$, of the calculated 
$\mathbf{Q}^{T}$ and $\mathbf{R}^{T}$ matrices was done, we did not receive $\mathbf{P}^{T}$ but a matrix where $\mathbf{P}^{T}$'s rows were
permuted. Another approach is to use singular value decomposition,
decomposing $\mathbf{P}$ as $\mathbf{P}=\mathbf{W}\mathbf{D}\mathbf{V}^{\ast}$, where $\mathbf{W}$ and $\mathbf{V}$ are
unitary, $\mathbf{D}$ is a diagonal matrix with positive real values and $^{\ast}$ is the conjugate transpose. 

It was noticed using \texttt{R}'s \texttt{svd()} function that if $\mathbf{P}$'s columns are of unit length
and for each complex column it has, it also has its conjugate then all values of $\mathbf{W}$ 
are real (so we have an orthogonal matrix and using Givens rotations we can parametrize it
by $k_{Y}(k_{Y}-1)/2$ numbers) and that for every complex row of $\mathbf{V}$, $\mathbf{V}$ also contains its conjugate.
It is obvious that with these conditions we will get a matrix $\mathbf{P}$ of the desired properties, but
it is not so obvious that this is a situation which will exist for every $\mathbf{P}$. For this
we would need to prove a statement as below.
\begin{prop}
If $\mathbf{P}$ is invertible and such that its columns are of unit length ($<x,x>=x^{T}x$=1) 
and for every complex column it also contains its conjugate then there exists
a singular value decomposition $\mathbf{P}=\mathbf{W}\mathbf{D}\mathbf{V}^{\ast}$ of it such that $\mathbf{W}$ contains
only real entries and for each complex row of $\mathbf{V}$, $\mathbf{V}$ also contains its conjugate.
\end{prop}
Since $\mathbf{V}$ is unitary it is parametrizable by $k_{Y}(k_{Y}-1)/2$ complex Givens rotations (a discussion
of the appropriate definition of a complex Givens rotation is in \cite{DBinJDemWKah}). But this would
give us $k_{Y}(k_{Y}-1)$ parameters. One should then utilize the fact that for each complex row we have its
complex conjugate, and parametrize $\mathbf{V}^{T}$. 

The poor man's approach is to work
on writing out the formulae for the series of Givens rotations and then using this system of linear
equations (it is linear in $\mathbf{V}$'s elements, since columns are of unit length, there
are $k_{Y}^{2}-k_{Y}$ of them) to represent the $\mathbf{V}$ matrix.
Other options of parametrizing unitary matrices are in \emph{e.g.} \cite{CJar06}, \cite{CJar05}, \cite{PDit94}, \cite{PDit03} but
they are of limited interest for now since they consider an arbitrary unitary matrix.

It remains to deal with the singular value matrix $\mathbf{D}=\mathtt{diag(d_{i},k_{Y},k_{Y})}$. Once we know
$\mathbf{W}=[w_{ij}]$ and $\mathbf{V}=[v_{ij}]$ and use the fact that $\mathbf{P}$'s columns are of unit length and polar decomposition 
I believe that we can discard $\mathbf{D}$ as I believe it should be calculable from the other matrices.

In polar decomposition $\mathbf{P}=\mathbf{U}\mathbf{F}$, where $\mathbf{U}$ is unitary while $\mathbf{F}$ is positive--definite Hermitian, which makes it symmetric 
with positive real values on the diagonal. We have the relations $\mathbf{F}=\mathbf{V}\mathbf{D}\mathbf{V}^{\ast}$ and $\mathbf{U}=\mathbf{W}\mathbf{V}^{\ast}$.
Doing the multiplications $\mathbf{P}=\mathbf{W}\mathbf{D}\mathbf{V}^{\ast}$ and $\mathbf{F}=\mathbf{V}\mathbf{D}\mathbf{V}^{\ast}$ we get that for each $i$,
\begin{displaymath}
\sum_{k=1}^{n}\sum_{l=1}^{n} v_{ik} v_{il} d_{k} d_{l} \sum_{r=1}^{n} \bar{v}_{kr} \bar{v}_{lr} = 1
\end{displaymath}
subject to the constraints
\begin{displaymath}
\begin{array}{c}
d_{i} \in \mathbb{R} \\
d_{i} > 0 \\
\sum_{k=1}^{n} v_{ik} \bar{v}_{ki} d_{k} > 0,
\end{array}
\end{displaymath}
the last constraint coming from that $\mathbf{F}$'s diagonal elements must be real. It remains to be studied how
this equation behaves, how many solutions does it have, do the constraints force one solution and 
does a solution exist for arbitrary orthogonal $\mathbf{U}$ and unitary $\mathbf{V}$ pairs.

\subsection{Matrix parametrization implemented in package}
An important part of the program is to allow the user to choose a desired type of $\mathbf{F}$/$\mathbf{A}$ matrix (size $k_{Y}$ by $k_{Y}$),
the following ones are currently implemented,\\
\texttt{SingleValueDiagonal}\\
A diagonal matrix with just a single value along the diagonal is represented by this number.
\\
\texttt{Diagonal}\\
A diagonal matrix is represented by a vector which contains this diagonal.
\\
\texttt{Symmetric}\\
A symmetric matrix is represented by a vector which contains its upper triangular part, in \texttt{R} code \\ \texttt{A[upper.tri(A,diag=T)]<-v}.
\\
\texttt{SymmetricPositiveDefinite}\\
A symmetric positive definite
is represented exactly as in \texttt{ouch} (using \texttt{ouch}'s \texttt{sym.par()}), 
that is by a vector which contains its Cholesky decomposition.
In \texttt{R} code \texttt{X[lower.tri(X,diag=T)]<-v;A=X\%*\%t(X)}.
\\
\texttt{TwoByTwo}\\
An arbitrary $2 \times 2$ matrix is stored as \texttt{A<-matrix(v,2,2)}.
\\
\texttt{UpperTri}\\
An upper triangular matrix is stored as \texttt{A[upper.tri(A,diag=T)]<-v}.
\\
\texttt{LowerTri}\\
A lower triangular is stored as \texttt{A[lower.tri(A,diag=T)]<-v}.
\\
\texttt{DecomposablePositive}\\
We now consider storing a 
decomposable matrix with real positive eigenvalues. It is stored in a vector of 
length $k_{Y}^2$. The first $k_{Y}$ values store the logarithm of the eigenvalues. The remaining
$k_{Y}^2-k_{Y}$ numbers code the eigenvector matrix $\mathbf{P}$. The matrix is coded by $\mathbf{P}$'s QR decomposition
which is unique as $\mathbf{P}$ is invertible and $\mathbf{R}$'s diagonal is assumed to be positive. The coding
is done in the following manner, $\mathbf{P}=\mathbf{Q}\mathbf{R}$, where $\mathbf{Q}$ is an orthogonal matrix and $\mathbf{R}$ is upper triangular. 
We know that $\mathbf{R}$'s columns are of unit length. This is because the eigenvectors ($\mathbf{P}$'s columns)
are of unit length and $\mathbf{Q}$ is orthogonal. Therefore since $\mathbf{R}$'s diagonal is positive
it is determined by the values of $\mathbf{R}$ above the diagonal. $\mathbf{Q}$ is coded by $k_{Y}(k_{Y}-1)/2$ Givens
rotations. $\mathbf{R}$'s columns are coded in the following manner, we know every value has to be between $[-1,1]$.
\begin{algorithm}{
$\mathbf{R}_{ij}$ from $v$}
\begin{algorithmic}
\IF {$i==1$ and $j==1$} 
\STATE $\mathbf{R}_{11}=1$
\ELSIF{$j>i$}
\IF{$i==1$} 
\STATE $\mathbf{R}_{ij}=\frac{2}{\pi}\mathrm{atan}(v)$
\ELSE
\STATE $a=\sqrt{1-\sum_{k=1}^{i-1}\mathbf{R}_{kj}^2}$
\STATE $\mathbf{R}_{ij}=\frac{2a}{\pi}\mathrm{atan}(v)$
\ENDIF
\ELSIF {$j==i$}
\STATE $\mathbf{R}_{ii}=\sqrt{1-\sum_{k=1}^{i-1}\mathbf{R}_{ki}^2}$
\ELSE
\STATE $\mathbf{R}_{ij}=0$
\ENDIF
\STATE \textbf{return} $\mathbf{R}_{ij}$
\end{algorithmic}
\end{algorithm}                                
We can see that this gives, that the matrix is parametrized by $k_{Y}^{2}$ values.
\\
\texttt{DecomposableNegative}\\
The procedure is exactly the same as in the decomposable positive case, except that the eigenvalues are stored as the logarithm of them negated.
\\
\texttt{DecomposableReal}\\
The procedure is exactly the same as in the decomposable positive case, except that the eigenvalues are stored directly.
\\
\texttt{Invertible}\\
The storing of the matrix as decomposable real does not guarantee it to be invertible (it could have an eigenvalue of $0$)
and more importantly does not allow for a complex eigendecomposition. Storing an
invertible matrix is done by QR decomposition, $\mathbf{F}$/$\mathbf{A}=\mathbf{Q}\mathbf{R}$. $\mathbf{Q}$ as before is stored by $k_{Y}(k_{Y}-1)/2$ Givens rotations. 
$\mathbf{R}$'s diagonal is stored as its logarithm (to force it be non--zero, for it to be invertible, and positive as assumed) and the rest are stored as,
\texttt{R[upper.tri(R,diag=F)]<-v}.
\\~\\
In addition the user can force all elements of the parametrized matrix's diagonal to be positive/negative. This is done after transforming the parameter vector into 
the matrix by exponentiating the diagonal of $\mathbf{A}$.
The matrix $\mathbf{\Sigma}^{yy}$ can be parametrized as either \texttt{SingleValueDiagonal}, \texttt{Diagonal}, \texttt{Symmetric}, 
\texttt{UpperTri}m \texttt{LowerTri} or \texttt{Any}, \\ $\mathbf{\Sigma}^{yy}$ 
\texttt{<-matrix(v,nrow=}$k_{Y}$\texttt{,ncol=}$k_{Y}$\texttt{)}.

\newpage
\bibliography{mvslouch,compbio,sde,appendixMError,lic,LinearAlgebra,phdProjBishopsWarblers,matrix,compmethod}

\begin{thebibliography}{10}

\bibitem{MAicCRit}
M.~Ackin and C.~Ritenbaugh.
\newblock Analysis of multivariate reliability structures and the induceed bias
  in linear model estimation.
\newblock {\em Statistics in Medicine}, 16:1647--1661, 1996.

\bibitem{TAndIOlkLUnd}
T.~W. Anderson, I.~Olkin, and L.~G. Underhill.
\newblock Generation of random orthogonal matrices.
\newblock {\em SIAM Journal on Scientific and Statistical Computing},
  8(4):625--629, 1987.

\bibitem{CBal1967I}
C.~S. Ballantine.
\newblock Products of positive definite matrices {I}.
\newblock {\em Pacific Journal of Mathematics}, 23(3):427--433, 1967.

\bibitem{CBal1968II}
C.~S. Ballantine.
\newblock Products of positive definite matrices {II}.
\newblock {\em Pacific Journal of Mathematics}, 24(1):7--17, 1968.

\bibitem{CBal1968III}
C.~S. Ballantine.
\newblock Products of positive definite matrices {III}.
\newblock {\em Journal of Algebra}, 10(1):174--182, 1968.

\bibitem{CBal1970IV}
C.~S. Ballantine.
\newblock Products of positive definite matrices {IV}.
\newblock {\em Linear Algebra and Its Applications}, 3(1):79--114, 1970.

\bibitem{DBinJDemWKah}
D.~Bindel, J.~Demmel, and W.~Kahan.
\newblock On computing {G}ivens rotations reliably and efficiently.
\newblock {\em ACM Transactions on Mathematical Software}, 28(2):206--238,
  2002.

\bibitem{BoxJen}
G.~Box and G.~M. Jenkins.
\newblock {\em Time Series Analysis}.
\newblock Holden--Day, San Francisco, 1970.

\bibitem{JBuo}
J.~P. Buonaccorsi.
\newblock {\em Measurement Error Models, Methods and Applications}.
\newblock CRC Press, 2010.

\bibitem{MButAKinOUCH}
M.~A. Butler and A.~A. King.
\newblock Phylogenetic comparative analysis: a modelling approach for adaptive
  evolution.
\newblock {\em The American Naturalist}, 164(6):683--695, 2004.

\bibitem{NDraHSmi}
N.~R. Drapper and H.~Smith.
\newblock {\em Applied Regression Analysis}.
\newblock Wiley, New York, 1998.

\bibitem{LEva}
L.~C. Evans.
\newblock {\em An Introduction to Stochastic Differential Equations Version
  1.2}.
\newblock Department of Mathematics UC Berkley.

\bibitem{JFel85}
J.~Felsenstein.
\newblock Phylogenies and the comparative method.
\newblock {\em The American Naturalist}, 125(1):1--15, 1985.

\bibitem{RFrePHarMPag2002}
R.~P. Freckleton, P.~H. Harvey, and M.~Pagel.
\newblock Phylogenetic analysis and comparative data: A test and review of
  evidence.
\newblock {\em The American Naturalist}, 160(6), 2002.

\bibitem{WFul1987}
W.~A. Fuller.
\newblock {\em Measurement Error Models}.
\newblock Wiley, 1987.

\bibitem{WFul1995}
W.~A. Fuller.
\newblock Estimation in the presence of measurment error.
\newblock {\em International Statistical Review}, 63(2):121--147, 1995.

\bibitem{CGar}
C.~W. Gardiner.
\newblock {\em Handbook of Stochastic Methods}.
\newblock Springer, Berlin, 2004.

\bibitem{TGarADicCJanJJon93}
T.~Garland, A.~W. Dickerman, C.~M. Janis, and J.~A. Jones.
\newblock Phylogenetic analysis of covariance by computer simulation.
\newblock {\em Systematic Biology}, 42(3):265--292, 1993.

\bibitem{LGle1992}
L.~J. Gleser.
\newblock The importance of assessing measurement reliability in multivariate
  regression.
\newblock {\em Journal of the American Statistical Association},
  87(419):696--707, 1992.

\bibitem{GGolCvanLoan}
G.~H. Golub and C.~F. {Van Loan}.
\newblock {\em Matrix Computations}.
\newblock The Johns Hopkins University Press, Baltimore, 1989.

\bibitem{AGra1989}
A.~Grafen.
\newblock The phylogenetic regression.
\newblock {\em Philosophical Transactions of the Royal Society of London B},
  326(1233):119--157, 1989.

\bibitem{GGriDSti}
G.~R. Grimmett and D.~R. Stirzaker.
\newblock {\em Probability and Random Processes}.
\newblock Oxford University Press, Oxford, 2001.

\bibitem{JGur53}
J.~Gurland.
\newblock Distribution of quadratic forms and ratios of quadratic forms.
\newblock {\em The Annals of Mathematical Statistics}, 24(3):416--427, 1953.

\bibitem{THan97}
T.~F. Hansen.
\newblock Stabilizing selection and the comparative analysis of adaptation.
\newblock {\em Evolution}, 51(5):1341--1351, 1997.

\bibitem{EMarTHan96}
T.~F. Hansen and E.~P. Martins.
\newblock Translating between microevolutionary process and macroevolutionary
  patterns: the correlation structure of interspecific data.
\newblock {\em Evolution}, 50(4):1404--1417, 1996.

\bibitem{THanJPieSOrzSLOUCH}
T.~F. Hansen, J.~Pienaar, and S.~H. Orzack.
\newblock A comparative method for studying adaptation to randomly evolving
  environment.
\newblock {\em Evolution}, 62:1965--1977, 2008.

\bibitem{PHarMPag}
P.~H. Harvey and M.~Pagel.
\newblock {\em The Comparative Method in Evolutionary Biology}.
\newblock Oxford University Press, Oxford, 1991.

\bibitem{RHorCJoh}
R.~A. Horn and C.~R. Johnson.
\newblock {\em Matrix Analysis}.
\newblock Cambridge University Press, Cambridge, 1985.

\bibitem{JHueBRan}
J.~P. Huelsenbeck and B.~Rannala.
\newblock Detecting correlation between characters in a comparative analysis
  with uncertain phylogeny.
\newblock {\em Evolution}, 57(6):1237--1247, 2003.

\bibitem{CJar06}
C.~Jarslkog.
\newblock Recursive parametrization and invariant phases of unitary matrices.
\newblock {\em Journal of Mathematical Physics}, 46(1), 2005.

\bibitem{CJar05}
C.~Jarslkog.
\newblock A recursive parametrization of unitary matrices.
\newblock {\em Journal of Mathematical Physics}, 46, 2005.

\bibitem{CJoh70}
C.~R. Johnson.
\newblock Positive definite matrices.
\newblock {\em The American Mathematical Monthly}, 77(3):259--264, 1970.

\bibitem{ALabJPieTHan}
A.~Labra, J.~Pienaar, and T.~F. Hansen.
\newblock Evolution of thermal physiology in \emph{Liolaemus} lizards:
  Adaptation, phylogenetic inertia, and niche tracking.
\newblock {\em The American Naturalist}, 174(2):204--220, 2009.

\bibitem{EMarTHan97}
E.~P. Martins and T.~F. Hansen.
\newblock Phylogenis and the comparative method: a general approach to
  incorporating phylogenetic information into the analysis of interspecific
  data.
\newblock {\em The American Naturalist}, 149(4):646--667, 1997.

\bibitem{CMolCVanLoan}
C.~Moler and C.~F. {Van Loan}.
\newblock Nineteen dubious ways to compute the exponential of a matrix,
  twenty--five years later.
\newblock {\em SIAM Review}, 45(1):1--37, 2003.

\bibitem{MPag1993}
M.~Pagel.
\newblock Seeking the evolutionary regression coefficient: An analysis of what
  comparative methods measure.
\newblock {\em Journal of Theoretical Biology}, 164:191--205, 1993.

\bibitem{MPag1994}
M.~Pagel.
\newblock Detecting correlated evolution on phylogenies: A general method for
  the comparative analysis of discrete characters.
\newblock {\em Proceedings of the Royal Society B}, 255:37--45, 1994.

\bibitem{MPag19992}
M.~Pagel.
\newblock Inferring the historical patterns of biological evolution.
\newblock {\em Nature}, 401:877--884, 1999.

\bibitem{MPag19991}
M.~Pagel.
\newblock The maximum likelihood approach to reconstructing ancestral character
  states of discrete characters on phylogenies.
\newblock {\em Systematic Biology}, 48(3):612--622, 1999.

\bibitem{EPar2006}
E.~Paradis.
\newblock {\em Analysis of Phylogenetics and Evolution with R}.
\newblock Springer, New York, 2006.

\bibitem{EParJCla2002}
E.~Paradis and J.~Claude.
\newblock Analysis of comparative data using generalized estimating equations.
\newblock {\em Journal of Theoretical Biology}, 218:175--185, 2002.

\bibitem{JPieKBarTHanKVoj}
J.~Pienaar, K.~Bartoszek, T.~F. Hansen, and K.~Voje.
\newblock Overview of comparative methods for studying adaptation on adaptive
  landscapes.
\newblock in prep.

\bibitem{JPinDBat}
J.~C. Pinheiro and D.~M. Bates.
\newblock Unconstrained parametrizations for variance--covariance matrices.
\newblock {\em Statistics and Computing}, 6:289--296, 1996.

\bibitem{FPlaCBonJGai}
F.~Plard, C.~Bonenfant, and J.~M. Gaillard.
\newblock Revisiting the allometry of antlers among deer species: male--male
  sexual competition as a driver.
\newblock {\em Oikos}, 120:601--606, 2011.

\bibitem{R}
{R Development Core Team}.
\newblock {\em R: A Language and Environment for Statistical Computing}.
\newblock R Foundation for Statistical Computing, Vienna, Austria, 2010.
\newblock {ISBN} 3-900051-07-0.

\bibitem{MSmi89}
M.~D. Smith.
\newblock On the expectation of a ratio of quadratic forms in normal variables.
\newblock {\em Journal of Multivariate Analysis}, 31:244--257, 1989.

\bibitem{MSmi93}
M.~D. Smith.
\newblock Expectations of a ratios of quadratic forms in normal variables:
  evaluating some top--order invariant polynomials.
\newblock {\em Australian Journal of Statistics}, 35(3):271--282, 1993.

\bibitem{MSmi96}
M.~D. Smith.
\newblock Comparing approximations to the expectations of a ratio of quadratic
  forms in normal variables.
\newblock {\em Econometric Reviews}, 15(1):81--95, 1996.

\bibitem{GSte}
G.~W. Stewart.
\newblock The efficient generation of random orthogonal matrices with an
  application to condition estimators.
\newblock {\em SIAM Journal of Numerical Analysis}, 17(3):403--409, 1980.

\bibitem{PDit94}
P.~Di\c t\u a.
\newblock On the parametrization of unitary matrices by the moduli of their
  elements.
\newblock {\em Communications in Mathematical Physics}, 159:581--591, 1994.

\bibitem{PDit03}
P.~Di\c t\u a.
\newblock Factorization of unitary matrices.
\newblock {\em Journal of Physics A: Mathematical and General}, 36:2781--2789,
  2003.

\bibitem{CVanLoan77}
C.~F. {Van Loan}.
\newblock The sensitivity of the matrix exponential.
\newblock {\em SIAM Journal on Numerical Analysis}, 14(6):971--981, 1977.

\bibitem{CVanLoan78}
C.~F. {Van Loan}.
\newblock Computing integrals involving the matrix exponential.
\newblock {\em IEEE Transactions on Automatic Control}, 23(3):395--404, 1978.

\bibitem{AWen}
A.~Wentzell.
\newblock {\em Course in the Theory of Stochastic Processes}.
\newblock McGraw--Hill Inc, 1981.

\bibitem{DWil}
D.~Williams.
\newblock {\em Probability with Martingales}.
\newblock Cambridge University Press, Cambridge, 1991.

\bibitem{PWu1988}
P.~Y. Wu.
\newblock Products of positive semidefinite matrices.
\newblock {\em Linear Algebra and its Applications}, 111:53--61, 1988.

\end{thebibliography}
\bibliographystyle{plain}

%\fullpage{1-42}{./Papers/HansenBartoszekMain.pdf}
%\fullpage{./Papers/HansenBartoszekFigures.pdf}
%\fullpage{./Papers/HansenBartoszekSupplementaryA.pdf}
%\fullpage{./Papers/FiguresSupplementaryfileA.pdf}
%\fullpage{./Papers/HansenBartoszekSupplementaryB.pdf}
%\fullpage{./Papers/mvslouch.pdf}
%\fullpage{./Papers/mvslouchsupplement.pdf}

%\include{introduction}
%\part{PAPERS}
%\include{paperI}
%\include{chapterI}
%\include{paperII}
%\include{chapterII}
%\printglossary
%\maketitle

%\bibliographystyle{newapa}
%\bibliographystyle{apalike}
%\bibliographystyle{plainnat}
%\bibliography{bibpercolation}

\vspace{1cm}
\end{document}